\documentclass[acmsmall, screen]{acmart}

\AtBeginDocument{}

\setcopyright{cc}
\setcctype{by}
\acmDOI{10.1145/3720421}
\acmYear{2025}
\acmJournal{PACMPL}
\acmVolume{9}
\acmNumber{OOPSLA1}
\acmArticle{87}
\acmMonth{4}

\usepackage{fancyvrb}

\usepackage{framed}

\usepackage{enumitem}
\setlist[itemize]{leftmargin=0.2in, rightmargin=0.1in, itemsep=0.06in, topsep=0.04in}

\usepackage{wrapfig}

\usepackage{mathpartir}

\usepackage{pifont}\newcommand{\cmark}{\ding{51}}
\newcommand{\xmark}{\ding{55}}

\usepackage{multirow}

\graphicspath{{images/}}

\definecolor{BLUE}{RGB}{0, 0, 255} \definecolor{hiddenArgsGray}{RGB}{150, 150, 150}

\newcommand{\parahead}[1]
  {\paragraph{\textbf{#1}}}

\newcommand{\refAppendixHtmlFlatNested}
  {\autoref{sec:html-flat-nested}~\cite{Hass}}
\newcommand{\refAppendixTranslationToCss}
  {\autoref{sec:translation-to-css}~\cite{Hass}}
\newcommand{\refAppendixStyleComputation}
  {\autoref{sec:style-computation}~\cite{Hass}}
\newcommand{\refAppendixHassExamples}
  {\autoref{sec:hass-examples}~\cite{Hass}}

\newcommand{\hass}
{\textsc{Hass}}
\newcommand{\hasskell}
{\textsc{Hasskell}}

\newcommand{\vsCode}{VS Code}
\newcommand{\codeMirror}{CodeMirror}

\newcommand{\htmlTag}[1]
  {\texttt{<#1>}}
\newcommand{\htmlAttr}[1]
{\texttt{#1}}
\newcommand{\cssAttr}[1]
{\texttt{#1}}
\newcommand{\cssClass}[1]
  {\texttt{#1}}

\newcommand{\set}[1]{\{#1\}}

\newcommand{\narrowcdots}
  {\cdot \hspace{-0.03in} \cdot \hspace{-0.03in} \cdot}
\newcommand{\narrowldots}
  {...} 

\newcommand{\varCls}{cls}
\newcommand{\varDataCon}{C}
\newcommand{\varExp}{e}
\newcommand{\varPat}{p}
\newcommand{\varGuardedRule}{\mathit{rule}}
\newcommand{\varBasicSelector}{b} \newcommand{\varNodeSelector}{s}
\newcommand{\varPathSelector}{S}
\newcommand{\varStyleSheet}
{\mathit{sheet}}
\newcommand{\varAttrs}{\mathit{styles}}
\newcommand{\varAttrName}{\mathit{attr}}
\newcommand{\varAttrVal}{\mathit{val}}
\newcommand{\varClasses}{\mathit{classes}}

\newcommand{\varTypeCon}{D}
\newcommand{\varTypeContext}{\Sigma} \newcommand{\varVar}{x}

\newcommand{\varDoc}{d}
\newcommand{\varStr}{\mathit{str}}
\newcommand{\varVal}{v}
\newcommand{\varEnv}{E}
\newcommand{\varPath}{\pi}

\newcommand{\varInternalRule}{\widehat{\varGuardedRule}}
\newcommand{\varPathRule}{\widehat{\varPathSelector}}
\newcommand{\varNodeRule}{\widehat{\varNodeSelector}}
\newcommand{\varPatternRule}{\widehat{\varPat}}

\newcommand{\varEnvStyles}
{\Pi}

\newcommand{\varNamedStyles}
{\mathcal{X}}

\newcommand{\styleSetOf}[1]
  {\varNamedStyles(#1)}

\newcommand{\applyStyles}[2]{\applyTwo{}{#1}{#2}}

\newcommand{\smallSep}
  {\hspace{0.02in}}

\newcommand{\sepPremise}{\hspace{0.20in}}
\newcommand{\hsepRule}{\hspace{0.20in}}

\newcommand{\vsepRuleHeight}{0.12in}
\newcommand{\vsepRule}{\vspace{\vsepRuleHeight}}

\newcommand{\dataApp}[2]{#1\smallSep#2}

\newcommand{\emptyEnv}{-}
\newcommand{\envBind}[2]{#1 \mapsto #2}
\newcommand{\envExtend}[2]{#1, #2}

\newcommand{\valTrue}{\texttt{True}}
\newcommand{\valRoot}{\varVal_{\mathit{root}}}

\newcommand{\leaf}[1]{\dataApp{\texttt{Leaf}}{#1}}
\newcommand{\node}[4]
{\dataApp{\texttt{Node}}{(#1, #3, #2, #4)}}

\newcommand{\noPath}{\texttt{Nothing}}
\newcommand{\justPath}[1]{\dataApp{\texttt{Just}}{#1}}

\newcommand{\pathOf}[1]{\dataApp{\texttt{path}}\!(#1)}
\newcommand{\classesOf}[1]{\dataApp{\texttt{classes}}\!(#1)}
\newcommand{\stylesOf}[1]{\dataApp{\texttt{styles}}\!(#1)}
\newcommand{\childrenOf}[1]{\dataApp{\texttt{children}}\!(#1)}

\newcommand{\copyUpdate}[3]
  {#1 \set{#2 = #3}}
\newcommand{\copyUpdateTwo}[5]
  {#1 \set{#2 = #3, #4 = #5}}

\newcommand{\clsSelector}[1]{.#1}
\newcommand{\namedPattern}[2]{#1\smallSep@\smallSep#2}
\newcommand{\hasType}[2]{#1\!::\!#2}
\newcommand{\keepOut}
{\bot}

\newcommand{\patSelector}[2]{{#1}{#2}}
\newcommand{\compoundSelector}[2]{{#1}{#2}}
\newcommand{\varPattern}[2]{\hasType{#1}{#2}}

\newcommand{\rangeN}[3]
{\langle#3\rangle\smallSep^{#1 \in [#2]}}
\newcommand{\rangeZeroN}[3]
{\langle#3\rangle\smallSep^{#1 \in [0,\hspace{0.01in}#2]}}
\newcommand{\rangeTwoN}[3]
  {\langle#3\rangle\smallSep^{#1 \in [2,\hspace{0.01in}#2]}}

\newcommand{\varOp}{\odot}
\newcommand{\opDescendant}
{>^+}
\newcommand{\opChild}{>}
\newcommand{\opSibling}{\sim}
\newcommand{\opNextSibling}{+}

\newcommand{\mergeTwo}[3]{\ensuremath{\mathit{merge}_{#1}(\hspace{-0.005in}#2,#3)}}
\newcommand{\applyTwo}[3]{\ensuremath{\mathit{apply}_{#1}(\hspace{-0.005in}#2,#3)}}

\newcommand{\mergeDocs}[2]
{\mergeTwo{}{#1}{#2}}

\newcommand{\topSelector}{\top}
\newcommand{\nodeRule}[2]{#1\ \set{#2}}
\newcommand{\pathRule}[3]{#1 #2 #3}
\newcommand{\guardedRule}[2]{#1\ \texttt{if}\ #2}
\newcommand{\guardedRuleArrow}[3]{\guardedRule{#1}{#2}\rightarrow{#3}}
\newcommand{\namedStyles}[4]{\rangeN{#1}{#2}{#3\ \set{#4}}}
\newcommand{\andStyles}[2]{\nodeRule{#1}{#2}}

\newcommand{\sepmidsep}
  {\hspace{0.10in}|\hspace{0.10in}}
\newcommand{\syntaxRow}[3]
  {\textbf{#1} & \ensuremath{#2} & \ensuremath{::=} & \ensuremath{#3}}
\newcommand{\nextRow}
  {\\[2pt]}
\newcommand{\nextRowMoreSpace}
  {\\[8pt]}

\newcommand{\JudgementBox}[1]{{
\fbox{\ensuremath{#1}}
}}
\newcommand{\relDescription}[1]{\ensuremath{\textrm{\textbf{#1}}}}
\newcommand{\judgementHead}[2]
  {\ensuremath{\relDescription{#1}\hfill\JudgementBox{#2}}}
\newcommand{\judgementHeadThree}[3]
  {\ensuremath{\relDescription{#1}\ \textrm{{#2}}\hfill\JudgementBox{#3}}}

\newcommand{\ruleName}[1]{\mbox{\textsc{\begin{normalsize}#1\end{normalsize}}}}
\newcommand{\ruleNameFig}[1]{\textsc{\begin{scriptsize}[#1]\end{scriptsize}}}

\newcommand{\applyStyleSheet}[2]{\applyTwo{}{#1}{#2}}

 \newcommand{\outputPair}[2]{{#1}\dashv{#2}} 

\newcommand{\applyEquals}[3]{\applyTwo{}{#1}{#2}\rightsquigarrow{#3}}

\newcommand{\applyStyleSheetEquals}[3]{\applyTwo{}{#1}{#2}\rightsquigarrow{#3}}
\newcommand{\applyPathRuleEquals}[4]{\applyTwo{}{#1}{#2}\rightsquigarrow\outputPair{#3}{#4}}
\newcommand{\applyGuardedRuleEquals}[3]{\applyTwo{}{#1}{#2}\rightsquigarrow{#3}}

\newcommand{\evalsTo}[3]
  {\ensuremath{{#1}\vdash{#2}\hspace{0.01in}\Downarrow{#3}}}

\newcommand{\patternMatch}[5]{#3 \rhd_{#1} #2 \dashv #4;#5}
\newcommand{\selectorMatch}[5]{#3 \rhd (#1,#2) \dashv #4;#5}
\newcommand{\selectorMatchInputOnly}[3]{#3 \rhd (#1,#2)}

\newcommand{\hiddenArgs}[1]{\ensuremath{{\color{hiddenArgsGray}#1\vdash\ }}}
\newcommand{\hiddenArgsTwo}[2]{\ensuremath{{\color{hiddenArgsGray}#1;#2\vdash\ }}}

\newcommand{\figBubble}[1]{\raisebox{-0.03in}[0pt]{\includegraphics[width=0.14in]{letters/#1.pdf}}}
\newcommand{\refBubble}[1]
  {~\figBubble{#1}}

\newcommand{\refBubbleSmall}[1]
  {~\raisebox{-0.028in}[0pt]{\includegraphics[width=0.13in]{letters/#1.pdf}}}

\newcommand{\inlineFig}[1]{\begingroup\normalfont
\raisebox{-0.01in}{\includegraphics[height=1.2\fontcharht\font`\B]{#1}}
  \endgroup
}

\newcommand{\caseBubble}[1]{\raisebox{-0.03in}[0pt]{\includegraphics[height=0.13in]{letters/#1.pdf}}}

\newcommand{\sba}{\raisebox{0.005in}[0pt]{\includegraphics[width=0.15in]{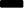}}}
\newcommand{\sbb}{\raisebox{-0.020in}[0pt]{\includegraphics[width=0.15in]{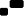}}}
\newcommand{\sbc}{\raisebox{-0.030in}[0pt]{\includegraphics[width=0.15in]{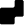}}}
 
\begin{document}

\title[Code Style Sheets]{Code Style Sheets: CSS for Code}

\author{Sam Cohen}
\orcid{0009-0008-9127-6518}
\email{samcohen@uchicago.edu}
\author{Ravi Chugh}
\orcid{0000-0002-1339-2889}
\email{rchugh@cs.uchicago.edu}
\affiliation{\institution{University of Chicago}
\city{Chicago}
  \state{IL}
  \country{USA}
  \postcode{60637}
}

\begin{abstract}

Program text is rendered using impoverished typographic styles.
Beyond choice of fonts and syntax-highlighting colors, code editors and related tools utilize very few text decorations.
These limited styles are, furthermore, applied in monolithic fashion, regardless of the programs and tasks at hand.

We present the notion of \emph{code style sheets} for styling program text. Motivated by analogy to cascading style sheets (CSS) for styling HTML documents, code style sheets provide mechanisms
for defining rules to select elements from an abstract syntax tree (AST) in order to style their corresponding visual representation.
Technically, our selector language generalizes essential constructs from CSS to a programming-language setting with algebraic data types (such as ASTs).
Practically, code style sheets allow ASTs to be styled granularly, based on semantic information---such as the structure of abstract syntax, static type information, and corresponding run-time values---as well as design choices on the part of authors and readers of a program.
Because programs are heavily nested in structure, a key aspect of our design is a layout algorithm that renders nested, multiline text blocks more compactly than in existing box-based layout systems such as HTML.

In this paper, we design and implement a code style sheets system for a subset of Haskell,
using it to illustrate several code presentation and visualization tasks.
These examples demonstrate that code style sheets provide a uniform framework for rendering programs in multivarious ways, which could be employed in future designs for text-based as well as structure editors.

\end{abstract}



\maketitle

\section{Introduction}
\label{sec:intro}

\begin{quote}

After three decades of continuous and in
some cases revolutionary development of
computer hardware and software systems,
one aspect of computer technology has
shown resilience to change: the presentation
of computer programs themselves...
[T]he typographic repertoire and appearance
of programs often remains little changed
from the manner in which teletypewriters
first printed out programs.

\begin{flushright}

---Marcus and Baecker (1982),
``On the Graphic Design of Program Text''~\citep{Marcus1982}

\end{flushright}

\end{quote}

\vspace{0.10in} 

Forty years on, the status quo described by Marcus and Baecker remains: The presentation of code comprises primarily the fixed-width fonts of mid-twentieth century terminals.
Granted, modern code editors now feature syntax highlighting---the coloring of tokens based on lexical and ``parts-of-speech''--type information---as well as wavy underlines\,\raisebox{-0.024in}{\inlineFig{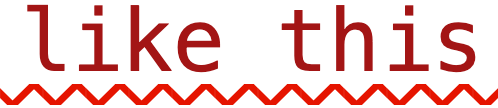}}and indicator glyphs~\inlineFig{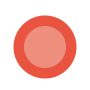}~to identify errors.
However, such basic kinds of information pale in comparison to the gamut of rich semantic information that might be highlighted through color and other visual channels.

\subsection{A Motivating Scenario}

Imagine a programming environment that offered a flexible framework for styling and rendering programs, incorporating both textual and graphical elements (e.g.,~blocks).
Multivarious, configurable kinds of information about programs could be conveyed as users and usage scenarios change.
For example, consider a programmer working to debug a ``CLOC'' program, written in Haskell, that aims to count the number of non-comment lines of code in a given source file.

\newcommand{\prefixOp}[1]{\texttt{(#1)}}
\newcommand{\bindOp}{\texttt{>\hspace{0in}>=}}
\newcommand{\lrComposeOp}{\texttt{>\hspace{0in}>\hspace{0in}>}}

\begin{figure}[b] \includegraphics[width=5.5in]{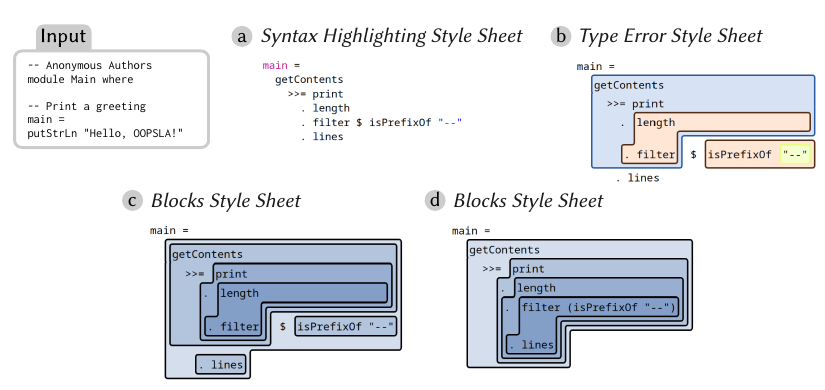}
  \caption{Count Lines of Code (CLOC) Program with Type Error and Blocks Style Sheets}
  \label{fig:cloc-blocks}
\end{figure}
 
\parahead{Step 1: Debugging a Binary Operation Error}

\autoref{fig:cloc-blocks}\refBubble{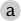} shows an initial version of the program, which fails to compile with a type error.
In addition to the error message itself (not shown), the compiler provides a \textit{code style sheet}, used to decorate the program with a color-coded visualization of the involved types and expressions (\autoref{fig:cloc-blocks}\refBubble{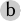}).
Unsure of why the function is failing, but beginning to suspect something to do with the usage of binary operators, the programmer decides to apply a ``blocks style sheet'' to help debug (\autoref{fig:cloc-blocks}\refBubble{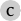}).
The programmer sees that the line of code \texttt{filter \$ isPrefixOf "-}\texttt{-"} is not contained within a box, suggesting that maybe the infix function application operator \texttt{(\$)} has lower precedence than the function composition operator \texttt{(.)}. So, they add parentheses around the call to \texttt{isPrefixOf}.
Edits made, the programmer views the blocks style sheet again to confirm that the change worked (\autoref{fig:cloc-blocks}\refBubble{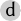}).

\parahead{Step 2: Debugging a Filter Predicate Error}

\begin{figure}[h]
  \includegraphics[width=5.5in]{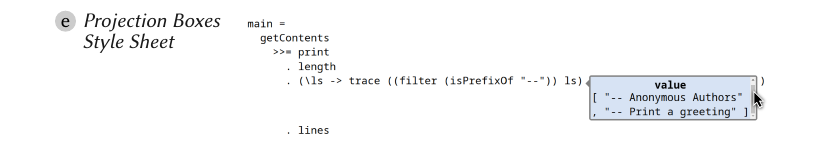}
  \caption{Projection Boxes Style Sheet}
  \label{fig:cloc-trace}
\end{figure}

Satisfied that the program now compiles, they run it but observe that the resulting number (2, not shown in \autoref{fig:cloc-blocks}) is not what they expect (4).
The programmer decides to add calls to \texttt{trace}---a standard debugging mechanism in Haskell---to inspect the run-time values that flow through the program.
The style sheet in \autoref{fig:cloc-trace}\refBubble{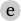} shows a projection boxes--like display \cite{Lerner2020} of values that flowed through the \texttt{trace} expression.
Seeing that the two resulting string values are commented (rather than non-commented) lines, the programmer quickly identifies a common (and forgivable) mistake:
misremembering the behavior of \texttt{filter}, which keeps (rather than discards) values that satisfy the given predicate.
To fix the issue, they compose the boolean predicate with \texttt{not} to negate its result, and re-run the tests (not shown).

\parahead{Step 3: Resolving a Point-Free Pipeline Warning}

\begin{wrapfigure}[23]{r}{2.2in}
  \vspace{-0.09in}
  \includegraphics[width=2.2in]{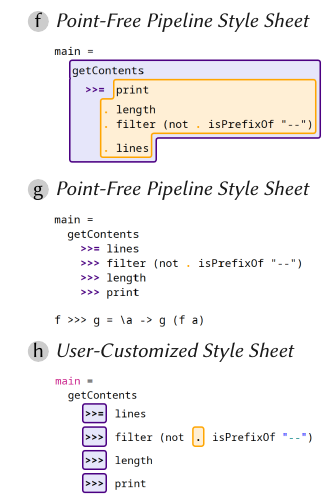}
  \caption{Point-Free Pipeline and User Customized Style Sheets}
  \label{fig:cloc-pipeline}
\end{wrapfigure}
 
The program meets its specification, but as depicted in \autoref{fig:cloc-pipeline}\refBubble{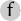} the linter applies a new style sheet to convey a warning: The program chains
together two binary operators, \prefixOp{\bindOp} and \texttt{(.)}, which propagate data in opposite ``directions.''
Whereas \mbox{\texttt{mx \bindOp{} f}} denotes (a kind of) left-to-right function application, \mbox{\texttt{f . g}} denotes right-to-left function composition.
The rendered style sheet chooses a different color for each of the two directions that data flows through the pipeline of operators, highlighting those expressions on either side of a \emph{transition} from left-to-right or right-to-left.

The programmer agrees that maintaining a single direction will be more readable, so they reorder functions using the left-to-right composition operator \prefixOp{\lrComposeOp}, resolving the warning as shown in \autoref{fig:cloc-pipeline}\refBubble{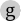}.

\parahead{Step 4: Updating User Preferences}

The programmer appreciates the suggestion about the flow of data through the operator pipeline:
When rewriting code for clarity, they lack a reliable heuristic for deciding about when to use left-to-right versus right-to-left function application and composition operators.
They believe proactively seeing the directionality information of these binary operators may help when reading and writing other code fragments.
So, they wish to repurpose some of the linter's styles to incorporate into their own, personal style sheet for styling by default---in the absence of overriding styles produced by the compiler or editor.

They identify the selector and attributes used by the point-free pipeline style sheet to effect the two-coloring.
They copy the rules into their default style sheet, modifying them slightly so that the styles are applied only to the relevant operators rather than the entire binary operation expressions.
The resulting styles for the final CLOC program are shown in \autoref{fig:cloc-pipeline}\refBubble{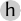}.

\subsection{Cascading Style Sheets (CSS) for Code}
\label{sec:intro-css-for-code}

\newcommand{\keyChallenge}[1]
  {#1}

\newcommand{\keyInsight}[1]
  {#1}

The motivating scenario---styling and rendering code according to circumstances and preferences---may seem familiar:
HTML documents, which encompass textual, graphical, and interactive elements, are separated into definitions of content and style; the rich system of \emph{cascading style sheets (CSS)}~\cite{css} styles the content according to a vast array of designer, user, and platform choices and preferences.
So, can't programs be displayed using HTML in a way that separates their content and styles?

Intuitively, the answer may seem to be clearly \emph{yes}. After all, a program is a tree (of abstract syntax); so, like HTML documents, why not simply write selectors that match paths into the tree and apply corresponding styles?
More detailed consideration, however, reveals two challenges that stand in the way when implementing program displays using existing approaches.

\parahead{Design Goal: Selecting AST Values, Styling the View}

In existing \emph{document} style sheet languages, such as CSS for HTML, style sheet rules select elements in a \emph{document} and defines style attributes for the display of those elements.
By analogy, in a \emph{code} style sheet language,
rules should select elements in the \emph{code (the AST)} in order to style their display.
\keyChallenge{
But unlike for \emph{document} style sheet languages, in which the document is a single object (a tree) which comprises content and is then decorated with styles,
there are \emph{two} objects (trees) involved inherent to the notion of a \emph{code} style sheet language: the AST and the document displaying it.
}

\begin{wrapfigure}[12]{r}{2.3in}
  \vspace{-0.09in} \includegraphics[width=2.3in]{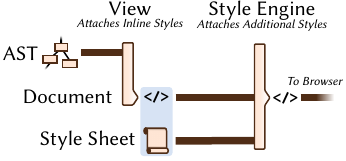}
  \caption{
Standard Pipeline for Styling Code.
  Style sheets refer to the \emph{document} being displayed,
  not the \emph{code} itself.
}
  \label{fig:architecture-standard}
\end{wrapfigure}
 To understand the hitch, consider a standard architecture for rendering code as depicted in \autoref{fig:architecture-standard};
we refer to ``(program) displays'' and ``documents'' synonymously, because HTML documents are a canonical platform for graphical interfaces.
The program display can be styled at two stages.
First, the system's \emph{view function} transforms a given AST into a document visualizing the program; the view function may attach {inline styles} directly to elements in the document.
Second, clients of the display---users and downstream tools---can define style sheet rules which restyle, or attach {additional styles} to, elements in the document.

In a standard architecture,
users and downstream tools are forced to restyle a \emph{proxy} of the original code, which furthermore may have discarded important information stored in the AST.
By comparison, code style sheets ought to refer to elements in the \emph{code} in order to style their display.
Existing tree transformation and query languages (e.g., XSLT~\cite{XSLT} and XQuery~\cite{XQuery}) may provide a starting point for defining a suitable language of selectors for AST elements.
But how should {AST} elements matched by code style sheet selectors be related to corresponding {document} elements to be styled?
\keyInsight{
Alas, in a standard architecture, there is no persistent connection between the input to the view function and the output document.
Our first key insight is the need to maintain \emph{provenance}~\cite{Cheney2009} that connects these ``source'' and ``view'' values, respectively.
}

\begin{wrapfigure}[13]{r}{2.4in}
  \vspace{-0.18in} \includegraphics[width=2.3in]{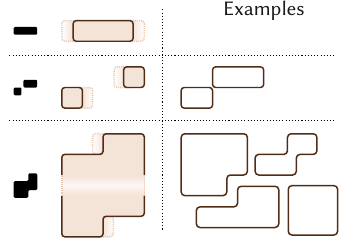}
  \vspace{0.075in} \caption{S-Blocks. Three types: single-line, two-line without overlap, and multiline.
    Shapes resemble selected-text regions in existing GUIs.
}
\label{fig:s-block}
\end{wrapfigure}
 \parahead{Design Goal: Compact Layout for Nested Text}

Once AST elements have been selected and their corresponding displays (documents) styled,
the second challenge for ``CSS for
code'' stems from the fact that program text comprises deeply nested and non-rectangular regions,
and these substructures should be subject to visual decoration as demonstrated in \autoref{fig:cloc-blocks}\refBubble{b}\refBubble{c}\refBubble{d} and \autoref{fig:cloc-pipeline}\refBubble{f}.
\keyChallenge{
Unfortunately, existing box-based layout engines, as for HTML and
CSS, offer essentially only rectangles for display.
Rendering the nested structure of ASTs naively using rectangles leads to displays that differ significantly from the raw, unstyled program text.
}
(To preview this inadequacy, see \autoref{fig:sandblocks} and \autoref{fig:html-flat-nested}\refBubble{b} in \autoref{sec:flat-nested}.)

\keyInsight{
Our second key insight is to identify a class of rectilinear shapes, called \emph{stylish blocks} or \emph{s-blocks} (\autoref{fig:s-block}),
which have the potential to
tightly wrap multiline, non-rectangular regions of text even when heavily nested.
}
S-blocks, which can be described as rectangles where the top-left and bottom-right corners may be missing, resemble how selected text is highlighted in many document-rendering interfaces. However, in our setting, we need an algorithm for computing \emph{nested} s-blocks, as demonstrated in \autoref{fig:cloc-blocks} and \autoref{fig:cloc-pipeline}, in order to more closely resemble the layout of unstyled text.

\subsection{Contributions}

\begin{figure}[t]
  \includegraphics[width=5.5in]{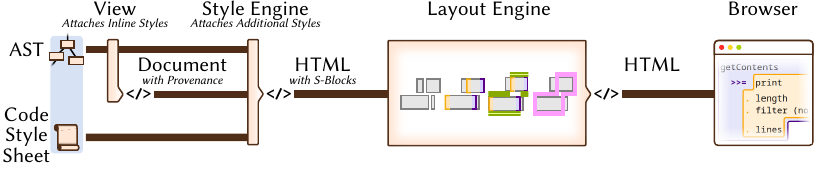}
  \caption{
    \hass{} Rendering Pipeline.
For a given program, first the \textbf{view function}
    generates a display with initial styles; the display is represented as a
    ``stylish text'' document, which retains provenance for subsequent
    restyling.
    Clients can customize the display by providing a code style sheet,
    evaluated by the \textbf{style engine}; the result resembles
    a fully styled HTML document but with s-blocks rather than (rectangular)
    blocks.
    The \textbf{layout engine} decomposes s-blocks into
    ordinary HTML for rendering.
}
  \label{fig:architecture-hass}
\end{figure}

We design and implement a code style sheets system that satisfies the two design goals above---connecting the AST with the document displaying it, and compactly laying out nested blocks of stylized text. Specifically, this paper makes the following technical contributions:

\begin{itemize}

\item
We present a code style sheet language in which rules select values in the \emph{code} (i.e., the AST) in order to style corresponding elements in the \emph{document} (the display) shown to the user.
The selector language adopts essential constructs from CSS, adding pattern-matching and predicates in a way that gives meaning to the idea of ``style sheets for \emph{code}.''
The key technical device underlying our language design is simple:
We require that functions generating program displays produce \emph{stylish text documents}, which we define to be similar to HTML documents but that additionally carry a form of \emph{dependency provenance}~\citep{TML} to track connections between how the input AST was transformed into the output document.
When evaluating a code style sheet to decorate the stylish text document with additional styles, our mostly standard tree-traversal semantics uses these dependency-provenance connections to refer back to the AST to evaluate patterns and predicates that appear in selectors.~(\autoref{sec:ppp}, \autoref{sec:style-sheets})

\item 
Second, we define a layout algorithm for computing arbitrarily nested \emph{stylish blocks} or \emph{s-blocks}, a class of rectilinear shapes which tightly wrap potentially-multiline, non-rectangular regions of text.
The algorithm translates each node in the stylish text document into multiple (ordinary) HTML elements,
appropriately positions and surrounds these elements with non-rectangular scalable vector graphics (SVGs),
and then relies on the browser's existing HTML layout engine to resolve CSS style attributes that do not need special attention by our layout algorithm.
Compared to standard box-based layout engines, ours lays out nested blocks of text more compactly, bearing closer resemblance to the original, unstyled program text.~(\autoref{sec:layout})

\item
To validate the feasibility and expressiveness of our design,
we implement a prototype system, called \hass{}, that supports code style sheets for a subset of Haskell;
the architecture of \hass{} is depicted in \autoref{fig:architecture-hass}.
We present a collection of examples in \hass{} which
demonstrate how a range of program visualizations---involving basic syntactic and structural information, as well as semantic type- and run-time information---can be phrased using our style sheet language.
Our implementation is used to generate all of the examples presented in this paper.
Before \hass{} can serve as a testbed for larger examples, however, our prototype needs to be further developed to offer additional syntactic conveniences and better performance.
~(\autoref{sec:hass}, \autoref{sec:evaluation})

\end{itemize}

\parahead{Scope}

These contributions constitute a first milestone towards our motivating goal---to develop a common framework for implementing and rendering a wide range of program visualizations.
Future work may carry this research program forward in several ways.
Our system---which starts with a ``CSS for code'' design analogy and which is also implemented using HTML and CSS---is just one way in which our general notion of code style sheets may be brought to life.
Second, the design and implementation of \hass{} presently do not support interactive features of HTML, such as event handling and CSS animations, which would be needed to integrate code style sheets into an editor.
Furthermore, the kinds of visual decorations that can be realized using our prototype are somewhat restricted;
the examples we present demonstrate the outer envelope of what visual styles can be easily achieved using our system.
Lastly, while this approach may eventually support experimentation to study how stylistic choices may facilitate program comprehension and other usability benefits, these are longer-term goals beyond the scope of this work.
These topics are discussed at length in \autoref{sec:discussion}.  
\section{Flat vs. Nested Displays}
\label{sec:flat-nested}

We began the paper by lamenting the lack of rich semantic information being visually communicated in modern code editors.
Before discussing our contributions as outlined above, in this section
we describe how more experimental designs offer a broader variety of decorations beyond mainstream code editors, yet fail to accomplish the motivating vision for our work---to provide a flexible framework for styling and rendering programs, incorporating both textual and graphical elements (e.g.,~blocks).
We also illustrate the specific layout challenges that stand in the way of this goal when using HTML and CSS as the rendering platform, along the way introducing basic HTML and CSS terminology and constructs that are needed in subsequent sections.\footnote{Throughout this paper, we use the existing acronym \emph{CSS} to abbreviate \emph{cascading style sheets}, but never for \emph{code style sheets}.}

\subsection{Text, Blocks, and Projections}

Some direct-manipulation code editors render widgets, such as boolean toggles~\inlineFig{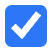}~and number sliders~\inlineFig{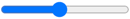}, for editing (``scrubbing''~\cite{ScrubbingCalculator, CSSTricksNumberScrubbing}) program literals through mouse-based interactions.
Live programming environments and projectional editors add visualizations of run-time values produced by the program.
(Program displays are sometimes referred to as ``projections''~\cite{Fowler08Projectional}.)
For example, Prong~\cite{Prong} renders data visualizations such as histograms~\inlineFig{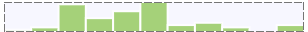} inline with code, whereas Omnicode~\cite{Omnicode} displays such visualizations in a separate pane.
In between these two extremes, projection boxes~\cite{Lerner2020} are tabular visualizations overlaid partially atop the program text.
Systems that incorporate the (graphical) decorations above succeed in augmenting the (textual) display of code.
However, these decorations are limited to appearing between lines of program text, between tokens of a single line, or entirely off to the side.
We refer to such displays as being ``flat'': whereas the program itself is a richly structured and deeply nested (abstract syntax) tree, the user interface has a rigid, often grid-like, structure, where styles and decorations take effect only at the leaves.

\begin{wrapfigure}[11]{r}{2.2in}
  \vspace{-0.02in} \includegraphics[width=2.2in]{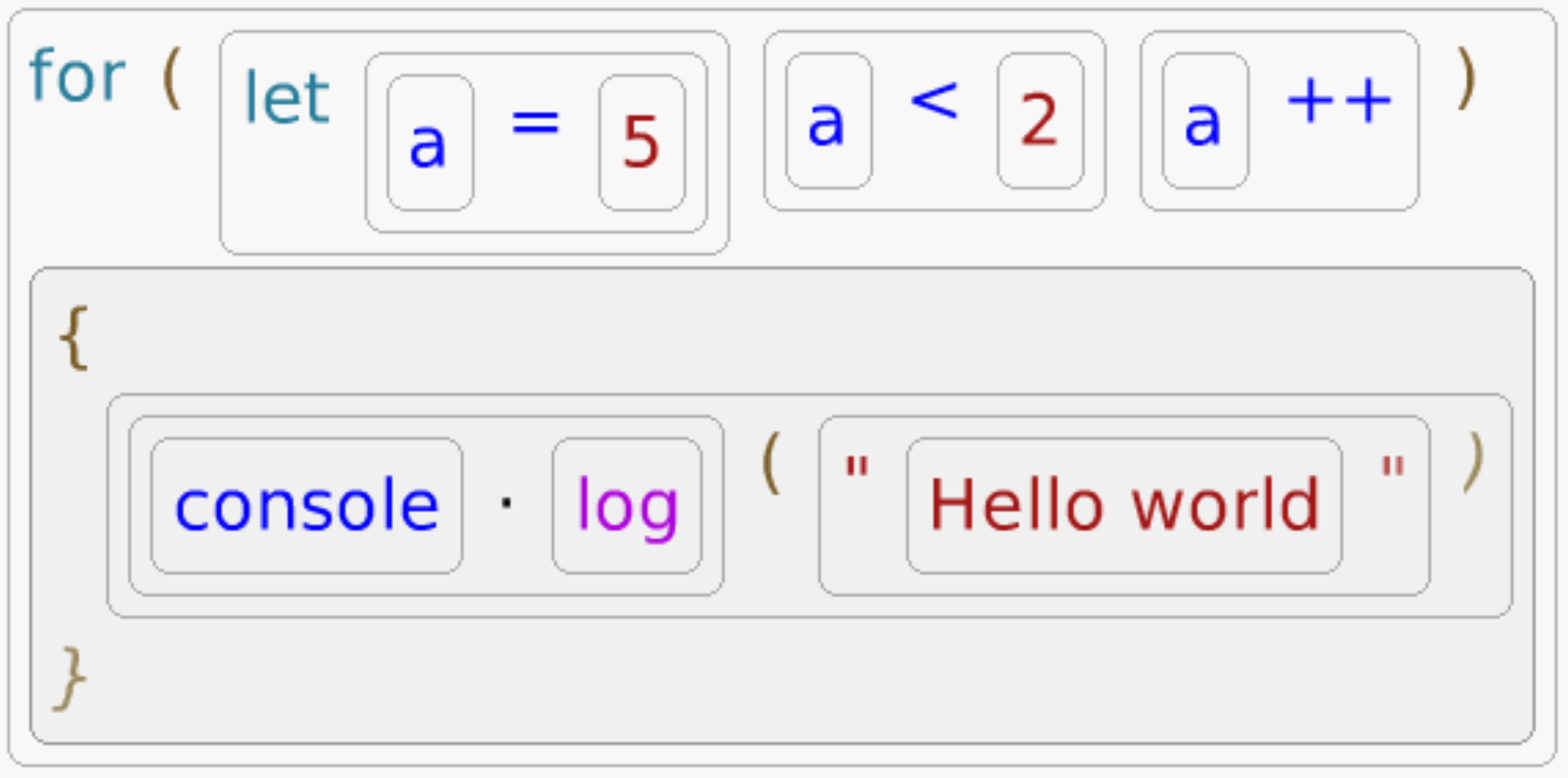}
  \caption{A fully ``nested'' display in the Sandblocks structure editor. Image: \citet[Fig.~12]{Beckmann2023} (\href{https://creativecommons.org/licenses/by/4.0/}{CC-BY 4.0}).}
  \label{fig:sandblocks}
\end{wrapfigure}
 Some editors forgo a global text buffer, in exchange allowing ``nested'' displays that reflect program structure (and which generally restrict text edits to leaves of the AST).
Block-based editors, such as Scratch~\cite{Resnick2009}, employ different shapes to indicate which code fragments can be ``snapped'' together to form valid syntax, whereas
other structure editors, such as Sandblocks~\cite{Beckmann2023}, employ rectangles uniformly for all syntactic substructures.
Blurring the boundaries between flat and nested displays, some projectional editors, such as MPS~\cite{MPS} and Hazel~\cite{Omar2021}, allow graphical elements to be placed within nested (i.e.,~non-leaf) fragments of program text.
However, these structure and projectional editors fail to lay out nested regions of textual and graphical elements in compact, ``natural-looking'' ways relative to the underlying, unstyled program text (cf.~\autoref{fig:sandblocks}).
We hypothesize that this chasm between \emph{text} and \emph{blocks} is one factor limiting the relevance and impact of structure editors.

\subsection{Tooling for Decorating Program Text}

How are program visualizations such as the above implemented?
There are essentially two strategies.
The first is through an extensible code editor, such as \vsCode{}~\cite{VSCode} or \codeMirror{}~\cite{CodeMirror}, that provides an API for customizing the display.
However, these editors offer flat, grid-based displays, which allow non-textual decorations only at the leaves of the AST.
Thus, only some of the visualizations shown above---where decorations appear between lines or between tokens---can be implemented through their APIs.
(Projection boxes~\cite{Lerner2020} are nominally implemented in \vsCode{}, but required forking the implementation because the decorations appear as a layer atop the code.)

Some visualizations, such as nested rectangles in Sandblocks and drag-and-drop blocks in Scratch, convey the underlying structure of the program by allowing elements with border properties to be arbitrarily nested, in a way that resembles the unstyled textual structure.
However, there does not exist a code rendering framework that exposes an API expressive enough to customize the display with these nested decorations.
Thus, the second implementation strategy is simply to implement a custom display tailored to the needs of the visualization; this is the approach taken by the systems above. (Granted, there are likely many reasons why these tools are implemented separately, in addition to different rendering needs.)
What is missing is a framework that can render fine-grained combinations of text and blocks, with a flexible interface for customizing the styling of program displays.
As we explain next, ordinary HTML and CSS do not constitute a suitable framework.

\begin{figure}[b]

\includegraphics[width=5.5in]{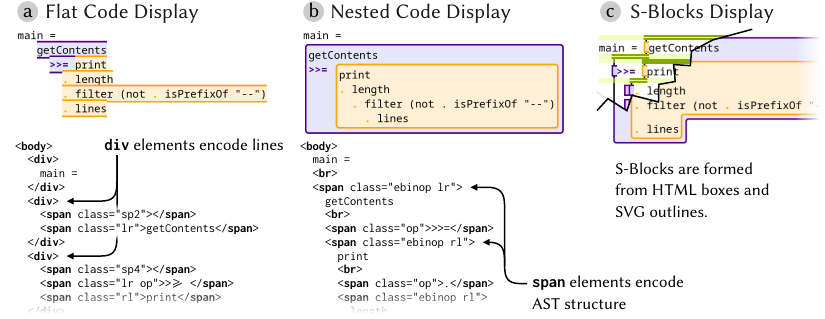}

\caption{
Displays with ``Flat'' and ``Nested'' HTML vs. S-Blocks.
Full HTML and CSS definitions in \refAppendixHtmlFlatNested{}.
}
\label{fig:html-flat-nested}
\end{figure}

\subsection{Flat and Nested Displays in HTML}
\label{sec:overview-layout}

\setlength\fboxsep{0pt}

\newcommand{\highlightCssTerm}[1]
{\colorbox{orange!20}{#1}}

How might the code display in \autoref{fig:cloc-pipeline}\refBubble{f} (the point-free pipeline warning) be attempted using ordinary HTML and CSS?
To demonstrate, \autoref{fig:html-flat-nested} outlines two approaches.
(For easy reference, different types of \highlightCssTerm{CSS selectors} are highlighted below when they are first introduced.)

\autoref{fig:html-flat-nested}\refBubble{a} outlines the structure of an HTML \emph{document} in which each line of program text is encoded as a \htmlTag{div} \emph{element} (or \emph{node}), each of which contains one or more \htmlTag{span} elements as children.
Roughly speaking, \htmlTag{div} and \htmlTag{span} elements denote subtrees to be treated as vertical or horizontal containers, or \emph{boxes}~\cite{mdnDocs}, for styling purposes.
The leaves of the document tree are \emph{text nodes} which denote content to be displayed.
Elements can be defined with a \htmlAttr{class} attribute containing one or more \emph{CSS classes} (i.e., names), for use by \emph{style sheets}
to identify sets of elements for styling.
For example, the following CSS rule (not shown in \autoref{fig:html-flat-nested}) uses a \highlightCssTerm{\emph{class selector}} written ``\cssClass{.sp2}'' to select all elements with the class \cssClass{sp2} and style them to be two characters wide:

{\small

\begin{Verbatim}[xleftmargin=\parindent]
.sp2 { display: inline-block; width: 2ch; }
\end{Verbatim}

}

\noindent
The CSS style sheet also defines rules (not shown) that use class selectors \cssClass{.op}, \cssClass{.lr}, and \cssClass{.rl} to style operators differently if they are read left-to-right or right-to-left.
The HTML display in \autoref{fig:html-flat-nested}\refBubble{a}, like modern text editors such as \vsCode{} and \codeMirror{}, is flat: the nesting structure of the abstract syntax tree (AST) is not maintained in the HTML document nor conveyed in its rendering---non-textual decorations may be rendered only for leaves of the AST.

Alternatively, \autoref{fig:html-flat-nested}\refBubble{b} outlines an HTML document that displays the same program, but where
some of the AST structure, namely, the parts pertaining to binary operation expressions, has been reflected in the document:
\htmlTag{span}s are now arbitrarily nested, and the \cssClass{lr} and \cssClass{rl} classes mark entire binary expressions, as opposed to just their leaf children, for styling purposes.
(Top-level \htmlTag{div}s have also been replaced in favor of \htmlTag{span}s, with explicit line breaks using \htmlTag{br} elements.)
Consider two CSS rules (not shown in \autoref{fig:html-flat-nested}) that contribute to the styling of this display:

{\small

\begin{Verbatim}[xleftmargin=\parindent]
.ebinop.lr     > .ebinop.rl  { border: 2px solid orange; margin: 2px; ... }
.ebinop.lr:has(> .ebinop.rl) { border: 2px solid indigo; margin: 2px; ... }
\end{Verbatim}

}

\noindent
The first rule uses a \highlightCssTerm{\emph{compound selector}} written ``\cssClass{.ebinop.lr}'' to select a node that is both a binary operation expression (\cssClass{binop}) and which is read left-to-right (\cssClass{lr}), and an \highlightCssTerm{\emph{immediate descendant selector}} written ``\texttt{> }\cssClass{.ebinop.rl}'' to select a child binary expression that is read right-to-left; the child is styled to have an orange border, margin, and padding.
The second rule uses a similar selector, but the effect of the \highlightCssTerm{\emph{pseudo-class predicate}} written ``\texttt{:has(...)}'' is that the \textit{parent} element (selected by \cssClass{.ebinop.lr}) is styled, as opposed to the child.
In this case, a purple border is applied to left-to-right elements with a right-to-left child.
(Related to but not demonstrated above: the \highlightCssTerm{\emph{descendant selector}}, written with a space instead of \texttt{>}, selects any descendant, not just children.)

Unfortunately, a code display that is structured, styled, and laid out using nested boxes---such as Sandblocks (cf. \autoref{fig:sandblocks})---faces a dilemma.
One option is to convey the structure of the program using nested boxes, which affords use of \cssAttr{padding}, \cssAttr{border}, and \cssAttr{margin} (as for \htmlTag{div} elements).
However, the resulting display (as shown in \autoref{fig:html-flat-nested}\refBubble{b}) introduces unnatural indentation after each \htmlTag{span}, since each element \emph{must} occupy a box.
Alternatively, we can render all document content \cssAttr{inline}, causing the text to be laid out according to its natural flow, but forgo the use of \cssAttr{padding}, \cssAttr{border}, and \cssAttr{margin}---this restricts styles which affect the size of an element to leaves of the AST, as in the flat approach (as shown in \autoref{fig:html-flat-nested}\refBubble{a}).

Rather than drawing a rectangle---either a \htmlTag{div}, or a \htmlTag{span} with \cssAttr{display} set to \cssAttr{inline-block}---around each multiline string, we would like to draw s-blocks (\autoref{sec:intro}, \autoref{fig:s-block}), which can be nested more compactly.
To generate such layouts, in \hass{} we decompose each s-block into smaller (rectangular) regions, each of which can be laid out using HTML's ordinary rules (in a manner similar to the ``Flat Code Display'' in \autoref{fig:html-flat-nested}\refBubble{a}).
The layout algorithm is presented in \autoref{sec:layout}, following the presentation of the code style sheet language in the next two sections.

\section{Patterns, Predicates, and Provenance}
\label{sec:ppp}

Compactly laying out nested blocks of text is one design goal for this work (cf.~\autoref{sec:intro-css-for-code}).
The other design goal---selecting elements from the \emph{AST}, and styling elements in the \emph{document}---follows from the fact that these two trees are distinct objects of interest for a \emph{code} style sheet language.

Next, we present a few small examples to introduce the structure of \hass{} style sheets (\autoref{sec:overview-hass-rules}) and our strategy for evaluating them (\autoref{sec:overview-provenance}).
The complete language design is formalized in \autoref{sec:style-sheets}.

\subsection{Overview of Style Sheet Syntax}
\label{sec:overview-hass-rules}

\begin{figure}[b]
  \includegraphics[width=5.4in]{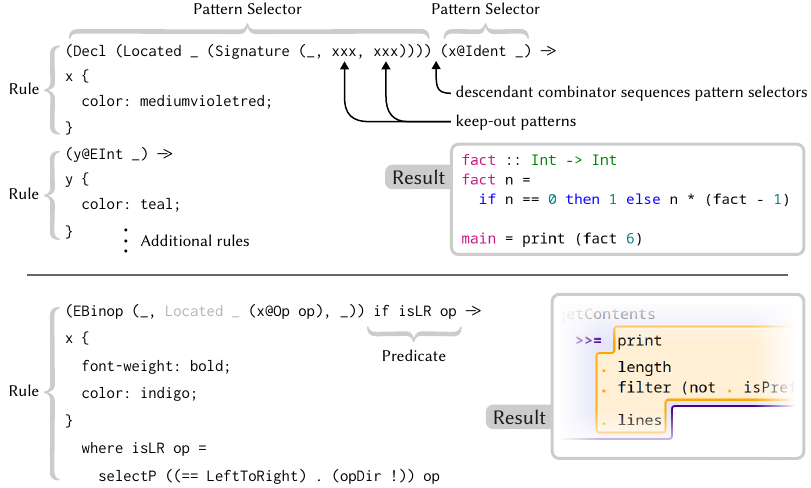}
  \caption{
Example Code Style Sheet Rules in \hass{}.
The top and bottom halves show fragments of the Syntax Highlighting and Point-Free Pipeline Style Sheets, respectively, presented in \autoref{fig:cloc-blocks}\refBubbleSmall{a} and \autoref{fig:cloc-pipeline}\refBubbleSmall{f}.
  }
  \label{fig:hass-stylesheet-overview}
\end{figure}

The basic structure of a \hass{} style sheet is demonstrated in \autoref{fig:hass-stylesheet-overview}.
Each rule comprises a selector and styles, separated by an arrow; the selector matches elements in the \emph{AST}, defining styles to the corresponding \emph{document} elements.

\parahead{Pattern Selectors}

Let us begin by considering the second rule in \autoref{fig:hass-stylesheet-overview}, which is simplest to explain:
it selects every integer literal (\texttt{EInt} expression) and styles their textual representations in \cssAttr{teal}.
Whereas in this rule the selector binds a single variable (\texttt{y}), in general a \hass{} selector can bind any number of AST subvalues to variables, each of which may be used to assign styles.

Whereas the \emph{pattern selector} above matches a single node in the AST, selectors may also match multiple nodes (a path) in the AST.
For example, the \emph{path selector} in the first rule of \autoref{fig:hass-stylesheet-overview} matches a pair comprising a node and its descendant, by separating two pattern selectors with the \emph{descendant combinator} (written with a space, like in CSS).
The first pattern selector, \texttt{Decl (... Signature (\_, xxx, xxx) ...)}, will match each top-level type signature declaration, and the second pattern selector will match each identifier (\texttt{Ident}) that appears as a descendant within the \texttt{Signature}'s first subvalue---but not within the second or third subvalues.
The latter restriction is implemented using the ``keep-out'' pattern \texttt{xxx} in \hass{}: It ensures that no descendant patterns will match by pruning the search tree rooted at \texttt{xxx}.
Once the parent node (\texttt{Signature}) and descendant node (\texttt{Ident}) are matched, the latter is styled in \cssAttr{mediumvioletred}.

\parahead{Predicates}

In addition to selecting paths in the AST through pattern matching, code visualizations will often pose other queries to decide what subvalues to style.
The third rule in \autoref{fig:hass-stylesheet-overview} evaluates a \emph{predicate} on every binary operation (\texttt{EBinop}) in the program, styling only those belonging to the particular set of operators that should by default be chained left-to-right.
Analogous rules (not shown) use predicates to select and style right-to-left operators, as well as nested binary operations that exhibit conflicting flows.

 \subsection{Overview of Style Sheet Semantics}
\label{sec:overview-provenance}

\newcommand{\emphTerm}[1]
{\textbf{#1}}

Once a selector matches a path of values in the AST, the next step is to style corresponding elements in the document.
But what does ``corresponding'' mean?

\setlength\fboxsep{0pt}

\newcommand{\diffAdd}[1]{\colorbox{yellow!20}{#1}}

\newcommand{\astNode}{\raisebox{0.003in}{\includegraphics[width=0.05in]{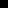}}}
\newcommand{\justNode}{\raisebox{0.003in}{\includegraphics[width=0.05in]{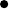}}}
\newcommand{\nothingNode}{\raisebox{0.003in}{\includegraphics[width=0.05in]{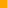}}}

\begin{figure}[b]
\includegraphics[width=5.5in]{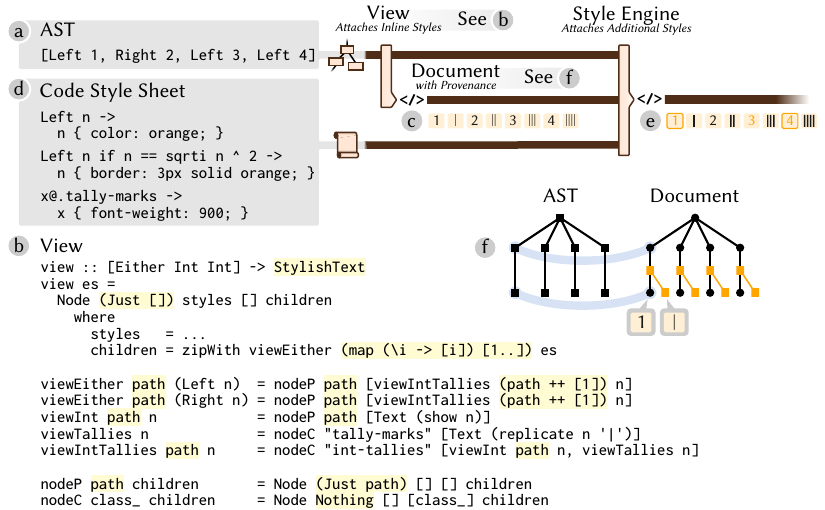}
\small

\caption{
Tally Marks Example.
Given the code string
\mbox{\small \texttt{[Left 1, Right 2, Left 3, Left 4]}} shown in \refBubbleSmall{a},
the view function in \refBubbleSmall{b}
generates the document with initial styles shown in \refBubbleSmall{c}.
Given the code style sheet in \refBubbleSmall{d},
the document is restyled as shown in \refBubbleSmall{e}.
Differences compared to a function generating an ordinary HTML document are highlighted \diffAdd{like this}.
The structure of the AST and document are shown in \refBubbleSmall{f};
AST nodes are depicted with black squares~\astNode{}, nodes with \texttt{Just} paths are depicted with black circles~\justNode{}, and nodes without paths are depicted with orange squares~\nothingNode{}.
}
\label{fig:tally-marks-combined}
\end{figure}

\parahead{Example: Tally Marks (AST Structure $\not\equiv$ Document Structure)}

To understand the task of connecting AST values to document elements, consider a simple ``AST'' comprising a list of integers each tagged either with the constructor \texttt{Left} or \texttt{Right},
and a view function that, given an AST, displays each integer followed by its representation using tally marks.
\autoref{fig:tally-marks-combined} diagrams how this visualization works, reusing the representation of the \hass{} rendering pipeline shown in \autoref{fig:architecture-hass}.
For example, given the ``program'' \texttt{[Left 1, Right 2, Left 3, Left 4]} shown in \autoref{fig:tally-marks-combined}\refBubble{a}, the document generated by the view function would have the eight adjacent elements shown in \autoref{fig:tally-marks-combined}\refBubble{c}.
(The implementation of the view function, shown in \autoref{fig:tally-marks-combined}\refBubble{b}, will be discussed below.)
After the display has been generated with initial styles, the user ought to be able to define a code style sheet to restyle it as desired---for example, the style sheet in \autoref{fig:tally-marks-combined}\refBubble{d}, which restyles \autoref{fig:tally-marks-combined}\refBubble{c} to look like \autoref{fig:tally-marks-combined}\refBubble{e}.
(The renderings of this particular stylish text document include only ordinary rectangles, not s-blocks. As a result, we omit the the ``Layout Engine'' and ``Browser'' components from \autoref{fig:architecture-hass}.)

This toy example demonstrates two noteworthy properties that distinguish it from the examples discussed in \autoref{sec:overview-hass-rules}:
first, the document structure is not the same as the AST structure, and
second, the document includes elements that are computed from the AST but are not directly represented as data in the AST (see \autoref{fig:tally-marks-combined}\refBubble{f}).
This is relevant because, in general, view functions may render additional information besides the program text.

\parahead{Paths from Root to Subvalues}

\autoref{fig:tally-marks-combined}\refBubble{d} shows three simple \hass{} rules for (re)styling the display.
We will start by considering the first two rules.
The effect of the first rule should be to color all \texttt{Left} numbers---namely, \texttt{1}, \texttt{3}, and \texttt{4}---orange, and the effect of the second rule should be to further style those \texttt{Left} numbers which are square---namely, \texttt{1} and \texttt{4}---with square borders.

Each selected number is a subvalue of the \emph{root} value (the list of tagged numbers), and can be described by a list of integer indices, called a \emph{path}, that define its location within the root value.
For example, the number \texttt{1} in the code is denoted by the path $[1,1]$, which says that \texttt{1} is the first subvalue of \texttt{Left 1}, which is the first subvalue of the root value (the list).
Likewise, the number \texttt{2} in the code is denoted by the path $[2, 1]$, the first subvalue of the \emph{second} subvalue of the root value, and so on.
If these paths were stored along with document nodes, then for styling purposes these document nodes could ``depend on'' the corresponding AST subvalues.

\parahead{Classes and Class Selectors}

Before turning to the specifics of how paths interact with documents, we will explain the third style sheet rule in \autoref{fig:tally-marks-combined}\refBubble{d}.
This rule selects each of the four tally-mark boxes using the class selector \cssClass{.tally-marks}, making each mark wider by setting its \cssAttr{font-weight}.
Pattern selectors in \hass{} do not help in this case, because the tally marks do not originate (appear as data) within the AST.
Rather, the view function generated the tally marks and prearranged those document nodes with the \cssClass{tally-marks} class to facilitate styling.
\hass{} includes ordinary CSS class selectors for use cases such as this, which may occur when the view function generates a display that is structurally different from the input AST.

\parahead{Stylish Text: Documents with Provenance}

In a standard architecture (\autoref{sec:intro}, \autoref{fig:architecture-standard}), the role of the view function is to produce an HTML document given an AST, which contains inline styles plus prearranged CSS classes for additional styling through style sheets.

In \hass{}, the view function produces a \emph{stylish text} document---phrased in \autoref{fig:stylish-text} using Haskell---which resembles a simplified HTML document.
Like an HTML document, each inner node carries a \verb+styles+ field---initial style attributes attached by the view, later extended with additional styles---and a \verb+classes+ field corresponding to CSS classes (for reasons described above).
Unlike an HTML document, each inner node optionally carries a path; the \texttt{path} field will be \texttt{Nothing} for elements like tally marks, which do not appear literally in the AST.

\begin{wrapfigure}[11]{r}{0.43\textwidth}
\vspace{-0.07in} \small
\begin{Verbatim}
data StylishText
  = Text String
  | Node {
      path     :: Maybe [Int],
      styles   :: [(String, String)],
      classes  :: [String],
      children :: [StylishText]
    }
\end{Verbatim}
\caption{Stylish Text Documents}
\label{fig:stylish-text}
\end{wrapfigure}
 \autoref{fig:tally-marks-combined}\refBubble{b} defines the tally-mark view function which transforms the AST (a list of tagged integers) into the stylish text document (a string leaf-tree with room for styles and classes at each node).
This function is written in ordinary Haskell and is an ordinary transformation from a structured data type (in this case, \texttt{[Either Int Int]}) to \texttt{StylishText}. Compared to a ``plain'' view function generating ordinary HTML (without provenance), consider the differences---highlighted \diffAdd{like this}---for generating a \texttt{StylishText} document with paths into the source value.
The top-level, root value is represented as a node with the root path \texttt{[]}.
Nodes are annotated with a path which is systematically extended from its parent (in the case of the root, \texttt{[1]} for the first child, \texttt{[2]} for the second child, and so on).
However, not all nodes in the document have AST provenance, namely, nodes created by \texttt{viewTallies} which contain tally-mark representations, and
nodes created by \texttt{viewIntTallies} which render \texttt{Int} values with both numeric and tally-mark representations.
Because these nodes do not have paths, they are marked with CSS classes for use in style sheets.

\parahead{Styling the View}

Akin to evaluating style sheets in existing style sheet languages for documents, in \hass{} the document is traversed while all rules are applied everywhere possible.
The fundamental difference stems from the addition of pattern selectors.
In CSS, the basic operation of selectors is to walk the \emph{(untyped) document} checking whether each document node matches particular \emph{classes} (i.e. \emph{tags}).
In \hass{}, selectors refer to any \emph{(richly-typed) AST value} that corresponds to a document node; if the node has a path into the AST, then the referenced AST subvalue is retrieved to perform the pattern matching and predicate evaluation.

The first two rules in \autoref{fig:tally-marks-combined}\refBubble{d} select AST values matching the pattern \texttt{Left n}---for example, the AST subvalues with paths [1] and [1,1]---
but the view function in \autoref{fig:tally-marks-combined}\refBubble{b} inserts document nodes between those displaying each \texttt{Left} constructor and its numeric child (\texttt{n}).
Nevertheless, because document nodes are tagged with paths, nodes corresponding to the pattern-matched values can be styled despite the additional structure between them;
see the thick blue lines in \autoref{fig:tally-marks-combined}\refBubble{b}, which indicate the paths [1] and [1,1].
The precise semantics of style sheet evaluation is defined in the next section.\footnote{
At this point, it may be tempting to ask:
Why does style computation need to explicitly maintain connections between AST and document?
To answer this question, recall that according to the architecture depicted in \autoref{fig:architecture-hass}, style sheets are applied \emph{after} the view function has been applied, but query the \emph{input} to view. (This is in contrast to \autoref{fig:architecture-standard}, for example, in which style sheets are applied after view, but cannot query the original AST.)
For this scheme to work, style computation must retain some information about the AST during styling.

If a style sheet is known ahead of time, it is possible to preemptively decorate a document with all of the semantic information one might need in order to query and style it.
However, this is not a practical approach for \emph{code} style sheets, since pattern selectors and predicates are essentially arbitrary computations, which would need to be evaluated for every value that might be scrutinized.
See \refAppendixTranslationToCss{} for more details about how attempted solutions in this vein fail to address the design goal for this work.
\label{footnote:translation-to-css}
}

\section{Style Sheet Language}
\label{sec:style-sheets}

Next, we define the syntax and semantics introduced by example in the previous section.

\begin{figure}[t]

\begin{tabular}{rrcl}

\multicolumn{4}{c}{
  \textbf{Datatypes}          $\varTypeCon$ \hspace{0.10in}
  \textbf{Constructors}       $\varDataCon$ \hspace{0.10in}
\textbf{Variables}          $\varVar$     \hspace{0.10in}
  \textbf{Strings        }    $\varAttrName, \varAttrVal, \varCls$
} \nextRowMoreSpace

\syntaxRow{Expressions}{\varExp}{
  \narrowcdots
} \nextRow

\syntaxRow{Values (of ADTs)}{\varVal}{
  \dataApp{\varDataCon}{\rangeZeroN{i}{n}{\varVal_i}}
} \nextRow

\syntaxRow{Documents}{\varDoc}{
  \leaf{\varStr} \sepmidsep
  \node{\varPath}
       {\rangeZeroN{j}{m}{\varCls_j}}
       {\varAttrs}
       {\rangeZeroN{i}{n}{\varDoc_i}}
} \nextRow

\syntaxRow{Paths}{\varPath}{
  \noPath \sepmidsep
  \justPath{\rangeZeroN{i}{n}{\mathit{idx}_i}}
} \nextRow

\syntaxRow{Style Attributes}{\varAttrs}{
  \rangeZeroN{i}{n}{ \varAttrName_i : \varAttrVal_i }
} \nextRowMoreSpace

\syntaxRow{Style Sheets}{\varStyleSheet}{
  \rangeZeroN{i}{n}{\varGuardedRule_i}
} \nextRow

\syntaxRow{Rules}{\varGuardedRule}{
  \guardedRuleArrow
{\varPathSelector}
    {\varExp}
    {\varNamedStyles}
  \hspace{0.13in}\textrm{where }
  \varNamedStyles \ensuremath{::=} {\namedStyles{j}{m}{\varVar_j}{\varAttrs_j}}
} \nextRow

\syntaxRow{(Path) Selectors}{\varPathSelector}{
\varNodeSelector \sepmidsep
  \pathRule{\varNodeSelector}{\varOp}{\varPathSelector}
  \hspace{0.22in}\textrm{where }
  \odot \in \set{\opChild,\opDescendant,\opSibling,\opNextSibling,\narrowldots}
} \nextRow

\syntaxRow{Node Selectors}{\varNodeSelector}{
  \compoundSelector{\varBasicSelector}{\rangeZeroN{i}{n}{\clsSelector{\varCls_i}}}
} \nextRow

\syntaxRow{Basic Selectors}{\varBasicSelector}{
\varPat
  \sepmidsep
\namedPattern{\varVar}{\clsSelector{\varCls}}
} \nextRow

\syntaxRow{Patterns}{\varPat}{
  \varPattern{\varVar}{\varTypeCon} \sepmidsep
  \namedPattern{\varVar}{\dataApp{\varDataCon}{\rangeN{i}{n}{\varPat_i}}} \sepmidsep
  \keepOut
} \nextRow

\end{tabular}
\caption{Syntax of Code Style Sheets in \hass{}}
\label{fig:external-syntax}
\end{figure}
 
\subsection{Syntax}
\label{sec:style-sheet-syntax}

\autoref{fig:external-syntax} defines the syntax of stylish text documents and style sheets in \hass{}.
Inner nodes of a stylish text document $\varDoc$ (of type \verb+StylishText+) maintain optional paths $\varPath$ into the original value $v$ of some algebraic datatype $\varTypeCon$ (such as an AST type), as described in the previous section.
Optional paths constitute a simple form of \emph{dependency provenance}~\cite{TML}, which allows output subvalues to be tagged with an arbitrary number of input subvalues; in our setting, document nodes are tagged with at most \emph{one} subvalue.

The overall structure of a style $\varStyleSheet$ in \hass{} bears close resemblance with the core CSS constructs reviewed (and \highlightCssTerm{highlighted}) in \autoref{sec:overview-layout}.
A $\varGuardedRule$, written $\guardedRuleArrow{\varPathSelector}{\varExp}{\varNamedStyles}$, begins with a \emph{path selector} $\varPathSelector$ and predicate $\varExp$, followed by named styles $\varNamedStyles$.
A path selector is a list of \emph{node selectors} $\varNodeSelector$ separated by tree combinators $\varOp$, such as the \emph{child} combinator ($\opChild$) and the \emph{descendant} combinator (represented as ``$\opDescendant$'' in \autoref{fig:external-syntax} but, following CSS, written with a space in our implementation).
Node selectors $\varNodeSelector$ are described below along with their semantics.
If the path selector matches a path of nodes in the document, the predicate $\varExp$---referring to AST values bound in the selector---is evaluated.
If true, each style set $\set{\varAttrs_j}$ in $\varNamedStyles$ is applied to document nodes whose paths correspond to values bound by $\varVar_j$ in the selector.
(Paths and selector variables can be viewed as interchangeable representations.)

\subsection{Semantics}
\label{sec:style-sheet-semantics}

\newcommand{\notHiddenArgs}[1]{\ensuremath{#1\vdash}}
\newcommand{\notHiddenArgsTwo}[2]{\ensuremath{#1;#2\vdash}}

Each rule in a style sheet is applied throughout a stylish text document $\varDoc$, producing a new document $\varDoc'$ that decorates the original with additional styles.
We define this semantics as a series of nondeterministic document transformations of the form
$
\notHiddenArgsTwo{\varNamedStyles}{\valRoot}\applyEquals{\cdot}{\varDoc}{\varDoc'}
$,
one for each syntactic component of a rule.
The named styles $\varNamedStyles$ (for a specific rule) and the root AST value $\valRoot$ are inputs to these transformations,
each of which we outline below.

\parahead{Basic Selectors: Patterns}

A primary goal for the \hass{} style sheet language is the ability to select and style document nodes by pattern matching the AST values being rendered.
For this, we define \emph{pattern selectors} using
the standard notion of pattern matching for destructuring values of algebraic datatypes (ADTs).
(Pattern selectors can be understood as generalizing CSS \emph{type selectors}, which pattern match only a node's ``tag,'' i.e., a string literal.)
The variable pattern $\varPattern{\varVar}{\varTypeCon}$ includes a type annotation to facilitate the implementation of predicates (\autoref{sec:hass}), and
the pattern $\namedPattern{\varVar}{\dataApp{\varDataCon}{\rangeN{i}{n}{\varPat_i}}}$ names and matches values with a particular constructed form.
We also introduce a special pattern, $\keepOut$ (written \texttt{xxx} in the implementation, as described in \autoref{sec:overview-hass-rules}) that never matches any values, which is useful for preventing certain subvalues from being further scrutinized.

The following rule defines how pattern selectors operate on a document node $\varDoc$:
$$
\inferrule*[lab=\ruleNameFig{ApplyPatternSelector}]
  {
   \varPath = \pathOf{\varDoc}
   \sepPremise
   \varVal = \varPath\valRoot
   \sepPremise
   \varPat \rhd_\varPath \varVal \dashv \varEnv;\varEnvStyles
   \sepPremise
   \varDoc' = \applyStyles{\varEnvStyles}{\varDoc}
  }
  {
   \notHiddenArgsTwo{\varNamedStyles}{\valRoot}
   \applyEquals{\varPat}{\varDoc}{\outputPair{\varDoc'}{\varEnv}}
  }
$$
First, the subvalue $\varVal$ corresponding to path $\varPath$ is extracted from the root value $\valRoot$;
if the path is \texttt{Nothing}, the path application (the second premise) is undefined.
Next, the value is pattern matched against $\varPat$, producing a value environment $\varEnv$, mapping variables to values, if successful.
Compared to an ordinary pattern matching definition, the operation here also takes a path input, which denotes the provenance of the value being scrutinized.
Here, the pattern matching definition also produces a \emph{style environment} $\varEnvStyles$, mapping paths $\varPath$ to style sets $\set{\varAttrs}$, as output.
This helps with the fact that,
because patterns are recursive, a single pattern selector may name multiple subvalues to be styled.
For example, the rule \texttt{pair@(Pair x y) -> pair \{...\} x \{...\} y \{...\}} defines three named style sets; a successful pattern match produces a style environment that maps the paths [], [1], and [2], respectively, to the given style sets for \texttt{pair}, \texttt{x}, and \texttt{y}.
The last step in \ruleName{ApplyPatternSelector} is to apply the style environment to the document: augmenting every subnode that has a path $\varPath' \in dom(\varEnvStyles)$ with the additional styles defined by $\varEnvStyles(\varPath')$.

\parahead{Basic Selectors: Classes}

Patterns and predicates are expressive programming-language tools for selecting AST values.
Occasionally, however, it is useful to select elements in the document without referring to the originating AST subvalue
(for example, the tally marks display in \autoref{sec:overview-provenance}).

\hass{} includes {class selectors} $\namedPattern{\varVar}{\clsSelector{\varCls}}$ to select elements in the document based on their class.
For document nodes $\varDoc$ which contain the class $\varCls$, the following rule augments their styles with $\styleSetOf{\varVar}$ in straightforward fashion:
$$
\inferrule*[lab=\ruleNameFig{ApplyClassSelector}]
  {
   \varCls \in \classesOf{\varDoc}
   \sepPremise
   \varDoc' =
     \copyUpdate{\varDoc}
                {\texttt{styles}}
                {\stylesOf{\varDoc} \cup \styleSetOf{\varVar}}
  }
  {
   \notHiddenArgsTwo{\varNamedStyles}{\valRoot}
   \applyEquals{\namedPattern{\varVar}{\clsSelector{\varCls}}}
               {\varDoc}
               {\varDoc'}
  }
$$

\parahead{Node and Path Selectors}

With the semantics for basic selectors in place, the definitions for node and path selectors are straightforward.
Analogous to ``compound selectors'' in CSS, node selectors
$\compoundSelector{\varBasicSelector}{\rangeZeroN{i}{n}{\clsSelector{\varCls_i}}}$
in \hass{} augment a basic selector with zero or more classes; compared to the \ruleName{ApplyPatternSelector} and \ruleName{ApplyClassSelector} rules above, this requires also checking that each $\varCls_i$ is in $\classesOf{\varDoc}$.
Analogous to ``complex selectors'' in CSS~\cite{mdnDocs}, the application of path selectors in \hass{} involves standard tree traversal semantics.

\parahead{Predicates}

Given our programming-language setting, it is natural is to allow \emph{arbitrary} predicates (i.e., boolean expressions $\varExp$) over the data values being matched (cf. bottom half of \autoref{fig:hass-stylesheet-overview}).
In \hass{}, predicates are evaluated as a side condition after the rule's selector has been matched:
$$
\inferrule*[lab=\ruleNameFig{ApplyRule}]
  {
   \notHiddenArgsTwo{\varNamedStyles}{\valRoot}
   \applyEquals{\varPathSelector}{\varDoc}{\outputPair{\varDoc'}{\varEnv}}
   \sepPremise
   \evalsTo{\varEnv}{\varExp}{\valTrue}
  }
  {
   \notHiddenArgs{\valRoot}
   \applyEquals{\guardedRuleArrow{\varPathSelector}{\varExp}{\varNamedStyles}}
               {\varDoc}
               {\varDoc'}
  }
$$

\parahead{Formal Definitions}

See \refAppendixStyleComputation{} for detailed definitions and discussion of the semantics outlined above.
To simplify the presentation, we desugar the \emph{external} syntax shown in \autoref{fig:external-syntax} into a closely related \emph{internal} syntax.
Whereas rules in the external syntax comprise a selector (and predicate) followed by a ``global'' set of named styles $\varNamedStyles$,
rules in the internal syntax attach style sets $\set{\varAttrs_j}$ ``locally'' with components of the selector where the names $\varVar_j$ are introduced.
On one hand, the external syntax is convenient for reading and writing; on the other hand, the internal syntax avoids the indirection through names, thus simplifying the formal definitions (there are no named styles $\varNamedStyles$ to thread through the definitions).

\section{Layout}
\label{sec:layout}

Given a stylish text document styled through a combination of inline styles and style sheets,
the final consideration is to render a display, where nested blocks of stylized text are laid out compactly.
As discussed in \autoref{sec:intro}, s-blocks (\autoref{fig:s-block}) may serve as the essential building block for satisfying this design goal, but s-blocks are not native to any rendering platform.

\parahead{Decomposing S-Blocks to HTML, CSS, and SVG}
Although our design does not necessitate any particular rendering target, we choose native HTML and CSS to avoid duplicating existing functionality.
Our approach is to decompose s-blocks into smaller rectangular regions, each of which can be laid out using HTML's ordinary rules (in a manner similar to the ``Flat Code Display'' in \autoref{fig:html-flat-nested}\refBubble{a}).
Some of these rectangles will represent the leaves of the document, in other words, its textual or visual \textit{content}.
But many of these rectangles will only allocate space for visualizing structure between pieces of content; we call these rectangles \textit{gadgets}.
The \hass{} layout algorithm translates s-blocks into a ``flat'' HTML document, to use the terminology of \autoref{sec:flat-nested}, comprised of these rectangles, which is part of what allows \hass{} layouts to preserve ``natural'' looking text flow.
Another important advantage of translating s-blocks into HTML is that many \hass{} style properties---for example, those concerned with fonts and colors---can be directly forwarded to the output document to be interpreted by CSS.

Some properties, however, require specific interpretations in our setting:
what we call ``border properties'' such as \texttt{padding}, \texttt{border}, and \texttt{margin}, and related ``border spacing'' properties such as \texttt{border-width}, \texttt{border-color}, and \texttt{background-color}.
We treat these properties separately because they are explicit inputs to the layout algorithm;
the size and position of s-blocks, along with the shape of their borders depend directly on the values of these properties.\footnote{
There are also some CSS properties which are not supported in \hass{}.
For example, because we treat each leaf in the document is an \texttt{inline-block}, overriding the \texttt{display} property would not produce correct layouts.
Furthermore, attributes that refer to user events or that trigger layout changes in the positioning of elements in the document would require periodic invocations of the layout algorithm.
These would be natural extensions to \hass{}, as discussed in \autoref{sec:limitations}.
}
Regarding s-block borders themselves, there is no straightforward way to represent them using HTML and CSS.
So, the layout algorithm generates SVG paths which are overlaid atop the generated HTML.
In effect, the output of the layout algorithm has two components: A flat HTML document with space allocated for structure-surfacing visualizations, and a set of SVG paths which should be rendered on top of the document.

\parahead{Fragments}
The leaves of the stylish text tree contain strings which form the textual (or visual) content of the document.
For the purposes of layout, we call each of these atomic strings \textit{fragments}.
In addition to its string content, each fragment also records its size as well as the amount of border spacing required on its left and right edges (these spacers are called \textit{horizontal gadgets}, or \textit{h-gadgets}).
With this terminology, we can more precisely phrase the goal of layout: to flatten the stylish text tree into a vector of fragments, where the position of each fragment is known.
Given such information, it is easy to compute the geometry of s-block borders.
Our algorithm computes s-blocks in three steps, visualized in \autoref{fig:layout-pipeline} and described below.

\subsection{Flattening: Stylish Text to Fragments}

The first step is to construct a vector of every text fragment in the tree, in the order that it appears in the document.
Each fragment's size is then determined by constructing a \texttt{<span>} with its content and styles, and querying a browser for the rendered size of the element.
We simultaneously construct a new tree, the \textit{layout tree}, which augments the stylish text tree by recording the start and end indices of all of the fragments beneath it (see the top row of \autoref{fig:layout-pipeline}).
This allows a structural traversal of the document (by traversing the layout tree and following the indices), or a linear traversal (by traversing the vector of fragments directly).

\subsection{Width Resolution}

The purpose of width resolution is to determine horizontal positions for all fragments. The width of each fragment is the width of its content (measured during the flattening step) plus the widths of any h-gadgets on its left and right edges.
The horizontal position of each fragment is determined by summing the widths of preceding fragments and h-gadgets on the same line.
Therefore, width resolution amounts to the placement of h-gadgets, thus
determining the positions of \textbf{\textit{vertical} lines} for all s-blocks (\autoref{fig:s-block}).

\begin{figure} \includegraphics[width=5.5in]{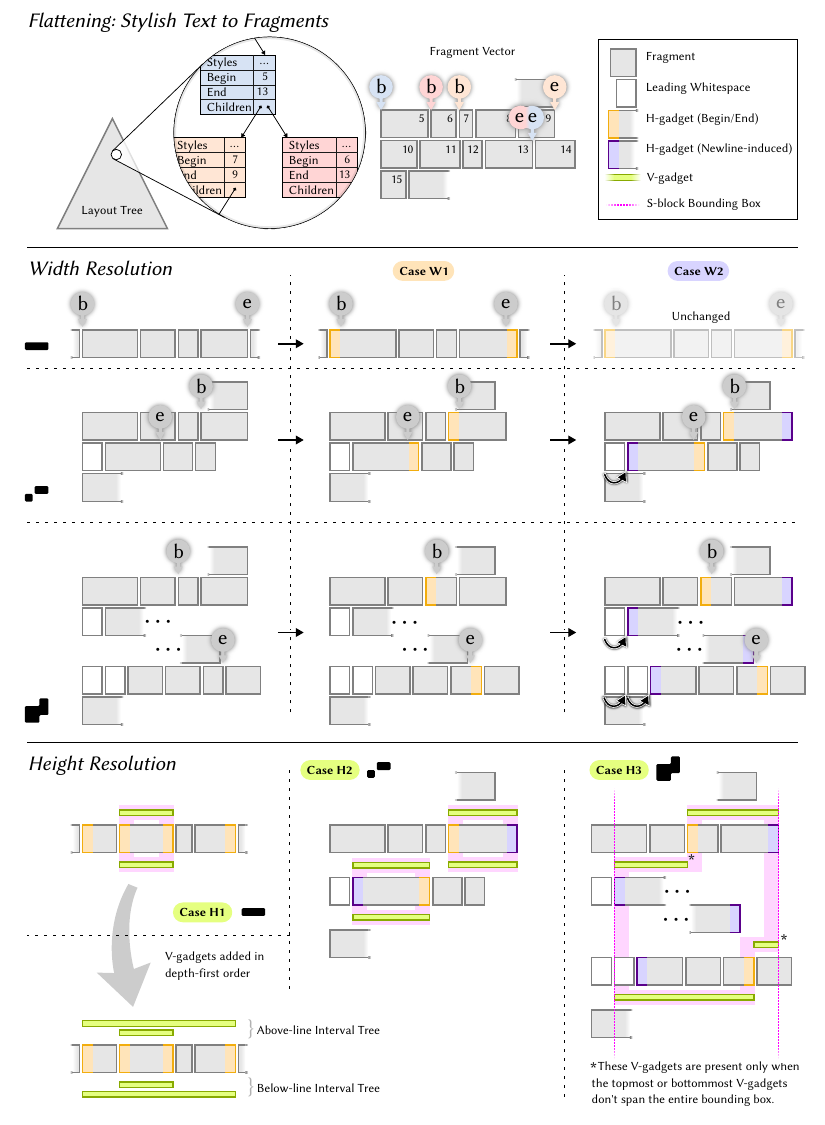}
  \caption{
  Three Steps of Layout: Flattening, Width Resolution, and Height Resolution
}
\label{fig:layout-pipeline}
\end{figure}

Due to the shape of s-blocks, we can enumerate all of the possible locations where horizontal padding (and thus, h-gadgets) can appear.
Recall that each subtree in the layout tree is annotated with the range of fragments below it; call this range $\mathcal{R}$.
H-gadgets can appear in one of four places within $\mathcal{R}$:
(i) before the first fragment in $\mathcal{R}$ (\autoref{fig:layout-pipeline} \caseBubble{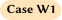}),
(ii) after the last fragment in $\mathcal{R}$ (\autoref{fig:layout-pipeline} \caseBubble{w1}),
(iii) on the last fragment of a line (\autoref{fig:layout-pipeline} \caseBubble{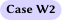}),
and (iv) on the first non-whitespace fragment of a new line (\autoref{fig:layout-pipeline} \caseBubble{w2}).
Once inserted in this way, the h-gadgets, along with their constituent fragments, can be horizontally positioned as outlined above.

\subsection{Height Resolution}

The job of height resolution is to ``connect the dots'' of the gadgets and fragments placed during width resolution, ensuring that there is inter-line space for any border spacing around a subtree of the layout tree.
As with width resolution, we can observe the conditions under which we must reserve vertical space above and below a subtree.
Here, the s-block restriction guarantees that only the topmost and bottommost two lines of any given subtree need to be allocated vertical space---it is never necessary to allocate vertical space within the interior of an s-block.
We insert \textit{vertical gadgets}, or \textit{v-gadgets}, to model vertical spacing; each v-gadget corresponds to a \textbf{\textit{horizontal} line} in \autoref{fig:s-block}.
Each type of s-block (\sba, \sbb, and \sbc) is handled as follows:

\begin{itemize}

\item[\sba] For subtrees that occupy a single line, v-gadgets are allocated above and below the line, from the beginning of the first fragment to the end of the last fragment. (\autoref{fig:layout-pipeline} \caseBubble{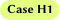})

\item[\sbb] For subtrees that occupy two lines but without any horizontal overlap between them, the procedure is the same as above but repeated for both disjoint parts. (\autoref{fig:layout-pipeline} \caseBubble{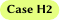})

\item[\sbc] For all other subtrees, we allocate v-gadgets: (i) above the first line from its start to the maximum rightward extent of any fragment in the subtree, (ii) above the second line from the maximum leftward extent of any fragment to the beginning of the first line, (iii) on the last line from the maximum leftward extent of any fragment to the end of the last line, and (iv) on the second-to-last line from the end of the last line to the maximum rightward extent of any fragment. Here, the s-block restriction ensures that there are a maximum of four places where vertical gadgets may be allocated, since an s-block needs vertical padding at most above its first two lines, and after its last two. (\autoref{fig:layout-pipeline} \caseBubble{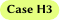})

\end{itemize}

The process of allocating v-gadgets proceeds in depth-first order. This ensures that for any two subtrees $\mathcal{A}$, and $\mathcal{B}$, if $\mathcal{B}$ is below $\mathcal{A}$, then $\mathcal{B}$ will have its v-gadgets created \textit{before} $\mathcal{A}$, ensuring that its borders are \textit{inside} those of $\mathcal{A}$.

Lastly, the quantities involved in height resolution (namely the maximum extents of the fragments in the subtree, and the extents of the four outermost lines) are exactly those needed to compute the s-block border for the subtree, should it require one.
SVG commands for drawing these borders are collected during the same traversal which is responsible for allocating v-gadgets and resolving their heights (see thick pink outlines in the bottom third of \autoref{fig:layout-pipeline}).

\section{Implementation}
\label{sec:hass}

Having presented our design, next we discuss notable details about our implementation.

\subsection{\hass{}: A Haskell Implementation of Code Style Sheets}

\hass{} is implemented in approximately 2.5K LOC of Haskell and 1.5K LOC of TypeScript. 

\parahead{Stylish Text (cf. \autoref{sec:overview-provenance}, \autoref{fig:stylish-text})}

The interface for view functions in our implementation is a type class called \texttt{Stylish} with a single member \texttt{showStylish :: (Path, a) -> StylishText}. This class resembles the standard \texttt{Show} class (which defines \verb+show :: a -> String+) but where inputs are tagged with dependency-provenance paths and the output is stylish text. (Notice how the function for generating tally marks in \autoref{fig:tally-marks-combined}\refBubble{b} can be made to fit into this interface.)
Compared to the definition of \texttt{StylishText} presented in \autoref{fig:stylish-text}, our implementation includes a second leaf constructor, \texttt{HtmlLeaf}, that
carries an HTML element represented as an uninterpreted \texttt{String}.
This enables, e.g., the Projection Boxes display in \autoref{fig:cloc-trace}\refBubble{e} to include a table.

For simplicity, \texttt{StylishText} models \verb+Path+s as lists of \verb+Int+eger projections from a root value, and \verb+showStylish+ functions must maintain and manipulate paths explicitly.
However, dependency provenance can be implemented internally and automatically by a tracing interpreter~\citep{TML}, an approach that has been used as the basis for interactive text-based structure editors~\citep{hempel2020tiny}.
A more full-featured implementation of \hass{} could also do this, which would allow ``plain'' view functions to be written, akin to the definition in \autoref{fig:tally-marks-combined}\refBubble{b} without the \diffAdd{highlighted} portions.

\parahead{Style Sheet Syntax and Semantics (cf. \autoref{sec:style-sheets})}

Unlike the direct semantics described in \autoref{sec:style-sheets},
\hass{} is implemented as an embedded domain-specific language in Haskell, so pattern selectors and predicates are represented as Haskell functions.
Path selectors may traverse arbitrarily deeply into subvalues being scrutinized.
So, we implement a \texttt{select} function (akin to the \texttt{mkT} function of ``scrap-your-boilerplate''~\citep{syb}) for lifting ordinary functions to more general types, relying on type-safe run-time casts via \texttt{Data.Typeable}.
The latter mechanism is also used to help implement predicate evaluation in our setting where selectors introduce value-bindings with heterogeneous types;
type-annotated variable patterns $\varPattern{\varVar}{\varTypeCon}$ help prevent ill-typed predicates.

The syntax for style sheets currently supported in our implementation deviates from the external syntax defined in \autoref{fig:external-syntax} (and demonstrated in \autoref{sec:evaluation}) in a couple ways. One difference is due to the fact that we have not implemented a standalone parser for style sheets.
Instead, we use Template Haskell~\cite{TemplateHaskell} and quasi-quoted strings~\cite{QuasiQuotes} to write style sheets (in Haskell source files) with syntax very similar to the external syntax---often this means simply surrounding rule definitions with \verb+[Stylesheet.rule| ... |]+, which tells Template Haskell which function to use to parse the contents.
Implementing a parser for standalone \hass{} style sheet files is an incidental engineering task for subsequent work.
A second, more important difference is that our implementation---in which rules are represented using using Haskell functions---allows style attribute values to be \emph{expressions} rather than string \emph{literals} as presented in \autoref{fig:external-syntax}.
This capability is utilized by a Type Error style sheet (discussed in \autoref{sec:evaluation}) to style each expression involved in the type error with a unique border and background color which matches the terminal output.
We have chosen to keep the presentation in \autoref{fig:external-syntax} simple by limiting expressions $\varExp$ to predicates, but incorporating them into style attributes as well would not pose problems: the semantics of code style sheets is largely agnostic to the syntax and semantics of expressions in the host language.

Beyond syntax,
another difference pertains to the nondeterminism of style computation.
To ensure that output styles do not depend on the order in which rules are evaluated, in our implementation each style attribute in a rule set is assigned a \emph{precedence} value to resolve conflicts.
The implementation also provides a simple way for elements to automatically inherit styles that are not set directly.
Future work should consider how to extend the definitions of \emph{cascading}, \emph{specificity}, \emph{initial values}, and \emph{inheritance}~\citep{mdnDocs} in CSS to our setting with ASTs and pattern selectors.

\parahead{Layout (cf. \autoref{sec:layout})}

After the document is fully decorated with styles, the layout algorithm runs on a Haskell server communicating with a TypeScript client to request fragment measurements from the browser.
After resolving all queries and calculating positions, the server sends the client the final HTML document for display.

\subsection{\hasskell{}: A \hass{}-Aware Implementation of a Haskell Subset}
\label{sec:hasskell}

We have implemented a front-end toolchain, called \hasskell{}, for a subset of Haskell, used to render the examples in \autoref{sec:evaluation}.
This consists of an additional 5.7K LOC of Haskell---including \hass{} style sheets (making up 1.2K LOC) for the examples in \autoref{sec:evaluation}, which are embedded as discussed above.

In contrast to an \emph{abstract} syntax tree, we use tree-sitter~\cite{tree-sitter, tree-sitter-haskell, haskell-tree-sitter} to obtain \emph{concrete} syntax trees---which contain location information for every element in the program, as well as all concrete characters (e.g., tokens, comments) needed to completely reconstruct the input program text.
Our translation from tree-sitter trees into a corresponding \texttt{Hasskell} data type includes only a few minor adjustments---to re-associate type-annotated expressions to better structure the data;
to explicitly track comments as a set of (located) strings; and
to restructure binary operations to respect associativity and precedence.

Thus equipped, our default \texttt{Stylish} instance for \texttt{Hasskell} is a mostly-boilerplate structural recursion.
The main non-trivial component is a helper function that, given a ``current'' position in the output buffer, systematically rebuilds spaces, newlines, and comments before emitting a syntax element at a specified ``start'' location.
(A couple examples below rely on custom instances, in order to augment the program text with additional elements; these are discussed at the end of \autoref{sec:evaluation}.)

\section{Examples}
\label{sec:evaluation}

To evaluate the expressiveness of our design, we present several examples built using our \hass{} implementation.
Figures depicting \hass{} (in \autoref{sec:intro}, \autoref{sec:overview-hass-rules}, and below) include excerpted style sheets that are unmodified except for whitespace for the purposes of formatting this paper and screenshots that are unmodified; as discussed below, the excerpt in \autoref{fig:even-odd-type-error} requires further explanation.

\newcommand{\exampleName}[1]
  {\textbf{#1}}

\parahead{Parse Information}

\begin{wrapfigure}[30]{r}{2.7in}
  \vspace{-0.04in} \includegraphics[width=2.7in]{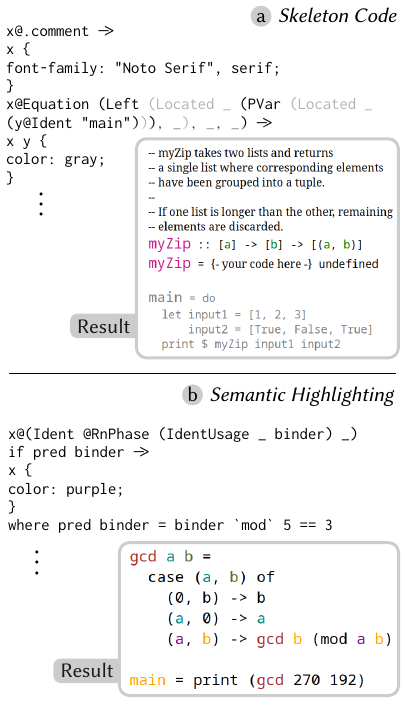}
  \caption{Parsing and Static Analysis Visualizations}
  \label{fig:skeleton-code-semantic-highlighting}
\end{wrapfigure}

\begin{figure}[b]
  \includegraphics[width=5.4in]{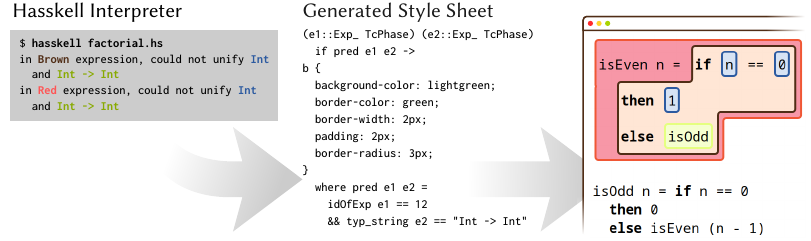}
\caption{Type Error Style Sheet.
The type checker generates style sheets in a generic way so that they work on any program.
This excerpt is an {instantiation} of the rules on one particular program,
but our prototype does not explicitly generate this file. Terminal interactions are depicted in gray.
}
\label{fig:even-odd-type-error}
\end{figure}

A hello-world example for \hass{} is the \exampleName{Syntax Highlighting} style sheet presented in \autoref{fig:hass-stylesheet-overview}.
While syntax highlighting is a general-purpose
visualization of program text, others are more {task-specific}.
\autoref{fig:skeleton-code-semantic-highlighting}\refBubble{a} presents a \exampleName{Skeleton Code} style sheet which
highlights one particular function among a skeleton of provided code,
``disabling'' it by styling it in gray.
Notice that the layout algorithm in \hass{} allows proportional fonts---in this case, to style comments---to be embedded within the document.

Sometimes parse information is needed but is not immediately clear from the program text itself.
One common scenario in Haskell involves debugging errors that stem from unexpected interaction between the precedence and associativity of the language's numerous infix operators.
 To ease this task, the \exampleName{Blocks} style sheet in \autoref{fig:cloc-blocks} uses border properties for each binary operation to convey the underlying structure.
(Similar styling, when applied to binding forms, has been referred to as ``scope highlighting''~\cite{Brown2023}.)

\parahead{Static Analysis Information}

Instead of coloring tokens based solely only on position in the parse tree, colors could indicate other semantically meaningful information.
For example, the \exampleName{Point-Free Pipeline} style sheet (\autoref{fig:cloc-pipeline} and \autoref{fig:hass-stylesheet-overview}) warns against chaining binary operators with different ``directions.''
(Technically, this example requires only parse information, but here we include it under the umbrella of lightweight static analysis tools like linters.)

As another example, consider a simple analysis that determines the binding site of each variable in the program, and a corresponding \exampleName{Semantic Highlighting} style sheet which colors each variable introduction and usage with the same color.
The style sheet excerpt in \autoref{fig:skeleton-code-semantic-highlighting}\refBubble{b} uses a boolean predicate to select only those \texttt{Ident} constructors which resolve to a specific binding---indicated by its \texttt{IdentUsage} annotation---and styles them to match.\footnote{This is commonly referred to as ``semantic highlighting''~\cite{vscode_semantic_highlighting}, but clearly there are other types of \emph{semantic} information that could be conveyed and other visual means for \emph{highlighting} such information besides highlighting syntax.}

\begin{wrapfigure}[37]{r}{2.7in}
  \vspace{-0.09in}
  \includegraphics[width=2.7in]{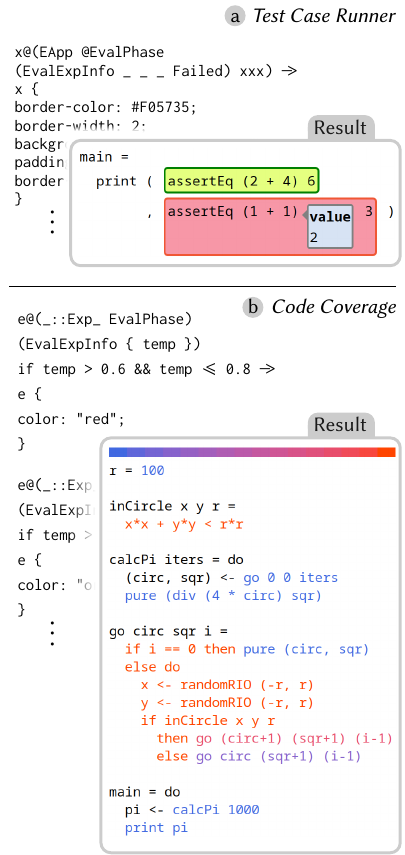}
\caption{Run-Time Visualizations}
  \label{fig:heat-map-assert}
\end{wrapfigure}
 The previous examples demonstrates that basic static analysis of the program can be surfaced through style sheets.
The next example demonstrates how to visualize type errors.
To start, the \hass{} type checker annotates expressions with inferred types, as well as unique identifiers which allow style sheets to style specific terms.
When type errors are emitted, the \hass{} interpreter generates a style sheet which (i) styles the offending expressions and (ii) highlights any subexpressions with a color corresponding to their type.

The \exampleName{Type Error} style sheet excerpt in \autoref{fig:even-odd-type-error} exemplifies how the latter goal is implemented: Selected expressions have type \mbox{\texttt{Int -> Int}} and are subexpressions of the expression with id 12.
Inspired by the error messages in Pyret~\citep{Pyret}, the colors used in the interpreter's terminal output correspond to those used in the style sheet.
The rendered display in \autoref{fig:even-odd-type-error} highlights a particularly attractive feature of this kind of visualization, which is that type errors that correspond to overlapping regions of the program can be more easily disambiguated using s-blocks, as opposed to the squiggle underline often seen in text editors and IDEs.
Future work could explore whether s-blocks would be a useful visual aid for explaining the provenance of type errors~\citep{Bhanuka2023,Zhao2024}.

\parahead{Run-Time Information}

In addition to static information, style sheets can access ASTs annotated with run-time information.
\autoref{fig:cloc-trace} shows a \exampleName{Projection Boxes}--style visualization~\cite{Lerner2020} with scrollable tables next to calls to the \texttt{trace} function (customized to record run-time environments).

In the \exampleName{Test Case Runner} style sheet shown in \autoref{fig:heat-map-assert}\refBubble{a}, the AST is annotated with a tag for each \texttt{assertEq} expression: \texttt{Passed} if the result of the expression is \texttt{True}, and \texttt{Failed} otherwise.
A pattern selects all function applications which contain this \texttt{Failed} tag, styling them red.
A custom implementation of \texttt{showStylish} injects a projection boxes--like view on all failing tests, showing the actual value of the expression inline with the program text.

While \autoref{fig:heat-map-assert}\refBubble{a} shows specific run-time values, the \exampleName{Heat Map} style sheet shown in \autoref{fig:heat-map-assert}\refBubble{b} shows how we can render a summary view of run-time information in the AST.
In this example, each expression in the program is colored according to how many times it was evaluated (normalized to the expression which was evaluated most).
The style sheet has a selector for each color and predicates are used color each expression's text based in its ``temperature.''

\parahead{Implementation Details}

Here we comment on several noteworthy aspects of these examples.
Complete style sheet definitions and additional details can be found in \refAppendixHassExamples{}. 

The Syntax Highlighting, Blocks, Point-Free Pipeline, and Skeleton Code style sheets work for all \texttt{Hasskell} programs.
The remaining views---Semantic Highlighting, Type Error, Projection Boxes, Test Case Runner, and Heat Map---work on a small \texttt{Hasskell} subset, for which we have implemented a variable resolution analysis, type checker, and a tracing interpreter.
Each of these phases adds analysis-specific information to the AST~\citep{Najd2016} (cf. \texttt{IdentUsage}, \texttt{TcExpInfo}, and \texttt{EvalExpInfo} in \autoref{fig:skeleton-code-semantic-highlighting}\refBubble{b}, \autoref{fig:even-odd-type-error}, and \autoref{fig:heat-map-assert}).
The code examples shown in \autoref{sec:intro} and \autoref{sec:evaluation} exercise all of the syntactic forms in the \texttt{Hasskell} subset for which we have currently implemented the aforementioned analyses.
Most of the example style sheets are \textit{program-agnostic}---without knowledge of the program that will be styled.
In contrast, the Type Error and Skeleton Code examples are \textit{program-specific}. Most of the examples rely on the standard \texttt{Stylish} instances, which emit only (styled) program text (cf.~\autoref{sec:hasskell}).
The Projection Boxes and Test Runner views, however, require separate \texttt{Stylish} instances---in these cases, to draw HTML tables in and around the program itself.
 
\section{Discussion}
\label{sec:discussion}

A motivation for this work is the lack of common frameworks for rendering programs with a flexible combination of nested textual and graphical elements.
The range of task-specific visualizations we presented in \hass{} suggest that our design is a fruitful step in this direction.

Regarding the first design goal (selecting AST values to style the view), the Projection Boxes, Test Runner, and Heat Map examples augment the (styled) program text with additional information (the rest do not), thus requiring an explicit data structure (stylish text) to track the connection between the two structurally different trees.
Regarding the second design goal (compactly rendering nested blocks of text), the Blocks, Point-Free Pipeline, Type Error, and Test Runner examples include nested multiline s-blocks; the remaining examples can be displayed using ordinary rectangles in a flat display.
These characteristics are summarized in \refAppendixHassExamples{}.

In terms of implementation effort, the style sheet fragments in \autoref{sec:evaluation}---and the summary statistics and full definitions in \refAppendixHassExamples{}---suggest that the implementation burden (for editor and tool builders) to achieve the example visualizations is reasonable,
but this needs to be investigated in future work.

\subsection{Limitations and Future Work}
\label{sec:limitations}

There are many ways to carry the ``CSS for code'' analogy further than considered in this paper.
Regarding examples, ours involve small programs and a small number of use cases.
We have not stretched the boundaries to determine what visualizations can and cannot be phrased in our framework.
Regarding implementation, our prototype is slow---0.76 seconds to run the syntax highlighting style sheet on a 129-line file, which exercises all of the syntactic forms in \hasskell{}---which must be optimized for more practical use.
Beyond these are considerations regarding the code style sheets design itself.

\parahead{Additional Style Sheet Mechanisms}

Our design includes the bare minimum from CSS: class selectors and path (``complex'') selectors.
Other constructs may be valuable to consider in our setting.
We discussed how the \cssClass{:has} pseudo-class in CSS can help select multi-node paths;
it can also select nodes that are \emph{not} on the path of nodes being styled.
For example, the selector ``\verb+.a:has(.c) .b+'' describes every path from a class-\cssClass{a} node to a class-\cssClass{b} node, where the former has some descendant with class \cssClass{c}.
\hass{} does not currently have a way to constrain paths in this way, though it could be useful. CSS also includes pseudo-\emph{elements} for displaying content not stored in the source value.
For example, setting the \cssAttr{content} attribute of the pseudo-element \verb+.test_case::before+ (or \verb+.test_case::after+) to be \texttt{"}\cmark\texttt{"} or \texttt{"}\xmark\texttt{"} would insert the given text before (or after) the given \cssClass{test\_case} element; this could be useful in a Test Runner style sheet (cf.~\autoref{fig:heat-map-assert}\refBubble{a}) rather than explicitly adding such annotations to the AST in a custom \texttt{Stylish} instance.

Furthermore, the relative simplicity of CSS has led to the design of many higher-level style sheet languages---such as Sass~\citep{sass}, Less~\citep{less}, and Stylus~\citep{stylus}---which offer variables, functions, etc. and compile to CSS.
Tree transformation languages, such as XSLT~\citep{XSLT}, XQuery~\citep{XQuery}, and XDuce~\citep{Hosoya2003}, offer expressive building blocks for querying richly-structured data, which may be useful to incorporate into a code style sheets language.
Going beyond syntactic querying constructs, one can also imagine a code style sheets design that relies on a relational query language.

\parahead{Additional Layout Constraints}

Our layout algorithm computes and renders s-blocks, which encapsulate multiline regions of text more compactly than rectangles.
There are additional visual constraints we might wish s-block layouts to satisfy.
Perhaps layout should aim for consistent line height rather than minimizing the height of each line, or ensure that horizontally aligned elements in the source text remain aligned after layout---which could help reduce the amount of vertical and horizontal whitespace shown in \autoref{fig:cloc-blocks}\refBubble{a}\refBubble{b}\refBubble{c}.
A layout system that considers additional relationships---perhaps building on CCSS~\citep{Badros1999}, a system that extends CSS with \emph{constraints}~\citep{Cassowary}---might produce more desirable program displays.
As the visual goals for layout become more complex, it may be useful to extend the work by \citet{Panchekha2018} that formalizes accessibility and usability specifications for CSS layouts with such constraints.

Beyond rectilinear shapes (rectangles in HTML, s-blocks in \hass{}, ``ragged'' blocks in Deuce~\cite{Hempel2018}), one can imagine other straight-edge shapes (e.g., convex hulls) or more amorphous contours that may be useful for displaying programs---even if they are less of a good fit with existing HTML and SVG layout systems.
(FFL~\citep{Wu2023} and Penrose~\citep{Ye2020}, for example, offer non-HTML based mechanisms for separating the definition of formulas and other diagrams from their styling.)

\parahead{Additional HTML Features}

View functions in \hass{} must be implemented to generate stylish text, which in our implementation (a) supports HTML only at the leaves of the document tree and (b) does not support interactive features of HTML and CSS.
As a result, it would be difficult to use \hass{} to display graphical decorations such as, say, arrows between s-blocks or tree-diagram representations of the program text.
However, the design is compatible with future extensions to more fully integrate textual, visual, and interactive elements.

The first step is to allow HTML elements to appear anywhere within a stylish text document rather than just the leaves.\footnote{
Because our current implementation treats \texttt{HtmlLeaf} strings as uninterpreted, the Projection Boxes and Test Runner examples (\refAppendixHassExamples{}) also include a raw CSS style sheet to define class selectors to style elements within the HTML table. This is another incidental limitation of current implementation.
}
We could extend the \texttt{StylishText} data type (\autoref{sec:hass}) such that each \verb+Node+---which stores a \texttt{path}, \texttt{classes}, \texttt{styles}, and \texttt{children}---also carries an HTML \texttt{tag} and \texttt{otherAttributes} relevant to that kind of HTML node.
In this more general setting, s-blocks can be understood as just another CSS layout algorithm (like \texttt{block}, \texttt{inline}, \texttt{inline-block}, or \texttt{flex}) that authors of view functions and authors of style sheets could decide to use.\footnote{
The term \emph{polyfill}~\cite{mdnDocs}---``a piece of code...used to provide modern functionality on older browsers that do not natively support it''---aptly describes our layout engine:
No browser (old or new) natively supports the \cssAttr{s-block} attribute,
so our algorithm (is a piece of code that) decomposes s-blocks into native elements (HTML and SVG) that can be drawn.
}

Once HTML elements are integrated, the next step would be to extend the client/server architecture of \hass{} to make the resulting views interactive.
One mechanism for interactivity provided by CSS is pseudo-classes---for example, the pseudo-class selector \texttt{button:hover} will select \htmlTag{button} elements only when actively hovered by the mouse.
Browser events such as these could be detected and forwarded to the \hass{} server for re-rendering---though it would be valuable to pursue incremental versions of the layout algorithm for efficiency.
Beyond these CSS features, a more fundamental source of interactivity in HTML are event handlers; these are discussed below.

\subsection{Structure and Projectional Editors}

We framed the context for this work in \autoref{sec:intro} and \autoref{sec:flat-nested} around the challenges of incorporating a richer variety of colors and typographical styles into program text, and integrating program text with graphical interfaces that display (i.e., project) information about the structure and behavior of the program.
Next, we describe how code style sheets relates to the design of graphical code editors.

\parahead{Text and Blocks}

In terms of layout,
designs for \emph{structure editors} are multivarious, ranging from \emph{block-} and \emph{frame-based editors} such as Scratch~\citep{Resnick2009}, Snap~\citep{Snap}, and Greenfoot~\citep{Kolling2015} that display programs largely as visual structures with fragments of text within, to \emph{heterogeneous}~\citep{Erwig1995} and \emph{projectional}~\citep{Fowler08Projectional} editors such as Barista~\citep{Ko2006}, DrRacket~\citep{Andersen2020}, Hazel~\citep{Omar2021}, and MPS~\cite{MPS} that display programs largely as text containing embedded GUIs.
Some editors also render \emph{program visualizations} inline with the program text~\cite{Hoffswell2018, Prong, Rauch2019, Omar2021, MPS}, overlaid on top~\cite{Lerner2020}, or off to the side~\cite{Omnicode}.

Existing structure and projectional editors do not compactly layout nested combinations of textual and graphical elements.
For example, Sandblocks~\citep{Beckmann2023} renders fully ``nested'' displays (\autoref{fig:sandblocks}) where each substructure is subject to additional indentation, unlike the underlying, naturally formatted text.
To avoid inefficient use of visual space, other structure editors forgo the use of styles that affect vertical positioning.
For example, \citet[Fig.~5]{TeenTylr} present an example where, in their recent version of Hazel's structure editor, horizontal decorations indicate structural information about the AST. Notice, however, that these decorations do not alter vertical border spacing of nested elements because the text is fundamentally aligned to a grid as usual in a conventional, ``flat'' display.
Similarly, \citet[Sec.~5.3]{Omar2021} explain that the layout of \emph{live literals} in Hazel---which are interactive, user-defined GUIs that can be displayed in place of textual literals in the program---``relies fundamentally on character counts.''
That is, the placement of graphical elements assumes that all textual elements are, again, arranged in a grid-like fashion.
(Regarding flat displays, although it does not display program text directly, Microsoft's Language Server Protocol (LSP) \cite{LSP} provides an interface that can be used to visualize the output of various program analyses.)

For such graphical editors to scale for more practical use, it seems necessary to more smoothly integrate graphical elements (e.g.~blocks) and text.
Some tools, such as Blockly~\citep{Blockly}, Pencil Code~\citep{PencilCode, Droplet}, MakeCode~\citep{MakeCode}, and App Inventor~\citep{AppInventorBlocks1, AppInventorBlocks2}, nominally convert between blocks and text. However, merely switching between distinct nested and flat displays does not seem to be an optimal way to bridge the divide.
In contrast, the visual and layout properties of s-blocks may provide a foundation for styling nested combinations of textual and graphical elements with greater harmony.

\parahead{Interactive GUIs and Text Editing}

Having discussed the \emph{graphical} aspects of graphical editors, next we discuss their characteristics as \emph{editors}.
Structure editors generally forgo a global text buffer, instead offering structural program transformations---through keyboard- and/or mouse-based interactions---with ordinary text editing restricted to delimited substructures.

In addition to built-in structure-editing interactions, several editors allow libraries and users to define custom, type-specific interfaces---referred to as \emph{views}~\cite{Ko2006}, \emph{palettes}~\cite{ActiveCodeCompletion}, \emph{livelits}~\cite{Omar2021}, and \emph{projections}~\cite{MPS}---for displaying, and often editing, the program.
In future editors designed with code style sheets, s-blocks could be equipped with event handlers which modify the AST.
(Event handlers would also provide means for interacting with the server in order to edit and save the code files themselves.)
Building on \hass{}, the simplest approach would be to allow event handlers defined as uninterpreted JavaScript code strings in \texttt{otherAttributes} (cf.~\autoref{sec:limitations}) to directly transform the AST.
A more principled approach could use the same language as that for the implementation of the code style sheets system---or even the object language that is being styled in the system.
There is a rich design space for implementing editor actions---for documents in general~\citep{Crichton2024} and programs in particular~\citep{Omar2017,Omar2021}---that could be exposed through s-blocks in the user interface.

Beyond interactivity through built-in and custom structure-editing interactions, some hybrid editors aim to provide natural support for text editing across syntactic subtrees.
For example, Hazel's recent structure editor, tylr~\cite{TeenTylr}, allows text-edits that temporarily violate, and later restore, structural properties.
In similar spirit, Sandblocks~\cite{Beckmann2023} supports familiar text-editing behaviors for copy-pasting, navigation, and code completion, and also employs heuristics for reconciling text-edits that do not form valid syntactic structures.
Future work could integrate such techniques with code style sheets---allowing text-editing interactions within and across s-blocks, preserving or modifying their visual styles in accordance with ongoing changes to the program structure.

\parahead{Other Considerations}

Several factors are crucial to the design of structure editors.
For example, the Cornell Program Synthesizer~\cite{cornellProgramSynthesizer}, MPS~\cite{Voelter2012, Voelter2016GrammarCells, Voelter2016IncA, Voelter2018, Voelter2019}, and Hazel~\cite{Omar2017, HazelnutLive} provide incremental static analyses for programs undergoing structural edits; Hazel, furthermore, partially evaluates programs with holes (through a form of ``incremental'' evaluation).
In addition, the CPS, MPS, and Sandblocks are language-agnostic: they are equipped to derive structure-editing interfaces from grammars.
These concerns are largely orthogonal to visual presentation as addressed in our work.

\subsection{Program Visualization}

In a taxonomy by \citet{VPandPV}, ``visual programming'' refers to systems that allow users to edit programs using more graphical representations than text---structure and projectional editors, as described above, fall into this category---whereas ``program visualization'' refers to systems that graphically render information about programs but where programs are written using conventional text editors.
Using their categorization and terminology, \hass{} is a system for defining program visualizations that illustrate \emph{code} (as opposed to \emph{data} or \emph{algorithms}) and are \emph{static} (as opposed to \emph{dynamic}).
Further refining their categorization of code visualizations, \hass{} supports \emph{program-text visualizations}, in which information is conveyed by graphically decorating the textual representation of the code as opposed to, say, a flowchart, or control- and data-flow graphs.
(COMEX~\cite{COMEX} is one example framework for rendering the latter kinds of code visualizations.)

\subsection{Usability}

Like the related works above, ultimately, the idea of code style sheets---to enable richer displays of programs besides flat, syntax-highlighted text---aims to improve some aspect of the user's experience.
Future work is needed to evaluate the usability of our code style sheets system, both for authors and users of code editors and visualization tools.
(Developing a ``DOM inspector for \hass{}'' could be one important tool to support the user-customization workflow sketched in \autoref{sec:intro}.)

The preliminary usability evaluations about alternatives to standard syntax highlighting
leave plenty of room for studying the potential benefits, if any, of such choices~\citep{Tapp1994,Oliveira2020,Jacques2015,Talsma2021,Weintrop2019,Brown2023}.
As \citet{Brown2023} emphasize, it may turn out that syntax highlighting or other variations may not have demonstrable benefits, noting  ``a parallel issue of font selection for general text readability; although there are various design guidelines on when to use serif or sans serif fonts, many studies find no effect of font serifs, or lack thereof.''
Intuitively, however, fundamentally different layout choices seem well worth considering,
as \citet{Baecker1988} suggests, to make code ``more readable, more comprehensible, more vivid, more appealing, more memorable, more useful, [and] more maintainable.''
A flexible framework for rendering programs with a combination of textual and graphical decorations may help facilitate this endeavor.

\begin{acks}

The authors would like to thank Anton Outkine for discussions about the layout algorithm, as well as Andrew McNutt and anonymous reviewers for many suggestions that improved this paper.

\end{acks}
 
\clearpage
\subsection*{Data Availability Statement}

Our prototype code style sheets system, \hass{}, together with the examples described in the paper and included in \refAppendixHassExamples{}, is available at \url{https://github.com/sbcohen2000/code-style-sheets-artifact} and in an archival repository~\cite{HassArtifact}.

\hass{} contains three \texttt{StylishText} (\autoref{sec:hass}) instances for \hasskell{} programs:
one ``default'' instance for an AST which supports most Haskell syntax;
one ``default'' instance for an AST that supports a small subset of Haskell syntax for which we have implemented the program analyses that are demonstrated in the examples; and
one custom instance that extends the latter with support for rendering projection boxes.
\hass{} implements the style computation algorithm (\autoref{sec:style-sheets}, \refAppendixStyleComputation{})---with incidental differences as discussed in \autoref{sec:hass}---and layout algorithm (\autoref{sec:layout}) and a web server so that the examples may be compiled and viewed in a web browser.

When implementing \hass{}, we prioritized features which would be important for constructing examples in the paper.
As a result, a user of \hass{} should be able to easily reproduce all of the examples shown.
However, they may have more difficulty constructing their own examples, since many ``ease of use'' features are not implemented.
Such limitations are discussed throughout the paper (\autoref{sec:hass}, \autoref{sec:evaluation}, \autoref{sec:limitations}, \refAppendixHassExamples{}).
 
\bibliographystyle{ACM-Reference-Format}
\bibliography{css}


\begin{thebibliography}{77}


\ifx \showCODEN    \undefined \def \showCODEN     #1{\unskip}     \fi
\ifx \showISBNx    \undefined \def \showISBNx     #1{\unskip}     \fi
\ifx \showISBNxiii \undefined \def \showISBNxiii  #1{\unskip}     \fi
\ifx \showISSN     \undefined \def \showISSN      #1{\unskip}     \fi
\ifx \showLCCN     \undefined \def \showLCCN      #1{\unskip}     \fi
\ifx \shownote     \undefined \def \shownote      #1{#1}          \fi
\ifx \showarticletitle \undefined \def \showarticletitle #1{#1}   \fi
\ifx \showURL      \undefined \def \showURL       {\relax}        \fi
\providecommand\bibfield[2]{#2}
\providecommand\bibinfo[2]{#2}
\providecommand\natexlab[1]{#1}
\providecommand\showeprint[2][]{arXiv:#2}

\bibitem[Acar et~al\mbox{.}(2013)]%
        {TML}
\bibfield{author}{\bibinfo{person}{Umut~A. Acar}, \bibinfo{person}{Amal Ahmed},
  \bibinfo{person}{James Cheney}, {and} \bibinfo{person}{Roly Perera}.}
  \bibinfo{year}{2013}\natexlab{}.
\newblock \showarticletitle{A Core Calculus for Provenance}.
\newblock \bibinfo{journal}{\emph{Journal of Computer Security}}
  (\bibinfo{year}{2013}).
\newblock


\bibitem[Andersen et~al\mbox{.}(2020)]%
        {Andersen2020}
\bibfield{author}{\bibinfo{person}{Leif Andersen}, \bibinfo{person}{Michael
  Ballantyne}, {and} \bibinfo{person}{Matthias Felleisen}.}
  \bibinfo{year}{2020}\natexlab{}.
\newblock \showarticletitle{{Adding Interactive Visual Syntax to Textual
  Code}}.
\newblock \bibinfo{journal}{\emph{Object Oriented Programming Systems Languages
  {\&} Applications {OOPSLA}}} (\bibinfo{year}{2020}).
\newblock
\href{https://doi.org/10.1145/3428290}{doi:\nolinkurl{10.1145/3428290}}


\bibitem[Badros et~al\mbox{.}(1999)]%
        {Badros1999}
\bibfield{author}{\bibinfo{person}{Greg~J. Badros}, \bibinfo{person}{Alan
  Borning}, \bibinfo{person}{Kim Marriott}, {and} \bibinfo{person}{Peter
  Stuckey}.} \bibinfo{year}{1999}\natexlab{}.
\newblock \showarticletitle{{Constraint Cascading Style Sheets for the Web}}.
  In \bibinfo{booktitle}{\emph{Symposium on User Interface Software and
  Technology (UIST)}}.
\newblock
\href{https://doi.org/10.1145/320719.322588}{doi:\nolinkurl{10.1145/320719.322588}}


\bibitem[Badros et~al\mbox{.}(2001)]%
        {Cassowary}
\bibfield{author}{\bibinfo{person}{Greg~J. Badros}, \bibinfo{person}{Alan
  Borning}, {and} \bibinfo{person}{Peter~J. Stuckey}.}
  \bibinfo{year}{2001}\natexlab{}.
\newblock \showarticletitle{{The Cassowary Linear Arithmetic Constraint Solving
  Algorithm}}.
\newblock \bibinfo{journal}{\emph{ACM Transactions on Computer-Human
  Interaction}} (\bibinfo{year}{2001}).
\newblock
\href{https://doi.org/10.1145/504704.504705}{doi:\nolinkurl{10.1145/504704.504705}}


\bibitem[Baecker(1988)]%
        {Baecker1988}
\bibfield{author}{\bibinfo{person}{Ronald Baecker}.}
  \bibinfo{year}{1988}\natexlab{}.
\newblock \showarticletitle{{Enhancing Program Readability and
  Comprehensibility with Tools for Program Visualization}}. In
  \bibinfo{booktitle}{\emph{International Conference on Software Engineering
  (ICSE)}}.
\newblock
\urldef\tempurl%
\url{https://dl.acm.org/doi/10.5555/55823.55858}
\showURL{%
\tempurl}


\bibitem[Ball et~al\mbox{.}(2019)]%
        {MakeCode}
\bibfield{author}{\bibinfo{person}{Thomas Ball}, \bibinfo{person}{Abhijith
  Chatra}, \bibinfo{person}{Peli de Halleux}, \bibinfo{person}{Steve Hodges},
  \bibinfo{person}{Micha\l{} Moskal}, {and} \bibinfo{person}{Jacqueline
  Russell}.} \bibinfo{year}{2019}\natexlab{}.
\newblock \showarticletitle{{Microsoft MakeCode: Embedded Programming for
  Education, in Blocks and TypeScript}}. In
  \bibinfo{booktitle}{\emph{Proceedings of the 2019 ACM SIGPLAN Symposium on
  SPLASH-E}}.
\newblock
\href{https://doi.org/10.1145/3358711.3361630}{doi:\nolinkurl{10.1145/3358711.3361630}}


\bibitem[Bau(2024)]%
        {Droplet}
\bibfield{author}{\bibinfo{person}{Anthony Bau}.}
  \bibinfo{year}{2014--2024}\natexlab{}.
\newblock \bibinfo{title}{{Droplet: Blocks and Text Together}}.
\newblock
\urldef\tempurl%
\url{https://droplet-editor.github.io/}
\showURL{%
\tempurl}


\bibitem[Bau et~al\mbox{.}(2015)]%
        {PencilCode}
\bibfield{author}{\bibinfo{person}{David Bau}, \bibinfo{person}{D.~Anthony
  Bau}, \bibinfo{person}{Mathew Dawson}, {and} \bibinfo{person}{C.~Sydney
  Pickens}.} \bibinfo{year}{2015}\natexlab{}.
\newblock \showarticletitle{{Pencil Code: Block Code for a Text World}}. In
  \bibinfo{booktitle}{\emph{International Conference on Interaction Design and
  Children (IDC)}}.
\newblock
\href{https://doi.org/10.1145/2771839.2771875}{doi:\nolinkurl{10.1145/2771839.2771875}}


\bibitem[Beckmann et~al\mbox{.}(2023)]%
        {Beckmann2023}
\bibfield{author}{\bibinfo{person}{Tom Beckmann}, \bibinfo{person}{Patrick
  Rein}, \bibinfo{person}{Stefan Ramson}, \bibinfo{person}{Joana Bergsiek},
  {and} \bibinfo{person}{Robert Hirschfeld}.} \bibinfo{year}{2023}\natexlab{}.
\newblock \showarticletitle{{Structured Editing for All: Deriving Usable
  Structured Editors from Grammars}}. In \bibinfo{booktitle}{\emph{{Conference
  on Human Factors in Computing Systems (CHI)}}}.
\newblock
\href{https://doi.org/10.1145/3544548.3580785}{doi:\nolinkurl{10.1145/3544548.3580785}}


\bibitem[Bhanuka et~al\mbox{.}(2023)]%
        {Bhanuka2023}
\bibfield{author}{\bibinfo{person}{Ishan Bhanuka}, \bibinfo{person}{Lionel
  Parreaux}, \bibinfo{person}{David Binder}, {and}
  \bibinfo{person}{Jonathan~Immanuel Brachth\"{a}user}.}
  \bibinfo{year}{2023}\natexlab{}.
\newblock \showarticletitle{{Getting into the Flow: Towards Better Type Error
  Messages for Constraint-Based Type Inference}}.
\newblock \bibinfo{journal}{\emph{Proceedings of the ACM on Programming
  Languages (PACMPL)}} \bibinfo{number}{OOPSLA} (\bibinfo{year}{2023}).
\newblock
\href{https://doi.org/10.1145/3622812}{doi:\nolinkurl{10.1145/3622812}}


\bibitem[{Brown PLT}(2018)]%
        {Pyret}
\bibfield{author}{\bibinfo{person}{{Brown PLT}}.}
  \bibinfo{year}{2018}\natexlab{}.
\newblock \bibinfo{title}{{Picking Colors for Pyret Error Messages}}.
\newblock
\urldef\tempurl%
\url{https://blog.brownplt.org/2018/06/11/philogenic-colors.html}
\showURL{%
\tempurl}


\bibitem[Brunsfeld et~al\mbox{.}(2024)]%
        {tree-sitter}
\bibfield{author}{\bibinfo{person}{Max Brunsfeld} {et~al\mbox{.}}}
  \bibinfo{year}{2013--2024}\natexlab{}.
\newblock \bibinfo{title}{{Tree-Sitter}}.
\newblock
\urldef\tempurl%
\url{https://tree-sitter.github.io/tree-sitter/}
\showURL{%
\tempurl}


\bibitem[Catlin et~al\mbox{.}(2006)]%
        {sass}
\bibfield{author}{\bibinfo{person}{Hampton Catlin}, \bibinfo{person}{Natalie
  Weizenbaum}, {and} \bibinfo{person}{Chris Eppstein}.}
  \bibinfo{year}{2006}\natexlab{}.
\newblock \bibinfo{title}{{Sass: Syntactically Awesome Style Sheets}}.
\newblock
\urldef\tempurl%
\url{https://sass-lang.com/}
\showURL{%
\tempurl}


\bibitem[Chadha(2014)]%
        {AppInventorBlocks1}
\bibfield{author}{\bibinfo{person}{Karishma Chadha}.}
  \bibinfo{year}{2014}\natexlab{}.
\newblock \bibinfo{title}{{Improving the Usability of App Inventor through
  Conversion Between Blocks and Text}}.
\newblock
\urldef\tempurl%
\url{https://repository.wellesley.edu/object/ir495}
\showURL{%
\tempurl}


\bibitem[Cheney et~al\mbox{.}(2009)]%
        {Cheney2009}
\bibfield{author}{\bibinfo{person}{James Cheney}, \bibinfo{person}{Laura
  Chiticariu}, {and} \bibinfo{person}{Wang-Chiew Tan}.}
  \bibinfo{year}{2009}\natexlab{}.
\newblock \showarticletitle{{Provenance in Databases: Why, How, and Where}}.
\newblock \bibinfo{journal}{\emph{Foundations and Trends in Databases}}
  (\bibinfo{year}{2009}).
\newblock
\href{https://doi.org/10.1561/1900000006}{doi:\nolinkurl{10.1561/1900000006}}


\bibitem[Cohen(2025)]%
        {HassArtifact}
\bibfield{author}{\bibinfo{person}{Sam Cohen}.}
  \bibinfo{year}{2025}\natexlab{}.
\newblock \bibinfo{title}{{\hass{}: Code Style Sheets Artifact}}.
\newblock
\href{https://doi.org/10.5281/zenodo.14920421}{doi:\nolinkurl{10.5281/zenodo.14920421}}


\bibitem[Cohen and Chugh(2025)]%
        {Hass}
\bibfield{author}{\bibinfo{person}{Sam Cohen} {and} \bibinfo{person}{Ravi
  Chugh}.} \bibinfo{year}{2025}\natexlab{}.
\newblock \bibinfo{title}{{Code Style Sheets: CSS for Code (Extended Version of
  OOPSLA 2025 Paper)}}.
\newblock
\href{https://doi.org/10.48550/arXiv.2502.09386}{doi:\nolinkurl{10.48550/arXiv.2502.09386}}


\bibitem[Coyier(2020)]%
        {CSSTricksNumberScrubbing}
\bibfield{author}{\bibinfo{person}{Chris Coyier}.}
  \bibinfo{year}{2020}\natexlab{}.
\newblock \bibinfo{title}{{Number Scrubbing}}.
\newblock
  \bibinfo{howpublished}{\url{https://css-tricks.com/number-scrubbing/}}.
\newblock


\bibitem[Crichton and Krishnamurthi(2024)]%
        {Crichton2024}
\bibfield{author}{\bibinfo{person}{Will Crichton} {and}
  \bibinfo{person}{Shriram Krishnamurthi}.} \bibinfo{year}{2024}\natexlab{}.
\newblock \showarticletitle{{A Core Calculus for Documents}}.
\newblock \bibinfo{journal}{\emph{Proceedings of the ACM on Programming
  Languages (PACMPL)}} \bibinfo{number}{POPL} (\bibinfo{year}{2024}).
\newblock


\bibitem[Das et~al\mbox{.}(2023)]%
        {COMEX}
\bibfield{author}{\bibinfo{person}{Debeshee Das}, \bibinfo{person}{Noble~Saji
  Mathews}, \bibinfo{person}{Alex Mathai}, \bibinfo{person}{Srikanth
  Tamilselvam}, \bibinfo{person}{Kranthi Sedamaki}, \bibinfo{person}{Sridhar
  Chimalakonda}, {and} \bibinfo{person}{Atul Kumar}.}
  \bibinfo{year}{2023}\natexlab{}.
\newblock \showarticletitle{{COMEX: A Tool for Generating Customized Source
  Code Representations}}. In \bibinfo{booktitle}{\emph{Conference on Automated
  Software Engineering (ASE)}}.
\newblock
\href{https://doi.org/10.1109/ASE56229.2023.00010}{doi:\nolinkurl{10.1109/ASE56229.2023.00010}}


\bibitem[Erwig and Meyer(1995)]%
        {Erwig1995}
\bibfield{author}{\bibinfo{person}{Martin Erwig} {and} \bibinfo{person}{Bernd
  Meyer}.} \bibinfo{year}{1995}\natexlab{}.
\newblock \showarticletitle{{Heterogeneous Visual Languages: Integrating Visual
  and Textual Programming}}.
\newblock \bibinfo{journal}{\emph{Symposium on Visual Languages (VL)}}
  (\bibinfo{year}{1995}).
\newblock
\urldef\tempurl%
\url{https://api.semanticscholar.org/CorpusID:17004705}
\showURL{%
\tempurl}


\bibitem[Fowler(2008)]%
        {Fowler08Projectional}
\bibfield{author}{\bibinfo{person}{Martin Fowler}.}
  \bibinfo{year}{2008}\natexlab{}.
\newblock \bibinfo{title}{Projectional Editing}.
\newblock
\urldef\tempurl%
\url{https://martinfowler.com/bliki/ProjectionalEditing.html}
\showURL{%
\tempurl}


\bibitem[{Google}(2024)]%
        {Blockly}
\bibfield{author}{\bibinfo{person}{{Google}}.}
  \bibinfo{year}{2013--2024}\natexlab{}.
\newblock \bibinfo{title}{{Blockly}}.
\newblock
\urldef\tempurl%
\url{https://developers.google.com/blockly}
\showURL{%
\tempurl}


\bibitem[Harvey et~al\mbox{.}(2013)]%
        {Snap}
\bibfield{author}{\bibinfo{person}{Brian Harvey}, \bibinfo{person}{Daniel~D.
  Garcia}, \bibinfo{person}{Tiffany Barnes}, \bibinfo{person}{Nathaniel
  Titterton}, \bibinfo{person}{Daniel Armendariz}, \bibinfo{person}{Luke
  Segars}, \bibinfo{person}{Eugene Lemon}, \bibinfo{person}{Sean Morris}, {and}
  \bibinfo{person}{Josh Paley}.} \bibinfo{year}{2013}\natexlab{}.
\newblock \showarticletitle{{SNAP! (Build Your Own Blocks)}}. In
  \bibinfo{booktitle}{\emph{Technical Symposium on Computer Science Education
  (SIGCSE TS)}}.
\newblock
\href{https://doi.org/10.1145/2445196.2445507}{doi:\nolinkurl{10.1145/2445196.2445507}}


\bibitem[Haverbeke(2024)]%
        {CodeMirror}
\bibfield{author}{\bibinfo{person}{Marijn Haverbeke}.}
  \bibinfo{year}{2007--2024}\natexlab{}.
\newblock \bibinfo{title}{CodeMirror: Extensible Code Editor}.
\newblock
\urldef\tempurl%
\url{https://codemirror.net/}
\showURL{%
\tempurl}


\bibitem[Hempel and Chugh(2020)]%
        {hempel2020tiny}
\bibfield{author}{\bibinfo{person}{Brian Hempel} {and} \bibinfo{person}{Ravi
  Chugh}.} \bibinfo{year}{2020}\natexlab{}.
\newblock \showarticletitle{{Tiny Structure Editors for Low, Low Prices!
  (Generating GUIs from toString Functions)}}. In
  \bibinfo{booktitle}{\emph{Symposium on Visual Languages and Human-Centric
  Computing (VL/HCC)}}. IEEE, \bibinfo{pages}{1--5}.
\newblock
\href{https://doi.org/10.1109/VL/HCC50065.2020.9127256}{doi:\nolinkurl{10.1109/VL/HCC50065.2020.9127256}}


\bibitem[Hempel et~al\mbox{.}(2018)]%
        {Hempel2018}
\bibfield{author}{\bibinfo{person}{Brian Hempel}, \bibinfo{person}{Justin
  Lubin}, \bibinfo{person}{Grace Lu}, {and} \bibinfo{person}{Ravi Chugh}.}
  \bibinfo{year}{2018}\natexlab{}.
\newblock \showarticletitle{{Deuce: A Lightweight User Interface for Structured
  Editing}}. In \bibinfo{booktitle}{\emph{International Conference on Software
  Engineering (ICSE)}}.
\newblock
\href{https://doi.org/10.1145/3180155.3180165}{doi:\nolinkurl{10.1145/3180155.3180165}}


\bibitem[Hoffswell et~al\mbox{.}(2018)]%
        {Hoffswell2018}
\bibfield{author}{\bibinfo{person}{Jane Hoffswell}, \bibinfo{person}{Arvind
  Satyanarayan}, {and} \bibinfo{person}{Jeffrey Heer}.}
  \bibinfo{year}{2018}\natexlab{}.
\newblock \showarticletitle{{Augmenting Code with In Situ Visualizations to Aid
  Program Understanding}}. In \bibinfo{booktitle}{\emph{{Conference on Human
  Factors in Computing Systems (CHI)}}}.
\newblock
\showISBNx{9781450356206}
\href{https://doi.org/10.1145/3173574.3174106}{doi:\nolinkurl{10.1145/3173574.3174106}}


\bibitem[Holowaychuk et~al\mbox{.}(2010)]%
        {stylus}
\bibfield{author}{\bibinfo{person}{TJ Holowaychuk} {et~al\mbox{.}}}
  \bibinfo{year}{2010}\natexlab{}.
\newblock \bibinfo{title}{{Stylus}}.
\newblock
\urldef\tempurl%
\url{https://stylus-lang.com/}
\showURL{%
\tempurl}


\bibitem[Hosoya and Pierce(2003)]%
        {Hosoya2003}
\bibfield{author}{\bibinfo{person}{Haruo Hosoya} {and}
  \bibinfo{person}{Benjamin~C. Pierce}.} \bibinfo{year}{2003}\natexlab{}.
\newblock \showarticletitle{{XDuce: A Statically Typed XML Processing
  Language}}.
\newblock \bibinfo{journal}{\emph{ACM Transactions on Internet Technology}}
  (\bibinfo{year}{2003}).
\newblock
\href{https://doi.org/10.1145/767193.767195}{doi:\nolinkurl{10.1145/767193.767195}}


\bibitem[Huang and Turbak(2019)]%
        {AppInventorBlocks2}
\bibfield{author}{\bibinfo{person}{Ruanqianqian Huang} {and}
  \bibinfo{person}{Franklyn Turbak}.} \bibinfo{year}{2019}\natexlab{}.
\newblock \showarticletitle{{A Design for Bidirectional Conversion between
  Blocks and Text for App Inventor}}. In \bibinfo{booktitle}{\emph{Blocks and
  Beyond Workshop (B{\&}B)}}.
\newblock
\href{https://doi.org/10.1109/BB48857.2019.8941197}{doi:\nolinkurl{10.1109/BB48857.2019.8941197}}


\bibitem[Jacques and Kristensson(2015)]%
        {Jacques2015}
\bibfield{author}{\bibinfo{person}{Jason~T. Jacques} {and}
  \bibinfo{person}{Per~Ola Kristensson}.} \bibinfo{year}{2015}\natexlab{}.
\newblock \showarticletitle{{Understanding the Effects of Code Presentation}}.
  In \bibinfo{booktitle}{\emph{Workshop on Evaluation and Usability of
  Programming Languages and Tools (PLATEAU)}}.
\newblock
\href{https://doi.org/10.1145/2846680.2846685}{doi:\nolinkurl{10.1145/2846680.2846685}}


\bibitem[{JetBrains}(2024)]%
        {MPS}
\bibfield{author}{\bibinfo{person}{{JetBrains}}.}
  \bibinfo{year}{2011--2024}\natexlab{}.
\newblock \bibinfo{title}{{MPS (Meta Programming System)}}.
\newblock
\urldef\tempurl%
\url{https://en.wikipedia.org/wiki/JetBrains_MPS}
\showURL{%
\tempurl}


\bibitem[Kang and Guo(2017)]%
        {Omnicode}
\bibfield{author}{\bibinfo{person}{Hyeonsu Kang} {and}
  \bibinfo{person}{Philip~J Guo}.} \bibinfo{year}{2017}\natexlab{}.
\newblock \showarticletitle{{Omnicode: A Novice-oriented Live Programming
  Environment with Always-on Run-time Value Visualizations}}. In
  \bibinfo{booktitle}{\emph{{Symposium on User Interface Software and
  Technology (UIST)}}}.
\newblock
\href{https://doi.org/10.1145/3126594.3126632}{doi:\nolinkurl{10.1145/3126594.3126632}}


\bibitem[Ko and Myers(2006)]%
        {Ko2006}
\bibfield{author}{\bibinfo{person}{Amy~J. Ko} {and} \bibinfo{person}{Brad~A.
  Myers}.} \bibinfo{year}{2006}\natexlab{}.
\newblock \showarticletitle{{Barista: An Implementation Framework for Enabling
  New Tools, Interaction Techniques and Views in Code Editors}}. In
  \bibinfo{booktitle}{\emph{{Conference on Human Factors in Computing Systems
  (CHI)}}}.
\newblock
\href{https://doi.org/10.1145/1124772.1124831}{doi:\nolinkurl{10.1145/1124772.1124831}}


\bibitem[K\"{o}lling et~al\mbox{.}(2015)]%
        {Kolling2015}
\bibfield{author}{\bibinfo{person}{Michael K\"{o}lling}, \bibinfo{person}{Neil
  C.~C. Brown}, {and} \bibinfo{person}{Amjad Altadmri}.}
  \bibinfo{year}{2015}\natexlab{}.
\newblock \showarticletitle{{Frame-Based Editing: Easing the Transition from
  Blocks to Text-Based Programming}}. In \bibinfo{booktitle}{\emph{Workshop in
  Primary and Secondary Computing Education (WiPSCE)}}.
\newblock
\href{https://doi.org/10.1145/2818314.2818331}{doi:\nolinkurl{10.1145/2818314.2818331}}


\bibitem[L\"{a}mmel and Jones(2003)]%
        {syb}
\bibfield{author}{\bibinfo{person}{Ralf L\"{a}mmel} {and}
  \bibinfo{person}{Simon~Peyton Jones}.} \bibinfo{year}{2003}\natexlab{}.
\newblock \showarticletitle{{Scrap Your Boilerplate: A Practical Design Pattern
  for Generic Programming}}. In \bibinfo{booktitle}{\emph{International
  Workshop on Types in Languages Design and Implementation (TLDI)}}.
\newblock
\href{https://doi.org/10.1145/604174.604179}{doi:\nolinkurl{10.1145/604174.604179}}


\bibitem[Lerner(2020)]%
        {Lerner2020}
\bibfield{author}{\bibinfo{person}{Sorin Lerner}.}
  \bibinfo{year}{2020}\natexlab{}.
\newblock \showarticletitle{{Projection Boxes: On-the-fly Reconfigurable
  Visualization for Live Programming}}. In
  \bibinfo{booktitle}{\emph{{Conference on Human Factors in Computing Systems
  (CHI)}}}.
\newblock
\href{https://doi.org/10.1145/3313831.3376494}{doi:\nolinkurl{10.1145/3313831.3376494}}


\bibitem[Mainland(2007)]%
        {QuasiQuotes}
\bibfield{author}{\bibinfo{person}{Geoffrey Mainland}.}
  \bibinfo{year}{2007}\natexlab{}.
\newblock \showarticletitle{Why It's Nice to Be Quoted: Quasiquoting for
  Haskell}. In \bibinfo{booktitle}{\emph{Proceedings of the ACM SIGPLAN
  Workshop on Haskell Workshop}}.
\newblock
\href{https://doi.org/10.1145/1291201.1291211}{doi:\nolinkurl{10.1145/1291201.1291211}}


\bibitem[Marcus and Baecker(1982)]%
        {Marcus1982}
\bibfield{author}{\bibinfo{person}{A. Marcus} {and} \bibinfo{person}{R.
  Baecker}.} \bibinfo{year}{1982}\natexlab{}.
\newblock \showarticletitle{{On the Graphic Design of Program Text}}. In
  \bibinfo{booktitle}{\emph{Proceedings of Graphics Interface (GI)}}.
\newblock
\urldef\tempurl%
\url{http://graphicsinterface.org/wp-content/uploads/gi1982-45.pdf}
\showURL{%
\tempurl}


\bibitem[McNutt and Chugh(2023)]%
        {Prong}
\bibfield{author}{\bibinfo{person}{Andrew McNutt} {and} \bibinfo{person}{Ravi
  Chugh}.} \bibinfo{year}{2023}\natexlab{}.
\newblock \showarticletitle{{Projectional Editors for JSON-Based DSLs}}. In
  \bibinfo{booktitle}{\emph{Symposium on Visual Languages and Human-Centric
  Computing (VL/HCC)}}.
\newblock
\href{https://doi.org/10.1109/VL-HCC57772.2023.00015}{doi:\nolinkurl{10.1109/VL-HCC57772.2023.00015}}


\bibitem[Microsoft(2024b)]%
        {VSCode}
\bibfield{author}{\bibinfo{person}{Microsoft}.}
  \bibinfo{year}{2015--2024}\natexlab{b}.
\newblock \bibinfo{title}{Visual Studio Code}.
\newblock
\urldef\tempurl%
\url{https://code.visualstudio.com/}
\showURL{%
\tempurl}


\bibitem[Microsoft(2024a)]%
        {LSP}
\bibfield{author}{\bibinfo{person}{Microsoft}.}
  \bibinfo{year}{2016--2024}\natexlab{a}.
\newblock \bibinfo{title}{Language Server Protocol}.
\newblock
\urldef\tempurl%
\url{https://microsoft.github.io/language-server-protocol/}
\showURL{%
\tempurl}


\bibitem[{Microsoft}(2023)]%
        {vscode_semantic_highlighting}
\bibfield{author}{\bibinfo{person}{{Microsoft}}.}
  \bibinfo{year}{2023}\natexlab{}.
\newblock \bibinfo{title}{{Semantic Highlighting Overview (in VS Code)}}.
\newblock
\urldef\tempurl%
\url{https://github.com/microsoft/vscode/wiki/Semantic-Highlighting-Overview}
\showURL{%
\tempurl}


\bibitem[Moon et~al\mbox{.}(2023)]%
        {TeenTylr}
\bibfield{author}{\bibinfo{person}{David Moon}, \bibinfo{person}{Andrew Blinn},
  {and} \bibinfo{person}{Cyrus Omar}.} \bibinfo{year}{2023}\natexlab{}.
\newblock \showarticletitle{{Gradual Structure Editing with Obligations}}. In
  \bibinfo{booktitle}{\emph{Symposium on Visual Languages and Human-Centric
  Computing (VL/HCC)}}.
\newblock
\href{https://doi.org/10.1109/VL-HCC57772.2023.00016}{doi:\nolinkurl{10.1109/VL-HCC57772.2023.00016}}


\bibitem[Mozilla(2023)]%
        {mdnDocs}
\bibfield{author}{\bibinfo{person}{Mozilla}.} \bibinfo{year}{2023}\natexlab{}.
\newblock \bibinfo{title}{{MDN Web Docs}}.
\newblock
\urldef\tempurl%
\url{{ https://developer.mozilla.org/en-US/docs/Web;
  https://developer.mozilla.org/en-US/docs/Web/API/Document;
  https://developer.mozilla.org/en-US/docs/Web/API/;
  https://developer.mozilla.org/en-US/docs/Learn/CSS/Building_blocks/The_box_model;
  https://developer.mozilla.org/en-US/docs/Web/CSS/CSS_selectors/Selector_structure;
  https://developer.mozilla.org/en-US/docs/Web/CSS/CSS_selectors/Selectors_and_combinators;
  https://developer.mozilla.org/en-US/docs/Web/CSS/Attribute_selectors;
  https://developer.mozilla.org/en-US/docs/Web/CSS/Cascade;
  https://developer.mozilla.org/en-US/docs/Web/CSS/Specificity;
  https://developer.mozilla.org/en-US/docs/Web/CSS/initial_value;
  https://developer.mozilla.org/en-US/docs/Web/CSS/Inheritance;
  https://developer.mozilla.org/en-US/docs/Glossary/Polyfill; }}
\showURL{%
\tempurl}


\bibitem[Myers(1990)]%
        {VPandPV}
\bibfield{author}{\bibinfo{person}{Brad~A. Myers}.}
  \bibinfo{year}{1990}\natexlab{}.
\newblock \showarticletitle{{Taxonomies of Visual Programming and Program
  Visualization}}.
\newblock \bibinfo{journal}{\emph{Journal of Visual Languages \& Computing}}
  (\bibinfo{year}{1990}).
\newblock
\href{https://doi.org/10.1016/S1045-926X(05)80036-9}{doi:\nolinkurl{10.1016/S1045-926X(05)80036-9}}


\bibitem[Najd and Peyton~Jones(2016)]%
        {Najd2016}
\bibfield{author}{\bibinfo{person}{Shayan Najd} {and} \bibinfo{person}{Simon
  Peyton~Jones}.} \bibinfo{year}{2016}\natexlab{}.
\newblock \showarticletitle{{Trees that Grow}}.
\newblock \bibinfo{journal}{\emph{Journal of Universal Computer Science}}
  (\bibinfo{year}{2016}).
\newblock
\href{https://doi.org/10.3217/jucs-023-01-0042}{doi:\nolinkurl{10.3217/jucs-023-01-0042}}


\bibitem[Oliveira et~al\mbox{.}(2020)]%
        {Oliveira2020}
\bibfield{author}{\bibinfo{person}{Delano Oliveira}, \bibinfo{person}{Reydne
  Bruno}, \bibinfo{person}{Fernanda Madeiral}, {and} \bibinfo{person}{Fernando
  Castor}.} \bibinfo{year}{2020}\natexlab{}.
\newblock \showarticletitle{{Evaluating Code Readability and Legibility: An
  Examination of Human-Centric Studies}}. In
  \bibinfo{booktitle}{\emph{International Conference on Software Maintenance
  and Evolution (ICSME)}}.
\newblock
\href{https://doi.org/10.1109/ICSME46990.2020.00041}{doi:\nolinkurl{10.1109/ICSME46990.2020.00041}}


\bibitem[Omar et~al\mbox{.}(2021)]%
        {Omar2021}
\bibfield{author}{\bibinfo{person}{Cyrus Omar}, \bibinfo{person}{David Moon},
  \bibinfo{person}{Andrew Blinn}, \bibinfo{person}{Ian Voysey},
  \bibinfo{person}{Nick Collins}, {and} \bibinfo{person}{Ravi Chugh}.}
  \bibinfo{year}{2021}\natexlab{}.
\newblock \showarticletitle{{Filling Typed Holes with Live GUIs}}. In
  \bibinfo{booktitle}{\emph{Conference on Programming Language Design and
  Implementation (PLDI)}}.
\newblock
\href{https://doi.org/10.1145/3453483.3454059}{doi:\nolinkurl{10.1145/3453483.3454059}}


\bibitem[Omar et~al\mbox{.}(2019)]%
        {HazelnutLive}
\bibfield{author}{\bibinfo{person}{Cyrus Omar}, \bibinfo{person}{Ian Voysey},
  \bibinfo{person}{Ravi Chugh}, {and} \bibinfo{person}{Matthew~A. Hammer}.}
  \bibinfo{year}{2019}\natexlab{}.
\newblock \showarticletitle{Live {F}unctional {P}rogramming with {T}yped
  {H}oles}.
\newblock \bibinfo{journal}{\emph{Proceedings of the ACM on Programming
  Languages (PACMPL), Issue POPL}} (\bibinfo{year}{2019}).
\newblock
\href{https://doi.org/10.1145/3290327}{doi:\nolinkurl{10.1145/3290327}}


\bibitem[Omar et~al\mbox{.}(2017)]%
        {Omar2017}
\bibfield{author}{\bibinfo{person}{Cyrus Omar}, \bibinfo{person}{Ian Voysey},
  \bibinfo{person}{Michael Hilton}, \bibinfo{person}{Jonathan Aldrich}, {and}
  \bibinfo{person}{Matthew~A. Hammer}.} \bibinfo{year}{2017}\natexlab{}.
\newblock \showarticletitle{{Hazelnut: A Bidirectionally Typed Structure Editor
  Calculus}}. In \bibinfo{booktitle}{\emph{Symposium on Principles of
  Programming Languages (POPL)}}.
\newblock
\href{https://doi.org/10.1145/3009837.3009900}{doi:\nolinkurl{10.1145/3009837.3009900}}


\bibitem[Omar et~al\mbox{.}(2012)]%
        {ActiveCodeCompletion}
\bibfield{author}{\bibinfo{person}{Cyrus Omar}, \bibinfo{person}{YoungSeok
  Yoon}, \bibinfo{person}{Thomas~D. LaToza}, {and} \bibinfo{person}{Brad~A.
  Myers}.} \bibinfo{year}{2012}\natexlab{}.
\newblock \showarticletitle{{Active Code Completion}}. In
  \bibinfo{booktitle}{\emph{International Conference on Software Engineering
  (ICSE)}}.
\newblock
\href{https://doi.org/10.1109/ICSE.2012.6227133}{doi:\nolinkurl{10.1109/ICSE.2012.6227133}}


\bibitem[Panchekha et~al\mbox{.}(2018)]%
        {Panchekha2018}
\bibfield{author}{\bibinfo{person}{Pavel Panchekha}, \bibinfo{person}{Adam
  Geller}, \bibinfo{person}{Michael Ernst}, \bibinfo{person}{Zachary Tatlock},
  {and} \bibinfo{person}{Shoaib Kamil}.} \bibinfo{year}{2018}\natexlab{}.
\newblock \showarticletitle{{Verifying That Web Pages Have Accessible Layout}}.
  In \bibinfo{booktitle}{\emph{Conference on Programming Language Design and
  Implementation (PLDI)}}.
\newblock
\href{https://doi.org/10.1145/3192366.3192407}{doi:\nolinkurl{10.1145/3192366.3192407}}


\bibitem[Park et~al\mbox{.}(2023)]%
        {Brown2023}
\bibfield{author}{\bibinfo{person}{{Kang-il} Park}, \bibinfo{person}{Pierre
  Weill-Tessier}, \bibinfo{person}{Neil C.~C. Brown}, \bibinfo{person}{Bonita
  Sharif}, \bibinfo{person}{Nikolaj Jensen}, {and} \bibinfo{person}{Michael
  Kölling}.} \bibinfo{year}{2023}\natexlab{}.
\newblock \showarticletitle{{An Eye Tracking Study Assessing the Impact of
  Background Styling in Code Editors on Novice Programmers' Code
  Understanding}}. In \bibinfo{booktitle}{\emph{International Computing
  Education Research (ICER)}}.
\newblock
\href{https://doi.org/10.1145/3568813.3600133}{doi:\nolinkurl{10.1145/3568813.3600133}}


\bibitem[Rauch et~al\mbox{.}(2019)]%
        {Rauch2019}
\bibfield{author}{\bibinfo{person}{David Rauch}, \bibinfo{person}{Patrick
  Rein}, \bibinfo{person}{Stefan Ramson}, \bibinfo{person}{Jens Lincke}, {and}
  \bibinfo{person}{Robert Hirschfeld}.} \bibinfo{year}{2019}\natexlab{}.
\newblock \showarticletitle{Babylonian-style {P}rogramming - {D}esign and
  {I}mplementation of {A}n {I}ntegration of {L}ive {E}xamples {I}nto
  {G}eneral-purpose {S}ource {C}ode}.
\newblock \bibinfo{journal}{\emph{The Art, Science, and Engineering of
  Programming}} (\bibinfo{year}{2019}).
\newblock
\href{https://doi.org/10.22152/programming-journal.org/2019/3/9}{doi:\nolinkurl{10.22152/programming-journal.org/2019/3/9}}


\bibitem[Resnick et~al\mbox{.}(2009)]%
        {Resnick2009}
\bibfield{author}{\bibinfo{person}{Mitchel Resnick}, \bibinfo{person}{John
  Maloney}, \bibinfo{person}{Andr{\'e}s Monroy-Hern{\'a}ndez},
  \bibinfo{person}{Natalie Rusk}, \bibinfo{person}{Evelyn Eastmond},
  \bibinfo{person}{Karen Brennan}, \bibinfo{person}{Amon Millner},
  \bibinfo{person}{Eric Rosenbaum}, \bibinfo{person}{Jay Silver},
  \bibinfo{person}{Brian Silverman}, {et~al\mbox{.}}}
  \bibinfo{year}{2009}\natexlab{}.
\newblock \showarticletitle{{Scratch: Programming for All}}.
\newblock \bibinfo{journal}{\emph{Communications of the ACM (CACM)}}
  (\bibinfo{year}{2009}).
\newblock
\href{https://doi.org/10.1145/1592761.1592779}{doi:\nolinkurl{10.1145/1592761.1592779}}


\bibitem[Rix et~al\mbox{.}(2023)]%
        {haskell-tree-sitter}
\bibfield{author}{\bibinfo{person}{Rob Rix}, \bibinfo{person}{Ayman Nadeem},
  {et~al\mbox{.}}} \bibinfo{year}{2023}\natexlab{}.
\newblock \bibinfo{title}{{Haskell Bindings for Tree-Sitter}}.
\newblock
\urldef\tempurl%
\url{https://github.com/tree-sitter/haskell-tree-sitter}
\showURL{%
\tempurl}


\bibitem[Sellier and Fadeyev(2009)]%
        {less}
\bibfield{author}{\bibinfo{person}{Alexis Sellier} {and}
  \bibinfo{person}{Smitry Fadeyev}.} \bibinfo{year}{2009}\natexlab{}.
\newblock \bibinfo{title}{{Lean: Learner Style Sheets}}.
\newblock
\urldef\tempurl%
\url{https://lesscss.org/}
\showURL{%
\tempurl}


\bibitem[Sheard and Jones(2002)]%
        {TemplateHaskell}
\bibfield{author}{\bibinfo{person}{Tim Sheard} {and}
  \bibinfo{person}{Simon~Peyton Jones}.} \bibinfo{year}{2002}\natexlab{}.
\newblock \showarticletitle{Template Meta-Programming for Haskell}. In
  \bibinfo{booktitle}{\emph{Proceedings of the 2002 ACM SIGPLAN Workshop on
  Haskell}}. \bibinfo{publisher}{Association for Computing Machinery}.
\newblock
\href{https://doi.org/10.1145/581690.581691}{doi:\nolinkurl{10.1145/581690.581691}}


\bibitem[Szab\'{o} et~al\mbox{.}(2018)]%
        {Voelter2018}
\bibfield{author}{\bibinfo{person}{Tam\'{a}s Szab\'{o}},
  \bibinfo{person}{G\'{a}bor Bergmann}, \bibinfo{person}{Sebastian Erdweg},
  {and} \bibinfo{person}{Markus Voelter}.} \bibinfo{year}{2018}\natexlab{}.
\newblock \showarticletitle{{Incrementalizing Lattice-Based Program Analyses in
  Datalog}}.
\newblock \bibinfo{journal}{\emph{Proceedings of the ACM Programming Languages
  (PACMPL), Issue OOPSLA.}} (\bibinfo{year}{2018}).
\newblock
\href{https://doi.org/10.1145/3276509}{doi:\nolinkurl{10.1145/3276509}}


\bibitem[Szab\'{o} et~al\mbox{.}(2016)]%
        {Voelter2016IncA}
\bibfield{author}{\bibinfo{person}{Tam\'{a}s Szab\'{o}},
  \bibinfo{person}{Sebastian Erdweg}, {and} \bibinfo{person}{Markus Voelter}.}
  \bibinfo{year}{2016}\natexlab{}.
\newblock \showarticletitle{{IncA: A DSL for the Definition of Incremental
  Program Analyses}}. In \bibinfo{booktitle}{\emph{Conference on Automated
  Software Engineering (ASE)}}.
\newblock
\href{https://doi.org/10.1145/2970276.2970298}{doi:\nolinkurl{10.1145/2970276.2970298}}


\bibitem[Talsma et~al\mbox{.}(2021)]%
        {Talsma2021}
\bibfield{author}{\bibinfo{person}{Renske Talsma}, \bibinfo{person}{Erik
  Barendsen}, {and} \bibinfo{person}{Sjaak Smetsers}.}
  \bibinfo{year}{2021}\natexlab{}.
\newblock \showarticletitle{{Analyzing the Influence of Block Highlighting on
  Beginning Programmers' Reading Behavior Using Eye Tracking}}. In
  \bibinfo{booktitle}{\emph{Computer Science Education Research Conference
  (CSERC)}}.
\newblock
\href{https://doi.org/10.1145/3442481.3442505}{doi:\nolinkurl{10.1145/3442481.3442505}}


\bibitem[Tapp and Kazman(1994)]%
        {Tapp1994}
\bibfield{author}{\bibinfo{person}{R. Tapp} {and} \bibinfo{person}{R. Kazman}.}
  \bibinfo{year}{1994}\natexlab{}.
\newblock \showarticletitle{{Determining the Usefulness of Colour and Fonts in
  a Programming Task}}. In \bibinfo{booktitle}{\emph{Workshop on Program
  Comprehension (WPC)}}.
\newblock
\href{https://doi.org/10.1109/WPC.1994.341265}{doi:\nolinkurl{10.1109/WPC.1994.341265}}


\bibitem[Teitelbaum and Reps(1981)]%
        {cornellProgramSynthesizer}
\bibfield{author}{\bibinfo{person}{Tim Teitelbaum} {and}
  \bibinfo{person}{Thomas Reps}.} \bibinfo{year}{1981}\natexlab{}.
\newblock \showarticletitle{{The Cornell Program Synthesizer: A Syntax-Directed
  Programming Environment}}.
\newblock \bibinfo{journal}{\emph{Commun. ACM}} \bibinfo{volume}{24},
  \bibinfo{number}{9} (\bibinfo{date}{sep} \bibinfo{year}{1981}),
  \bibinfo{pages}{563–573}.
\newblock
\showISSN{0001-0782}
\href{https://doi.org/10.1145/358746.358755}{doi:\nolinkurl{10.1145/358746.358755}}


\bibitem[Victor(2011)]%
        {ScrubbingCalculator}
\bibfield{author}{\bibinfo{person}{Bret Victor}.}
  \bibinfo{year}{2011}\natexlab{}.
\newblock \bibinfo{title}{{Scrubbing Calculator}}.
\newblock
  \bibinfo{howpublished}{\url{http://worrydream.com/ScrubbingCalculator/}}.
\newblock


\bibitem[Voelter et~al\mbox{.}(2019)]%
        {Voelter2019}
\bibfield{author}{\bibinfo{person}{Markus Voelter}, \bibinfo{person}{Klaus
  Birken}, \bibinfo{person}{Sascha Lisson}, {and} \bibinfo{person}{Alexander
  Rimer}.} \bibinfo{year}{2019}\natexlab{}.
\newblock \showarticletitle{{Shadow Models: Incremental Transformations for
  MPS}}. In \bibinfo{booktitle}{\emph{Conference on Software Language
  Engineering (SLE)}}.
\newblock
\href{https://doi.org/10.1145/3357766.3359528}{doi:\nolinkurl{10.1145/3357766.3359528}}


\bibitem[Voelter and Pech(2012)]%
        {Voelter2012}
\bibfield{author}{\bibinfo{person}{Markus Voelter} {and}
  \bibinfo{person}{Vaclav Pech}.} \bibinfo{year}{2012}\natexlab{}.
\newblock \showarticletitle{{Language Modularity with the MPS Language
  Workbench}}. In \bibinfo{booktitle}{\emph{International Conference on
  Software Engineering (ICSE)}}.
\newblock


\bibitem[Voelter et~al\mbox{.}(2016)]%
        {Voelter2016GrammarCells}
\bibfield{author}{\bibinfo{person}{Markus Voelter}, \bibinfo{person}{Tam\'{a}s
  Szab\'{o}}, \bibinfo{person}{Sascha Lisson}, \bibinfo{person}{Bernd Kolb},
  \bibinfo{person}{Sebastian Erdweg}, {and} \bibinfo{person}{Thorsten Berger}.}
  \bibinfo{year}{2016}\natexlab{}.
\newblock \showarticletitle{{Efficient Development of Consistent Projectional
  Editors using Grammar Cells}}. In \bibinfo{booktitle}{\emph{Conference on
  Software Language Engineering (SLE)}}.
\newblock
\href{https://doi.org/10.1145/2997364.2997365}{doi:\nolinkurl{10.1145/2997364.2997365}}


\bibitem[Weintrop et~al\mbox{.}(2019)]%
        {Weintrop2019}
\bibfield{author}{\bibinfo{person}{David Weintrop}, \bibinfo{person}{Heather
  Killen}, \bibinfo{person}{Talal Munzar}, {and} \bibinfo{person}{Baker
  Franke}.} \bibinfo{year}{2019}\natexlab{}.
\newblock \showarticletitle{{Block-Based Comprehension: Exploring and
  Explaining Student Outcomes from a Read-Only Block-Based Exam}}. In
  \bibinfo{booktitle}{\emph{Technical Symposium on Computer Science Education
  (SIGCSE TS)}}.
\newblock
\href{https://doi.org/10.1145/3287324.3287348}{doi:\nolinkurl{10.1145/3287324.3287348}}


\bibitem[Winfrey et~al\mbox{.}(2024)]%
        {tree-sitter-haskell}
\bibfield{author}{\bibinfo{person}{Rick Winfrey}, \bibinfo{person}{Torsten
  Schmits}, {et~al\mbox{.}}} \bibinfo{year}{2017--2024}\natexlab{}.
\newblock \bibinfo{title}{{Haskell Grammar for Tree-Sitter}}.
\newblock
\urldef\tempurl%
\url{https://github.com/tree-sitter/tree-sitter-haskell}
\showURL{%
\tempurl}


\bibitem[{World Wide Web Consortium (W3C)}(2014)]%
        {XQuery}
\bibfield{author}{\bibinfo{person}{{World Wide Web Consortium (W3C)}}.}
  \bibinfo{year}{2014}\natexlab{}.
\newblock \bibinfo{title}{{XML Query (XQuery) 3.0}}.
\newblock
\urldef\tempurl%
\url{https://www.w3.org/XML/Query/}
\showURL{%
\tempurl}


\bibitem[{World Wide Web Consortium (W3C)}(2017)]%
        {XSLT}
\bibfield{author}{\bibinfo{person}{{World Wide Web Consortium (W3C)}}.}
  \bibinfo{year}{2017}\natexlab{}.
\newblock \bibinfo{title}{{XSL Transformations (XSLT) 3.0}}.
\newblock
\urldef\tempurl%
\url{https://www.w3.org/TR/xslt-30/}
\showURL{%
\tempurl}


\bibitem[{World Wide Web Consortium (W3C)}(2023)]%
        {css}
\bibfield{author}{\bibinfo{person}{{World Wide Web Consortium (W3C)}}.}
  \bibinfo{year}{2023}\natexlab{}.
\newblock \bibinfo{title}{{Cascading Style Sheets (CSS) 3}}.
\newblock
\urldef\tempurl%
\url{https://www.w3.org/Style/CSS/}
\showURL{%
\tempurl}


\bibitem[Wu et~al\mbox{.}(2023)]%
        {Wu2023}
\bibfield{author}{\bibinfo{person}{Zhiyuan Wu}, \bibinfo{person}{Jiening Li},
  \bibinfo{person}{Kevin Ma}, \bibinfo{person}{Hita Kambhamettu}, {and}
  \bibinfo{person}{Andrew Head}.} \bibinfo{year}{2023}\natexlab{}.
\newblock \showarticletitle{{FFL: A Language and Live Runtime for Styling and
  Labeling Typeset Math Formulas}}. In \bibinfo{booktitle}{\emph{Symposium on
  User Interface Software and Technology (UIST)}}.
\newblock
\href{https://doi.org/10.1145/3586183.3606731}{doi:\nolinkurl{10.1145/3586183.3606731}}


\bibitem[Ye et~al\mbox{.}(2020)]%
        {Ye2020}
\bibfield{author}{\bibinfo{person}{Katherine Ye}, \bibinfo{person}{Wode Ni},
  \bibinfo{person}{Max Krieger}, \bibinfo{person}{Dor Ma’ayan},
  \bibinfo{person}{Jenna Wise}, \bibinfo{person}{Jonathan Aldrich},
  \bibinfo{person}{Joshua Sunshine}, {and} \bibinfo{person}{Keenan Crane}.}
  \bibinfo{year}{2020}\natexlab{}.
\newblock \showarticletitle{{Penrose: From Mathematical Notation to Beautiful
  Diagrams}}.
\newblock \bibinfo{journal}{\emph{ACM Transactions on Graphics}}
  (\bibinfo{year}{2020}).
\newblock
\href{https://doi.org/10.1145/3386569.3392375}{doi:\nolinkurl{10.1145/3386569.3392375}}


\bibitem[Zhao et~al\mbox{.}(2024)]%
        {Zhao2024}
\bibfield{author}{\bibinfo{person}{Eric Zhao}, \bibinfo{person}{Raef Maroof},
  \bibinfo{person}{Anand Dukkipati}, \bibinfo{person}{Andrew Blinn},
  \bibinfo{person}{Zhiyi Pan}, {and} \bibinfo{person}{Cyrus Omar}.}
  \bibinfo{year}{2024}\natexlab{}.
\newblock \showarticletitle{{Total Type Error Localization and Recovery with
  Holes}}.
\newblock \bibinfo{journal}{\emph{Proceedings of the ACM on Programming
  Languages (PACMPL)}} \bibinfo{number}{POPL} (\bibinfo{year}{2024}).
\newblock
\href{https://doi.org/10.1145/3632910}{doi:\nolinkurl{10.1145/3632910}}


\end{thebibliography}

\vspace{0.09in}
\noindent
{\small Received 2024-10-11; accepted 2025-02-18}

\clearpage

\appendix

\renewcommand{\leftmark}{}
\renewcommand{\rightmark}{}
\pagestyle{headings}

\section{Flat and Nested Displays in HTML}
\label{sec:html-flat-nested}

The full HTML and CSS definitions for the examples presented in \autoref{sec:overview-layout}, \autoref{fig:html-flat-nested}\refBubble{a}\refBubble{b} are below.

\bigskip

\begin{figure}[h]

\scriptsize

\newcommand{\subheader}[1]
{\begin{flushleft}\textbf{#1} \\[0.05in]\end{flushleft}}

\begin{minipage}[t]{.42\textwidth} \subheader{HTML Document}
\begin{Verbatim}[commandchars=\\\{\}]
<body>
  <div>
    main =
  </div>
  <div>
    <span class="sp2"></span>
    <span class="lr">getContents</span>
  </div>
  <div>
    <span class="sp4"></span>
    <span class="lr op">>>= </span>
    <span class="rl">print</span>
  </div>
  <div>
    <span class="sp6"></span>
    <span class="op rl">. </span>
    <span class="rl">length</span>
  </div>
  <div>
    <span class="sp6"></span>
    <span class="op rl">. </span>
    <span class="rl">filter (not </span>
    <span class="op rl">. </span>
    <span class="rl"> isPrefixOf "--")</span>
  </div>
  <div>
    <span class="sp6"></span>
    <span class="op rl">. </span>
    <span class="rl">lines</span>
  </div>
</body>
\end{Verbatim}
\end{minipage} \begin{minipage}[t]{.57\textwidth} \subheader{CSS Style Sheet}
  \begin{minipage}[t]{0.5\textwidth}
\begin{Verbatim}
.op {
  font-weight: bold;
}
.lr {
  border: 2px solid indigo;
  border-left: none;
  border-right: none;
  background-color: lavender;
}
.rl {
  border: 2px solid orange;
  border-left: none;
  border-right: none;
  background-color: papayawhip;
}
.op.lr {
  color: indigo;
}
.op.rl {
  color: orange;
}
\end{Verbatim}
  \end{minipage}
  \begin{minipage}[t]{0.49\textwidth}
\begin{Verbatim}
.sp2, .sp4, .sp6 {
  display: inline-block;
}
.sp2 {
  width: 2ch;
}
.sp4 {
  width: 4ch;
}
.sp6 {
  width: 6ch;
}
body {
  font-family: monospace;
}
span {
  vertical-align: top;
}
\end{Verbatim}
  \end{minipage}
  \vspace{0.05in}
  \subheader{Rendered Display}
  \includegraphics[width=2.3in]{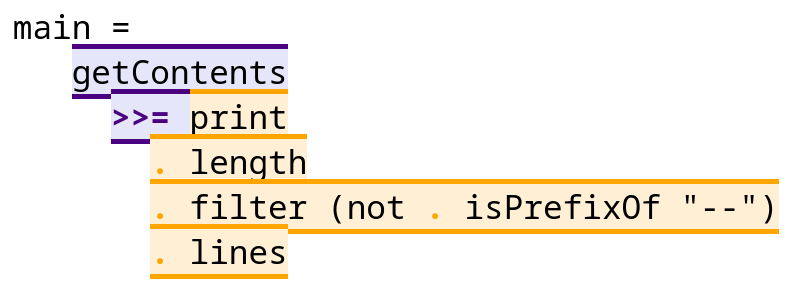}

\end{minipage} 

\bigskip
\bigskip
\begin{minipage}[t]{.42\textwidth} \subheader{HTML Document}
\begin{Verbatim}[commandchars=\\\{\}]
<body>
  main =
  <br>
  <span class="ebinop lr">
    getContents
    <br>
    <span class="op">>>=</span>
    <span class="ebinop rl">
      print
      <br>
      <span class="op">.</span>
      <span class="ebinop rl">
        length
        <br>
        <span class="op">.</span>
        <span class="ebinop rl">
          filter (<span class="ebinop rl">
            not . isPrefixOf "--"
          </span>)
          <br>
          <span class="op">.</span>
          lines
        </span>
      </span>
    </span>
  </span>
</body>
\end{Verbatim}
\end{minipage}\begin{minipage}[t]{.59\textwidth} \begin{minipage}[t]{.5\textwidth} \subheader{CSS Style Sheet}
\begin{Verbatim}
.ebinop.rl     > .ebinop.lr,
.ebinop.lr:has(> .ebinop.rl) {
    background-color: lavender;
    border: 2px solid indigo;
    margin: 2px;
    padding: 2px;
}
.ebinop.lr     > .ebinop.rl,
.ebinop.rl:has(> .ebinop.lr) {
    border: 2px solid orange;
    margin: 2px;
    padding: 2px;
    background-color: papayawhip;
}
.op {
  font-weight: bold;
}
\end{Verbatim}
  \end{minipage} \begin{minipage}[t]{.49\textwidth} \vspace{0.11in}
\begin{Verbatim}
.ebinop.rl > .op {
  color: orange;
}
.ebinop.lr > .op {
  color: indigo;
}
body {
    font-family: monospace;
}
span {
    display: inline-block;
    vertical-align: top;
    border-radius: 4px;
}
\end{Verbatim}
  \end{minipage}
  \vspace{0.05in}
  \subheader{Rendered Display}
  \includegraphics[width=2.4in]{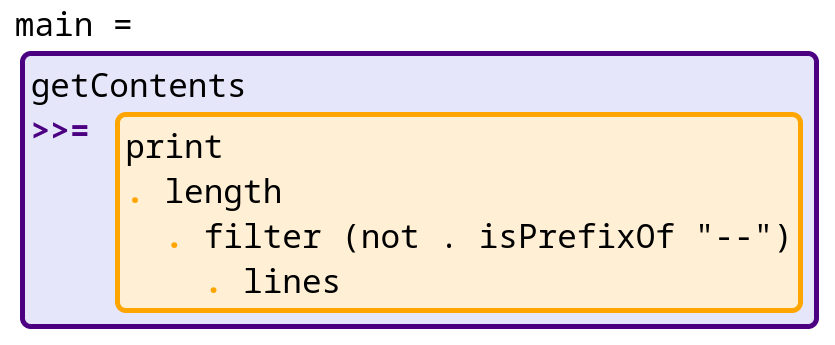}
\end{minipage}

\end{figure}
 
\clearpage
\section{On Translating \hass{} Style Sheets to CSS}
\label{sec:translation-to-css}

\setlength\fboxsep{0pt}

\newcommand{\diffReplaceAdd}[1]
{\colorbox{blue!20}{#1}}
\newcommand{\diffSubtract}[1]{\colorbox{red!20}{#1}}
\newcommand{\diffReplace}[2]
  {\diffSubtract{#1}\hspace{0.02in}$\rightsquigarrow$\hspace{0.02in}\diffReplaceAdd{#2}}

\begin{wrapfigure}[12]{r}{2.3in}
  \vspace{-0.09in} \includegraphics[width=2.3in]{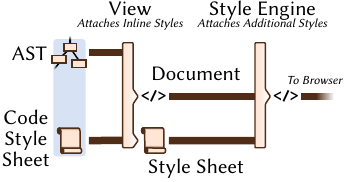}
  \caption{
  Hypothetical Pipeline: Translating code style sheets into CSS style sheets.
  }
  \label{fig:architecture-translation}
\end{wrapfigure}
 In \autoref{sec:overview-provenance} and \autoref{sec:style-sheets}, we define a style sheet semantics that explicitly maintains provenance (namely, paths) connecting the AST and the corresponding document being displayed.
As we said in Footnote~\ref{footnote:translation-to-css}, it may seem natural to ask whether \hass{} style sheets could instead by phrased in terms of ordinary CSS style sheets, operating on ordinary HTML documents without provenance.

\autoref{fig:architecture-translation} depicts a hypothetical architecture:
like in the \hass{} architecture (\autoref{fig:architecture-hass}), the input is a code style sheet, but
unlike in \hass{}, view functions generate HTML documents (without provenance) and user-facing \hass{} style sheets are translated to CSS style sheets (without pattern selectors or predicates).
We sketch a couple conspicuous steps in this direction, before quickly confronting the inherent limitations.

\parahead{Translating \hass{} Selectors to CSS Selectors}

Consider the three \hass{} style rules from \autoref{fig:tally-marks-combined}\refBubble{d} translated to CSS below;
the first rule demonstrates what we call a ``constructors-as-classes'' translation,
and the second rule demonstrates a ``predicates-as-classes'' translation.
In each case, information about selector computation---comprising pattern-matching and predicate evaluation---is encoded as data stored as CSS classes in the HTML document.

\begin{figure}[h]
\small

\begin{Verbatim}[commandchars=\\\{\}]
\diffReplace{Left n}{.left .int}                               \diffSubtract{->} \diffSubtract{n} \{ color: orange; \}
\diffReplace{Left n if n == sqrti n ^ 2}{.left.is-square .int} \diffSubtract{->} \diffSubtract{n} \{ border: 3px solid orange; \}
\diffSubtract{x@}.tally-marks \phantom{\diffReplace{X}{X}}                              \diffSubtract{->} \diffSubtract{x} \{ font-weight: 900; \}
\end{Verbatim}

\caption{
Hypothetical translation from \hass{} to CSS.
The diff ``\diffReplace{this text}{that}'' indicates that \diffSubtract{this text} in the original \hass{} be replaced with \diffReplaceAdd{that} in the target CSS.
The diff ``\diffSubtract{text}'' indicates \hass{} \diffSubtract{text} to be dropped.
}
\label{fig:tally-marks-css}
\end{figure}

\parahead{Constructors as CSS classes}

\begin{figure}[b]
\small
\begin{Verbatim}[commandchars=\\\{\}]
view                   :: [Either Int Int] -> Html
view es                =  Node [] [] (map viewEither es)
viewEither (Left n)    =  node ["left"]        [viewIntTallies n]
viewEither (Right n)   =  node ["right"]       [viewIntTallies n]
viewInt n              =  node ["int"]         [Text (show n)]
viewTallies n          =  node ["tally-marks"] [Text (replicate n '|')]
viewIntTallies n       =  node ["Int-tallies"] [viewInt n, viewTallies n]
node classes children  =  Node [] classes children
\end{Verbatim}

\caption{
Hypothetical view function for tally marks that systematically inserts CSS classes to track data constructors, types, and predicates.
}
\label{fig:view-tally-marks-css}
\end{figure}
 
How should view functions set up the ordinary HTML document so that these CSS selectors work?
First, as shown in \autoref{fig:view-tally-marks-css},
attach extra CSS classes to store information about data constructors and types for the particular AST subvalues for being displayed.
Notice how the HTML nodes corresponding to values constructed by \texttt{Left} and \texttt{Right} are decorated with classes \cssClass{left} and \cssClass{right}, respectively,
and those corresponding to integers (which are not created with data constructors) are decorated with the class \cssClass{int}.
Given a document annotated in this way,
the ordinary CSS selector \cssClass{.left .int} can be used to simulate the \hass{} selector for \texttt{Left} numbers.
\parahead{Predicates as CSS classes}

This ``constructors-as-classes'' approach helps when selectors are simple patterns, but what about more complicated predicates?
To support a rule like the second \hass{} selector, which matches \texttt{Left n} and then checks whether \verb+n == sqrti n ^ 2+, view functions could evaluate the given predicate on all \texttt{Left} numbers, adding a specially named class for nodes displaying square numbers:

{\small

\begin{Verbatim}[xleftmargin=\parindent]
viewEither (Left n) = node ["left", is_square n] [viewIntTallies n] where
  is_square n
    | n == sqrti n ^ 2 = "is-square"
    | otherwise        = ""
\end{Verbatim}

}

\noindent
Then, the ordinary CSS selector \cssClass{.left.is-square .int}---the second rule in that figure---could be used to select and style the appropriate nodes in the document.

Although suitable for toy examples, this ``predicates-as-classes'' approach has significant drawbacks.
First, it is \emph{rigid}, requiring that all predicates which user style sheets might reference be fixed in advance.
Furthermore, even if this rigidity were acceptable, the approach is also \emph{inefficient}: all predicates need to be evaluated in all (type-appropriate) positions throughout the AST.\footnote{
The obstacles faced when encoding constructors and predicates as classes are even steeper than already described.
Whereas ``single-path'' patterns---which name subvalues along a single path in the pattern tree, using wildcards everywhere else---can be encoded as CSS selectors following the approach using \texttt{:has(...)} pseudo-class predicates as shown in \autoref{sec:overview-layout}, it is less clear how to encode ``branching'' patterns (as arbitrary patterns are) without resorting to the predicate-encoding approach.
Speaking of which, the idea to evaluate all predicates everywhere through the AST is more inefficient than suggested: even with a fixed set of predicates, complex selectors may comprise an arbitrary number of basic selectors.
So, complex selectors would need to be enumerated upto a size that is sufficient for the particular (finite-size) AST value being scrutinized.
}

\clearpage
\section{\hass{} Style Sheet Language}
\label{sec:style-computation}

\setlength\fboxsep{2pt}

In \autoref{sec:overview-provenance} and \autoref{sec:style-sheets}, we presented the essential CSS features---class selectors, and path selectors built with tree combinators---and new mechanisms---pattern selectors and predicates---that comprise \hass{} style sheets.
Below, we provide additional discussion about the syntax and semantics.

\subsection{Selector Terminology}
\label{sec:selector-terminology}

The terminology we choose for selectors in \hass{} differs from the CSS terminology in MDN Web Docs~\cite{mdnDocs}, which itself deviates from the CSS standard~\cite{css}.
We forgo the terms ``complex'' and ``compound''---the semantics of selectors are not particularly complex, and many syntactic forms are compound in nature---in favor of ``path'' and ``node'' which describe the kinds of nodes selected by the given selectors.
\autoref{fig:selector-terminology} summarizes the varied CSS terminology and how ours relates.

\begin{figure}[h]
\centering
\newcommand{\theWinnerIs}[1]
  {\textbf{#1}}
\begin{tabular}{l|lll}
\hass{} &
CSS2$^*$ &
CSS3$^*$ &
\citet{mdnDocs} \\
\hline
\theWinnerIs{selector list} $\rangeN{i}{n}{\varPathSelector_i}$ & selector group  & selector group & \theWinnerIs{selector list} \\
(path) \theWinnerIs{selector} $\varPathSelector$ & \theWinnerIs{selector} & \theWinnerIs{selector} & complex selector \\
{node selector} $\varNodeSelector$ & simple selector & sequence of simple selectors & compound selector \\
N/A & " & simple selector & simple selector \\
\theWinnerIs{basic selector} $\varBasicSelector$ & " & " & \theWinnerIs{basic selector} \\
\end{tabular}
\caption{Selector Terminology.
($^*$\url{https://www.w3.org/TR/selectors-3/\#changesFromCSS2}) \\
Some ``non-basic, simple'' selectors in CSS are not currently modeled in \hass{} (hence the ``N/A''), including:
the universal selector, attribute selectors, id selectors, and ``type'' (i.e., tag) selectors (all of which are subsumed by class selectors, pattern selectors, and predicates); and
various pseudo-classes and pseudo-elements.
}
\label{fig:selector-terminology}
\end{figure}
 
Note that the \emph{selector list} (i.e., comma operator) from CSS is omitted from the syntax of \hass{} rules in \autoref{sec:style-sheets}, \autoref{fig:external-syntax}; this disjunction of selectors can be encoded by duplicating rules:
$$
\small
\underbrace{
\guardedRuleArrow
  {\overbrace{\rangeN{i}{n}{\varPathSelector_i}}^{\textrm{selector list}}}
  {\varExp}
  {\varNamedStyles}
}_{\textrm{rule (external syntax)}}
\hspace{0.10in}
\Longrightarrow
\hspace{0.10in}
\underbrace{
\guardedRuleArrow{\varPathSelector_1}{\varExp}{\varNamedStyles}
\hspace{0.25in} \ldots
\hspace{0.25in}
\guardedRuleArrow{\varPathSelector_n}{\varExp}{\varNamedStyles}
}_{\textrm{multiple rules (external syntax)}}
$$

\noindent
Each selector $\varPathSelector_i$ must agree on the names used in the top-level predicate $\varExp$.
It less important that each $\varPathSelector_i$ names and styles every member of the shared style environment $\varNamedStyles$.

\subsection{Internal Syntax}
\label{sec:style-sheets-internal}

\newcommand{\blah}[1]
{\raisebox{-4pt}{#1}} 

\newcommand{\desugarRule}[3]
  {\blah{\ensuremath{#1}} & \blah{$\Longrightarrow$} &
   \blah{\ensuremath{#2}} & \blah{\ensuremath{#3}} \\[6pt]}

\newcommand{\justInternal}[1]
  { & & \blah{\ensuremath{#1}} & \\[6pt]}

\newcommand{\desugarCategory}[4]
{\cellcolor[HTML]{dddddd} \blah{\textbf{#1}} \blah{\ensuremath{#2}} &
   \cellcolor[HTML]{dddddd} &
   \cellcolor[HTML]{dddddd} \blah{\textbf{#3}} \blah{\ensuremath{#4}} &
   \cellcolor[HTML]{dddddd} \\[6pt]\hline}

\newcommand{\nextCategory}
  {\hline\hline}

\newcommand{\arrowNamedStylesHighlight}
  {\fcolorbox{red}{yellow}{$\rightarrow\varNamedStyles\vphantom{(}$}}

\newcommand{\namedPatternHighlight}[2]
  {\fcolorbox{red}{yellow}{$#1\smallSep@\vphantom{(}$}\smallSep#2}

\newcommand{\andStylesHighlight}[2]
  {\nodeRule{#1}{\fcolorbox{blue}{yellow}{\ensuremath{#2}}}}

\begin{figure}[t]
\small

\setlength{\fboxrule}{1pt}\setlength{\fboxsep}{1pt}

\begin{tabular}{l|c|l|l}
\multicolumn{1}{c}{A} &
\multicolumn{1}{c}{B} &
\multicolumn{1}{c}{C} &
\multicolumn{1}{c}{D} \\
\multicolumn{1}{c}{\textbf{External Syntax}} &
\multicolumn{1}{c}{} &
\multicolumn{1}{c}{\textbf{Internal Syntax}} &
\multicolumn{1}{c}{(desugaring subterms)} \\[7pt]
\hline
\desugarCategory{Rule}{\varGuardedRule}{Rule}{\varInternalRule}
\desugarRule
  {
\guardedRule
    {\varPathSelector}
    {\varExp}
  \ \arrowNamedStylesHighlight
  }
  {
  \guardedRule
    {\varPathRule}
    {\varExp}
  }
  {\textrm{where } \varPathSelector \Longrightarrow \varPathRule
   \textrm{ and } \varNamedStyles = \namedStyles{j}{m}{\varVar_j}{\varAttrs_j}
}
\nextCategory
\desugarCategory{Path Sel.}{\varPathSelector}
                {Path Subrule}{\varPathRule}
\desugarRule
  {\varNodeSelector}
  {\varNodeRule}
  {\textrm{where } \varNodeSelector \Longrightarrow \varNodeRule
  }
\desugarRule
  {\pathRule{\varNodeSelector}{\varOp}{\varPathSelector}}
  {\pathRule{\varNodeRule}{\varOp}{\varPathRule}}
  {\textrm{where } \varNodeSelector \Longrightarrow \varNodeRule
   \textrm{ and } \varPathSelector \Longrightarrow \varPathRule
  }
\nextCategory
\desugarCategory{Node Sel.}{\varNodeSelector}
                {Node Subrule}{\varNodeRule}
\desugarRule
  {\compoundSelector{\varPat}{\rangeZeroN{i}{n}{\clsSelector{\varCls_i}}}}
  {\compoundSelector{\varPatternRule}{\rangeZeroN{i}{n}{\clsSelector{\varCls_i}}}}
  {\textrm{where } \varPat \Longrightarrow \varPatternRule
  }
\desugarRule
{\namedPatternHighlight{\varVar}{\rangeN{i}{n}{\clsSelector{\varCls_i}}}}
  {\andStylesHighlight
    {\rangeN{i}{n}{\clsSelector{\varCls_i}}}
    {\varAttrs}}
  {\textrm{where } \varAttrs = \styleSetOf{\varVar}}
\justInternal{\topSelector}
\nextCategory
\desugarCategory{Pattern}{\varPat}
                {Pattern Subrule}{\varPatternRule}
\desugarRule
  {\varPattern{\varVar}{\varTypeCon}}
  {\andStylesHighlight
    {\varPattern{\varVar}{\varTypeCon}}
    {\varAttrs}}
  {\textrm{where } \varAttrs = \styleSetOf{\varVar}}
\desugarRule
  {\namedPattern{\varVar}{\dataApp{\varDataCon}{\rangeN{i}{n}{\varPat_i}}}}
  {\andStylesHighlight
    {\namedPattern{\varVar}
                  {\dataApp{\varDataCon}{\rangeN{i}{n}{\varPatternRule_i}}}}
    {\varAttrs}}
  {\textrm{where } \varAttrs = \styleSetOf{\varVar}
   \textrm{ and } \rangeN{i}{n}{\varPat_i \Longrightarrow \varPatternRule_i}
  }
\desugarRule
  {\keepOut}
  {\keepOut}
  {}
\hline
\end{tabular}

\caption{
Code Style Sheets in \hass{}.
(A) External syntax.
(C) Internal syntax.
(ABCD) Desugaring.
}
\label{fig:desugaring}
\end{figure}
 
To simplify the presentation of the semantics, we desugar the external syntax (\autoref{fig:desugaring}a) into an internal syntax (\autoref{fig:desugaring}c).
Compared to \autoref{fig:external-syntax}, the definition of external syntax here does not include a separate syntactic category of basic selectors $\varBasicSelector$; these are inlined into the syntax of node selectors $\varNodeSelector$.
Rules in the external syntax consist of a selector and predicate, followed by style sets corresponding to names bound in the selector.

In contrast, in the internal syntax, these style sets are paired syntactically with individual components of the selector.
For example, consider the following path selector that uses class selectors for both parent and child:
$$
\small
\underbrace{
\overbrace{{\texttt{x@.cls1}\ \opChild \ \texttt{y@.cls2}}}^{\textrm{path selector}}\texttt{ -> }\overbrace{\texttt{x \{...red\} y \{...blue\}}}^{\textrm{named styles}}
}_{\textrm{rule (external syntax)}}
\hspace{0.10in}
\Longrightarrow
\hspace{0.10in}
\underbrace{
\overbrace{\texttt{.cls1 \{...red\}}}^{\textrm{node subrule}}\ \opChild\ \overbrace{\texttt{.cls2 \{...blue\}}}^{\textrm{node subrule}}
}_{\textrm{rule (internal syntax)}}
$$

\noindent
In the former, matched parent and child nodes in the document are bound to the names \texttt{x} and \texttt{y}, used to name corresponding styles.
In the latter, style sets are attached syntactically to individual components of the rule.
Whereas the former is more convenient for reading and writing, the latter avoids the indirection through names and simplifies formal definitions.

\parahead{Desugaring}

\autoref{fig:desugaring} defines the desugaring $\varGuardedRule \Longrightarrow \varInternalRule$ from external to internal rules.
The translation distributes the ``global'' set of named styles (highlighted red and yellow in \autoref{fig:desugaring}a) to the individual subrules (highlighted blue and yellow in \autoref{fig:desugaring}c) which introduced those names.
We use the term \emph{subrule} for all syntactic forms in the internal syntax because they describe both the selection and styling of document nodes.
Whereas class node subrules have no need for variables, pattern node subrules do for subsequent use in the top-level predicate of the rule.

\begin{figure}[t]

\judgementHead
  {Style Sheet Application}
  {\hiddenArgsTwo{\varTypeContext}{\valRoot}
   \applyStyleSheetEquals{\varStyleSheet}{\varDoc}{\varDoc'}}

\vsepRule 

$
\inferrule*[lab=\ruleNameFig{ApplyOne}]
  {\varInternalRule \in \varStyleSheet \sepPremise
   \applyGuardedRuleEquals{\varInternalRule}{\varDoc}{\varDoc'} \\\\
   \applyStyleSheetEquals{\varStyleSheet}{\varDoc}{\varDoc''}
  }
  {\applyStyleSheetEquals
    {\varStyleSheet}{\varDoc}
    {\mergeDocs{\varDoc'}{\varDoc''}}}
$
\hsepRule
$
\inferrule*[lab=\ruleNameFig{ApplyNone}]
{\rangeZeroN{i}{n}{\varDoc_i} = \childrenOf{\varDoc} \\\\
   \rangeN{i}{n}{\applyStyleSheetEquals{\varStyleSheet}{\varDoc_i}{\varDoc_i'}}
  }
  {\applyStyleSheetEquals
    {\varStyleSheet}{\varDoc}
    {\copyUpdate{\varDoc}{\texttt{children}}{\rangeZeroN{i}{n}{\varDoc_i'}}}}
$

\vsepRule

\judgementHead
  {Rule Application}
  {\hiddenArgsTwo{\varTypeContext}{\valRoot}
   \applyGuardedRuleEquals{\varInternalRule}{\varDoc}{\varDoc'}}

\vsepRule

$
\inferrule*[lab=\ruleNameFig{ApplyRule}]
  {\applyPathRuleEquals
    {\varPathRule}{\varDoc}
    {\varDoc'}{\varEnv} \sepPremise
   \evalsTo{\varEnv}{\varExp}{\valTrue}
  }
  {\applyGuardedRuleEquals
    {\guardedRule{\varPathRule}{\varExp}}{\varDoc}
    {\varDoc'}}
$

\vsepRule

\judgementHead
  {Document Merging}
  {\mergeDocs{\varDoc_1}{\varDoc_2}=\varDoc}

\vsepRule

$
\inferrule* {\varDoc_1 =
     \node{\varPath}{\varClasses}{\varAttrs_1}
       {\rangeZeroN{i}{n}{\varDoc_{1i}}}
   \\\\
   \varDoc_2 =
     \node{\varPath}{\varClasses}{\varAttrs_2}
       {\rangeZeroN{i}{n}{\varDoc_{2i}}}
   \\\\
   \varDoc =
     \node{\varPath}{\varClasses}{\varAttrs_1 \cup \varAttrs_2}
       {\rangeZeroN{i}{n}{\mergeDocs{\varDoc_{1i}}{\varDoc_{2i}}}}
  }
  {\mergeDocs{\varDoc_1}{\varDoc_2}={\varDoc}}
$

\vsepRule

\caption{Style Computation}
\label{fig:style-computation}
\end{figure}

\begin{figure}[h]

\judgementHeadThree
  {Path Subrule Application}
  {(selected rules)}
  {\hiddenArgsTwo{\varTypeContext}{\valRoot}
   \applyPathRuleEquals{\varPathRule}{\varDoc}{\varDoc'}{\varEnv}}

\vsepRule

$
\inferrule*[lab=\ruleNameFig{Node}]
  {\selectorMatch
     {\pathOf{\varDoc}}{\classesOf{\varDoc}}{\varNodeRule}
     {\varEnv}{\varEnvStyles}
  }
  {\applyPathRuleEquals
    {\varNodeRule}
    {\varDoc}
    {\applyStyles{\varEnvStyles}{\varDoc}}
    {\varEnv}}
$

\vsepRule

$
\inferrule*[lab=\ruleNameFig{Child}]
  {
   \applyPathRuleEquals{\varNodeRule}{\varDoc}{\varDoc_1}{\varEnv_1}
   \sepPremise
   \varDoc_j \in \childrenOf{\varDoc}
\sepPremise
   \applyPathRuleEquals
     {\varPathRule}{\varDoc_j}
     {\varDoc_j'}{\varEnv_2}
   \\\\
   \mathit{styles'} = \stylesOf{\varDoc_1}
   \sepPremise
   \mathit{children'} = \copyUpdate{(\childrenOf{\varDoc})}{j}{\varDoc_j'}
  }
  {\applyPathRuleEquals
    {\pathRule{\varNodeRule}{\opChild}{\varPathRule}}{\varDoc}
    {\copyUpdateTwo
       {\varDoc}
       {\texttt{styles}}{\mathit{styles'}}
       {\texttt{children}}{\mathit{children'}}}
    {\envExtend{\varEnv_1}{\varEnv_2}}}
$

\vsepRule

$
\inferrule*[lab=\ruleNameFig{Descendant1}]
  {\applyPathRuleEquals
    {\pathRule{\varNodeRule}{\opChild}{\varPathRule}}{\varDoc}
    {\varDoc'}{\varEnv}
  }
  {\applyPathRuleEquals
    {\pathRule{\varNodeRule}{\opDescendant}{\varPathRule}}{\varDoc}
    {\varDoc'}{\varEnv}}
$
\hsepRule
$
\inferrule*[lab=\ruleNameFig{Descendant2}]
  {\applyPathRuleEquals
    {\pathRule{\varNodeRule}
                 {\opChild}
                 {\pathRule{\topSelector}
                              {\opDescendant}
                              {\varPathRule}}}
    {\varDoc}
    {\varDoc''}{\varEnv}
  }
  {\applyPathRuleEquals
    {\pathRule{\varNodeRule}{\opDescendant}{\varPathRule}}{\varDoc}
    {\varDoc''}{\varEnv}}
$

\vsepRule

\vsepRule

\judgementHead
{Node Subrule Application}
  {\hiddenArgsTwo{\varTypeContext}{\valRoot}
   \selectorMatch{\varPath}{\varClasses}{\varNodeRule}{\varEnv}{\varEnvStyles}}

\vsepRule

$
\inferrule*[lab=\ruleNameFig{Class}]
  {\varClasses_1 \subset \varClasses_2
  }
  {\selectorMatch
{\varPath}
    {\varClasses_2}
    {\andStyles{.\varClasses_1}{\varAttrs}}
    {\emptyEnv}
    {(\envBind{\varPath}{\varAttrs})}}
$

\vsepRule
$
\inferrule*[lab=\ruleNameFig{Pattern}]
{\patternMatch{\varPath}{\varPath\valRoot}{\varPat}{\varEnv}{\varEnvStyles}
   \sepPremise
   \varClasses_1 \subset \varClasses_2
  }
  {\selectorMatch
{\varPath}
    {\varClasses_2}
    {\patSelector{\varPat}{.\varClasses_1}}
    {\varEnv}
    {\varEnvStyles}}
$
\hsepRule
$
\inferrule*[lab=\ruleNameFig{Top}]
  {
  }
  {\selectorMatch{\varPath}{\varClasses}{\topSelector}{\emptyEnv}{\emptyEnv}}
$

\vsepRule

\judgementHead
  {Pattern Subrule Application}
  {\hiddenArgs{\varTypeContext}
   \patternMatch{\varPath}{\varVal}{\varPatternRule}{\varEnv}{\varEnvStyles}}

\vsepRule

$
\inferrule*[lab=\ruleNameFig{Var}]
  {\varVal = \dataApp{\varDataCon}{\rangeZeroN{i}{n}{\varVal_i}}
   \sepPremise
\varDataCon \in \varTypeContext(\varTypeCon)
  }
  {\patternMatch
    {\varPath}
    {\varVal}
    {\andStyles{\varPattern{\varVar}{\varTypeCon}}{\varAttrs}}
    {(\envBind{\varVar}{\varVal})}
    {(\envBind{\varPath}{\varAttrs})}}
$

\vsepRule

$
\inferrule*[lab=\ruleNameFig{Constructor}]
  {\varVal = \dataApp{\varDataCon}{\rangeZeroN{i}{n}{\varVal_i}}
   \sepPremise
   \rangeZeroN{i}{n}{\patternMatch{\varPath.i}
                                  {\varVal_i}
                                  {\varPatternRule_i}
                                  {\varEnv_i}
                                  {\varEnvStyles_i}}
  }
  {\patternMatch
    {\varPath}
    {\varVal}
    {\andStyles
      {\namedPattern{\varVar}{\dataApp{\varDataCon}{\rangeZeroN{i}{n}{\varPatternRule_i}}}}
      {\varAttrs}}
      {(\envExtend{\rangeZeroN{i}{n}{\varEnv_i}}{\envBind{\varVar}{\varVal}})}
      {(\envExtend{\rangeZeroN{i}{n}{\varEnvStyles_i}}{\envBind{\varPath}{\varAttrs}})}
  }
$

\vsepRule

\caption{Style Computation (continued)}
\label{fig:style-computation-two}
\end{figure}

\begin{figure}[h]

\judgementHead
  {Style Application}
  {\applyStyles{\varEnvStyles}{\varDoc} = \varDoc'}

\vsepRule

$
\inferrule*
  {
   \varAttrs' = \stylesOf{\varDoc}\cup \varEnvStyles(\varPath)
   \sepPremise
   \mathit{children'} = \applyStyles{\varEnvStyles}{\childrenOf{\varDoc}}
  }
  {
   \applyStyles{\varEnvStyles}{\varDoc}
   =
   \copyUpdateTwo
     {\varDoc}
     {\texttt{styles}}{\varAttrs'}
     {\texttt{children}}{\mathit{children'}}
  }
$

\vsepRule

\judgementHead
  {Path Application}
  {\varPath\varVal = \varVal'}

$
(\justPath{[]})\varVal = \varVal
$
\\\vspace{0.03in}
$
(\justPath{\rangeN{i}{n}{\mathit{idx}_i}})
(\dataApp{\varDataCon}{\rangeZeroN{j}{m}{\varVal_j}})
  =
(\justPath{\rangeTwoN{i}{n}{\mathit{idx}_i}})
\varVal_{\mathit{idx}_1}
$

\vsepRule

\caption{Style Computation (continued)}
\label{fig:style-computation-three}
\end{figure}
 
\subsection{Semantics}

A style sheet is a list of rules to be applied throughout a stylish text document $\varDoc$, as defined in \autoref{fig:style-computation}, \autoref{fig:style-computation-two}, and \autoref{fig:style-computation-three}.
The $\applyStyleSheet{\varStyleSheet}{\varDoc}$ function produces new documents $\varDoc'$ that preserve the structure and content of the original but extend the style attributes attached to each node according to the style sheet rules.
The definition is highly nondeterministic:

\begin{itemize}

\item
The \ruleName{ApplyOne} rule chooses a rule to consider applying to the root node of $\varDoc$;
\ruleName{ApplyRule} applies the path subrule and, if successful, checks the predicate in the value environment $\varEnv$ (with bindings of the form $\envBind{\varVar}{\varVal}$) created by matching any patterns.
If this matches and produces a new document $\varDoc'$, the entire style sheet is applied to the original document again (to allow other rules a chance to also match the root of this tree); the two, structurally-equivalent but stylistically-different documents $\varDoc'$ and $\varDoc''$ are then merged.

\item
The \ruleName{ApplyNone} rule leaves the root of the given document $\varDoc$ unchanged and recursively applies the entire style sheet to each child.

\end{itemize}

The na\"ive definitions of $\applyStyleSheet{\varStyleSheet}{\varDoc}$ and the functions it depends do not serve as an efficient implementation, but they suffice for our goal: to concisely define the semantics of \hass{} style computation as the ordinary tree traversals of CSS but with added support for pattern-matching source values that were used to generate particular document nodes in the view.

\parahead{Node Subrule Application}

The \ruleName{Node} rule is responsible for checking whether a node subrule matches.
If so, this process returns a style environment $\varEnvStyles$ with bindings of the form $\envBind{\varPath}{\varAttrs}$.
This style environment may add styles to the root of the document $\varDoc$ and---as we will discuss more below---may also add styles to its descendants.
The function $\applyStyles{\varEnvStyles}{\varDoc}$ applies the styles; this definition is written with simple set-union, but it is easy to imagine a more structured representation that, for example, tracks the provenance and precedence of rule application to enable the ``CSS inspector'' tools in modern browsers.

Node subrule matching $\selectorMatchInputOnly{\varPath}{\varClasses}{\varNodeRule}$ takes both the path and classes attached to the given document node being scrutinized.
The \ruleName{Class} rule, for checking class selectors, ignores the path, checking only that the required classes are contained in the set of classes attached to the node.
The \ruleName{Pattern} rule, for checking pattern selectors, uses the path to retrieve corresponding subvalue, if any, in the root value to perform pattern matching; any classes are also then checked.
Pattern-matching produces a value environment, which is propagated as the output of the selector-matching ``predicate.''
It also takes a datatype context as an input, to require that variable patterns only match values of the appropriate type.

Compared to an ordinary pattern-matching operation, pattern subrule application includes an additional input (a path $\varPath$) and an additional output (a style environment $\varEnvStyles$).
The path $\varPath$ describes the provenance of $\varVal$ in the original AST value.
When the \ruleName{Var} rule successfully matches a value, the output style environment records a binding from the current path to the new styles to add.
When scrutinizing the $i$th subterm of a constructed value, the \ruleName{Constructor} rule uses the extended path $\varPath.i$ when evaluating the subpattern.
As described above, all of the styles in $\varEnvStyles$ are applied by \ruleName{Node} to the document node corresponding to the value at path $\varPath$.

Our formulation also includes a ``keep-out'' pattern $\keepOut$ that never matches (notice there is no rule for this pattern form).
This is useful, for example, to implement a selector that matches all variables in the right subvalue of an \texttt{EApp} expression (i.e., the function argument) but not the left.
Without a keep-out pattern, it is difficult to write such a selector because
there is no way for our path selectors to differentiate between subvalues.
This issue is ordinarily mitigated in HTML and CSS by inserting a marker element or extra class.
However, in our setting we prefer to avoid requiring that data types or corresponding stylish text documents be prearranged with additional data constructors or classes, respectively, to serve only this purpose.

\parahead{Pattern Subrule Application}

The last part of \autoref{fig:style-computation-two} to discuss is the handling of multi-node matches for tree combinators.
The child combinator, written ``$\opChild$'' as in CSS, is handled by the \ruleName{Child} rule, which matches the node subrule $\varNodeRule$ against the root of the given node and then the remaining path subrule $\varPathRule$ on (any) one of its (immediate) children.
The descendant combinator, which allows $\varNodeRule \opDescendant \varPathRule$ to be matched by paths of length one or more, is handled by two cases.
The \ruleName{Descendant1} rule handles the ``length exactly-one'' case by rephrasing the query using the child combinator.
The \ruleName{Descendant2} rule handles the ``length two-or-more'' case by naively extending the rule with an intermediary \emph{top selector} $\topSelector$; this will succeed for any node (via the \ruleName{Top} rule), effectively extending the obligation for matching $\varPathRule$ down to the grandchildren (or below).
As discussed earlier, a direct implementation of this approach would not be efficient, but it suffices for our purposes.
Other tree combinators can be implemented in similar ways; these details are elided in \autoref{fig:style-computation-two}.

\subsection{Non-Local Provenance}

The algorithm presented here is sufficient
for all of the examples presented in this paper, and many more. But
there are instances of \texttt{Stylish} that one might want to write
but cannot due to an assumption of the style computation algorithm---namely,
that our system prevents non-local references from the stylish text document
into the source value tree.
This manifests in the \ruleName{Node} rule:
the decoration of $\varDoc$ also produces a style environment $\varEnvStyles$
decorating other nodes, but $\varEnvStyles$ is applied only to (the subtree rooted at) $\varDoc$.
However, it is conceivable that a node styled by $\varEnvStyles$ (referenced by some path $\varPath'$) is not ``below'' $\varDoc$ (say, with path $\varPath$).
In other words, the parent reference $\varPath$ is not a prefix $\varPath'$.
One (particularly inefficient) way to allow $\varEnvStyles$ to style non-local nodes would be to rephrase all of the definitions---which are currently local document-to-document rewrites---as ``global'' versions, in which the transformation of any node generates a style environment to be applied to the root of the document.
We leave it to future work to consider whether or not there are realistic use cases to warrant such an approach.

\clearpage
\section{Style Sheets for Examples}
\label{sec:hass-examples}

\newcommand{\rot}[1]{\rotatebox{90}{#1}}
\newcommand{\No}{{\color{gray} No}}
\newcommand{\Yes}{\textbf{Yes}}
\newcommand{\hs}[1]{#1$^{\textrm{H}}$}
\newcommand{\css}[1]{#1$^{\textrm{CSS}}$}

\begin{figure}[h]
  \small
  {\renewcommand{\arraystretch}{2}\begin{tabular}{| r | c c c c | c c | c c c |}
    \hline
    &
    \multicolumn{4}{c|}{Parsing} &
    \multicolumn{2}{c|}{Static} &
    \multicolumn{3}{c|}{Run-Time} \\
    \hline
    &
    \rot{Blocks (\autoref{sec:listing-blocks})} &
    \rot{Skeleton Code (\autoref{sec:listing-skeleton-code})} &
    \rot{Point-Free Pipeline (\autoref{sec:listing-pipeline})} &
    \rot{Syntax Highlighting (\autoref{sec:listing-syntax-highlighting})} &
    \rot{Semantic Highlighting (\autoref{sec:listing-semantic-highlighting})\quad} &
    \rot{Type Error (\autoref{sec:listing-type-error})} &
    \rot{Projection Boxes (\autoref{sec:listing-projection-boxes})} &
    \rot{Test Runner (\autoref{sec:listing-test-runner})} &
    \rot{Heat Map (\autoref{sec:listing-heat-map})} \\
    \hline
    Style Sheet LOC (\texttt{.hs} file) & 22 & 52 & 78 & 98 & 78 & \hs{234} & \css{53} & \css{81} & 82 \\
    \hline
Visible S-Blocks & \Yes & \No & \Yes & \No & \No & \Yes & \No & \Yes$^{\sba}$ & \No \\
    \hline
AST Structure $\ncong$ Display Structure &
    \No & \No & \No & \No & \No & \No & $\Yes^\dag$ & $\Yes^\dag$ & $\Yes^\ddagger$ \\
    \hline
    Program-Specific Rules &
    \No & \Yes & \No & \No & \No & \Yes & \No & \No & \No \\
    \hline
    \hasskell{} Version &
    Full & Full & Full & Full & Tiny & Tiny & Tiny & Tiny & Tiny \\
    \hline
  \end{tabular}}
\caption{Characteristics of examples,
grouped by the type of information displayed and then ordered by LOC.
``H'' denotes a style sheet implemented in ``regular'' Haskell, rather than Template Haskell. ``CSS'' denotes two examples in which some rules are defined with ``raw'' CSS class selectors; these examples share a common 42 lines of CSS, embedded as string literals in Haskell.
\sba{}~denotes a nested display that includes only single-line, not multiline, s-blocks~(\autoref{fig:s-block}).
\dag{}~denotes two examples which share a custom \texttt{Stylish} instance with 24 additional LOC over the default (351 LOC) \texttt{Stylish} instance.
}
\label{fig:example-matrix}
\end{figure}

\parahead{Style Sheets in Template Haskell (TH)}

As explained in \autoref{sec:hass}, our current prototype does not implement a direct parser for standalone \verb+.hass+ style sheet files comprising the external syntax (\autoref{fig:external-syntax}).
Instead, style sheet definitions are embedded in Haskell files, often using Template Haskell which allows rule definitions to closely resemble the external syntax. For example: 

\vspace{0.08in}

\begin{center}
\begin{tabular}{rl}
External Syntax &
  {\small\verb+         pair@(Pair x y) -> pair {...} x {...} y {...}+}
\\[4pt]
Template Haskell &
  {\small\verb+[S.rule| pair@(Pair x y) -> pair {...} x {...} y {...} |]+}
\end{tabular}
\end{center}

\vspace{0.08in}

\noindent
Many of the Template Haskell definitions use selector lists (\autoref{sec:selector-terminology}) in order to reduce code duplication.
Besides the \texttt{Stylesheet.rule} parser, the TH definitions use native Haskell facilities, such as lists and boolean expressions, to encode the remainder of the style sheet syntax.

As described next, two examples---Type Error and Projection Boxes---are not specified using Template Haskell.

\parahead{Type Error Style Sheet: ``Regular'' Haskell (H)}

Our Template Haskell definitions constitute a compile-time desugaring from the external syntax into an internal representation for rules, defined using ordinary Haskell expressions.
In this internal representation, variables which bind subvalues for \emph{styling} are represented as paths (rather than variables), whereas variables binding subvalues for \emph{predicate evaluation} treated directly (as variables).
For example, the \texttt{Pair x y} example above would be desugared into the following expression:

\vspace{0.08in}

{\small
\begin{Verbatim}[xleftmargin=0.40in]
Rule
  [ PrimitiveSelector (-1)
      ([([], ...), ([1], ...), ([2], ...)]) -- styles for pair, x, y
      []                                    -- keepOutPaths
      (PatternSelector $                    -- selector
         \case
            pair@(Pair x y) -> Just [Binding pair, Binding x, Binding y]
             _              -> Nothing)
  ]
  (const True)                              -- predicate
\end{Verbatim}
}

\vspace{0.08in}

The Type Error style sheet is defined in Haskell using this internal representation directly.
As discussed in \autoref{sec:hass},
this is because the internal representation, though verbose, permits style attributes to be \emph{expressions}, rather than \emph{literals}.
This is used in the Type Error example to allow the color of backgrounds and borders to depend on the \textit{type} of an expression.

Alternatively, the \hasskell{} type checker could be reworked to generate a temporary Template Haskell \texttt{.hs} file, which is then used as input to style computation;
this would obviate the need to write the internal representation directly.
However, a more worthwhile engineering improvement is to simply implement a standalone parser for the external syntax---also allowing these \texttt{.hass} files to define expressions for style attributes (cf. \autoref{sec:hass}).

\parahead{Projection Boxes: ``Raw'' Style Sheet (CSS)}

Our implementation supports placing uninterpreted HTML strings at the leaves of the \texttt{StylishText} tree (cf. \texttt{HtmlLeaf} in \autoref{sec:hass} and \autoref{sec:limitations}).
This is what enables projection boxes to be rendered as HTML tables inline with the program text.
However, these HTML strings are ``opaque'' to \hass{} selectors. 

In order to style these tables, the custom \texttt{Stylish} instance, used by the Projection Boxes and Test Runner examples, attaches an ordinary CSS style sheet near the root of the \texttt{StylishText} tree.
This Projection Boxes style sheet is represented as a string literal (see \autoref{sec:listing-projection-boxes}).
The Test Runner example (\autoref{sec:listing-test-runner}) includes the same CSS rules among other (\hass{}) rules.

This limitation could be removed by replacing \texttt{HtmlLeaf} with a simple yet more structured representation of HTML, but still restricted to the leaves of the \texttt{StylishText} tree.
Going further would be to more fully integrate HTML and stylish text (cf. \autoref{sec:limitations}).

\clearpage

\subsection{Blocks (Figure 1)}
\label{sec:listing-blocks}
{\scriptsize
\begin{verbatim}
module Main where

import Common
import qualified Stylesheet as S (rule)

sBinops = [S.rule|
x@EBinop _ ->
x {
  border-width: 2;
  padding: 2;
  margin: 2;
  border-radius: 3;
  border-color: navy;
  background-color: rgba(44, 90, 160, 0.2);
}
|]

stylesheet :: [Rule]
stylesheet = sBinops

main :: IO ()
main = stylesheetMain stylesheet

\end{verbatim}}
\clearpage
\subsection{Skeleton Code (Figure 16)}
\label{sec:listing-skeleton-code}
{\scriptsize
\begin{verbatim}
module Main where

import Common
import qualified Stylesheet as S (rule)

selectMain = [S.rule|
x@Equation (Left (Located _ (PVar (Located _ (y@Ident "main"))), _), _, _) ->
x y {
  color: gray;
}
|]

selectMod = [S.rule|
x@ModDecl _ ->
x {
  color: gray;
}
|]

selectTVar = [S.rule|
TVar (Located _ t1) ->
t1 {
  color: green;
}
|]

selectDecls = [S.rule|
(Decl (Located _ (Signature (_, xxx, xxx)))) (x@Ident _),
(Decl (Located _ (Equation ((Left (_, xxx)), xxx, xxx)))) (x@Ident _) ->
x {
  color: mediumvioletred;
  font-size: 18px;
}
|]

selectComment = [S.rule|
x@.comment ->
x {
  font-family: "Noto Serif", serif;
}
|]

stylesheet :: [Rule]
stylesheet =
  selectDecls
  ++ selectMain
  ++ selectMod
  ++ selectTVar
  ++ selectComment

main :: IO ()
main = stylesheetMain stylesheet

\end{verbatim}}
\clearpage
\subsection{Point-Free Pipeline (Figure 3)}
\label{sec:listing-pipeline}
{\scriptsize
\begin{verbatim}
module Main where

import Common
import Data.Map (Map, (!))
import qualified Data.Map as Map
import qualified Stylesheet as S (rule)

data Dir = LeftToRight | RightToLeft | Don'tCare deriving Eq

opDir :: Map String Dir
opDir = Map.fromList
  [ (">>=", LeftToRight)
  , (">>>", LeftToRight)
  , (".",   RightToLeft)
  , ("==",  Don'tCare)
  , ("+",   Don'tCare)
  , (":",   Don'tCare) ]

sBinopTransitionLR = [S.rule|
x@(EBinop (_, Located _ (Op op1), Located _ (y@EBinop (_, Located _ (Op op2), _)))) if isDirUnEq op1 op2 ->
x {
  border-width: 2; padding: 2; margin: 2; border-radius: 3;
  border-color: indigo;
  background-color: lavender;
}
y {
  border-width: 2; padding: 2; margin: 2; border-radius: 3;
  border-color: orange;
  background-color: papayawhip;
}
|]
  where isDirUnEq op1 op2 =
          selectP ((== LeftToRight) . (opDir !)) op1
          && selectP ((== RightToLeft) . (opDir !)) op2

sBinopTransitionRL = [S.rule|
x@(EBinop (_, Located _ (Op op1), Located _ (y@EBinop (_, Located _ (Op op2), _)))) if isDirUnEq op1 op2 ->
x {
  border-width: 2; padding: 2; margin: 2; border-radius: 3;
  border-color: orange;
  background-color: papayawhip;
}
y {
  border-width: 2; padding: 2; margin: 2; border-radius: 3;
  border-color: indigo;
  background-color: lavender;
}
|]
  where isDirUnEq op1 op2 =
          selectP ((== RightToLeft) . (opDir !)) op1
          && selectP ((== LeftToRight) . (opDir !)) op2

sBinopLR = [S.rule|
(EBinop (_, Located _ (x@Op op), _)) if isLR op ->
x {
  font-weight: bold;
  color: indigo;
}
|]
  where isLR op = selectP ((== LeftToRight) . (opDir !)) op

sBinopRL = [S.rule|
(EBinop (_, Located _ (x@Op op), _)) if isRL op ->
x {
  font-weight: bold;
  color: orange;
}
|]
  where isRL op = selectP ((== RightToLeft) . (opDir !)) op

stylesheet :: [Rule]
stylesheet = sBinopTransitionLR
             ++ sBinopTransitionRL
             ++ sBinopLR
             ++ sBinopRL

main :: IO ()
main = stylesheetMain stylesheet

\end{verbatim}}
\clearpage
\subsection{Syntax Highlighting (Figure 10)}
\label{sec:listing-syntax-highlighting}
{\scriptsize
\begin{verbatim}
module Main where

import Common
import Data.Char (isUpper)
import qualified Stylesheet as S (rule)

selectConstant = [S.rule|
x@EInt _,
EChar (_, x, _),
EFloat (x, y) ->
x y {
  color: teal;
}
|]

selectString = [S.rule|
(EString (Located _ t1, _, Located _ t2)) x@.char ->
x {
  color: teal;
}
t1 t2 {
  color: blue;
}
|]

selectDecls = [S.rule|
(Decl (Located _ (Equation ((Left (_, xxx)), xxx, xxx)))) (x@Ident _),
(Decl (Located _ (Signature (_, xxx, xxx)))) (x@Ident _),
(ClassDecl _) (Signature (_, xxx, xxx)) (x@Ident _),
(ClassDecl _) (Equation ((Left (_, xxx)), xxx, xxx)) (x@Ident _),
(InstanceDecl _) (Equation ((Left (_, xxx)), xxx, xxx)) (x@Ident _) ->
x {
  color: mediumvioletred;
}
|]

selectConstructors = [S.rule|
v@(Ident nm) if pred nm ->
v {
  color: green;
}
|]
  where pred nm =
          selectP (isUpper . head) nm

selectTypeDecls = [S.rule|
TypeDecl (Located _ t, Located _ x, _, _, _),
NewtypeDecl (Located _ t, Located _ x, _, _, _),
DataDecl (Located _ t, Located _ x, _, _, _) ->
t {
  color: blue;
}
x {
  color: green;
}
|]

selectKeywords = [S.rule|
ImportDecl (Located _ t1, _, _, _, _),
ImportDecl (_, Just (Located _ t1), _, _, _),
ImportDecl (_, _, _, Just (Located _ t1, _), _),
ECase (Located _ t1, _, Located _ t2, _),
EIf (Located _ t1, _, Located _ t2, _, Located _ t3, _),
EDo (Located _ t1, _),
DoLet (Located _ t1, _),
ModDecl (Located _ t1, _, _, Located _ t2),
ClassDecl (Located _ t1, _, _, _, Located _ t2, _),
InstanceDecl (Located _ t1, _, _, _, Located _ t2, _),
Equation (_, _, Just (Located _ t1, _)),
InfixL (Located _ t),
InfixR (Located _ t) ->
t1 t2 t3 {
  color: blue;
}
|]

selectTypes = [S.rule|
TUnit (Located _ t1, Located _ t2),
TCon (Located _ t1),
TVar (Located _ t1),
TFun (_, Located _ t1, _) ->
t1 t2 {
  color: green;
}
|]

stylesheet :: [Rule]
stylesheet =
  selectConstant
  ++ selectDecls
  ++ selectTypeDecls
  ++ selectKeywords
  ++ selectTypes
  ++ selectString
  ++ selectConstructors

main :: IO ()
main = stylesheetMain stylesheet

\end{verbatim}}
\clearpage
\subsection{Semantic Highlighting (Figure 16)}
\label{sec:listing-semantic-highlighting}
{\scriptsize
\begin{verbatim}
module Main where

import TinyCommon
import qualified Stylesheet as S (rule)

nColors :: Int
nColors = 6

bindUse1 = [S.rule|
x@(Ident @RenamePhase (IdentBinding uniq) _) if pred uniq,
x@(Ident @RenamePhase (IdentUsage _ binder) _) if pred binder ->
x {
  color: brown;
}
|]
  where pred bind =
          selectP (\(TraceId id) -> id `mod` nColors == 0) bind

bindUse2 = [S.rule|
x@(Ident @RenamePhase (IdentBinding uniq) _) if pred uniq,
x@(Ident @RenamePhase (IdentUsage _ binder) _) if pred binder ->
x {
  color: darkcyan;
}
|]
  where pred bind =
          selectP (\(TraceId id) -> id `mod` nColors == 1) bind

bindUse3 = [S.rule|
x@(Ident @RenamePhase (IdentBinding uniq) _) if pred uniq,
x@(Ident @RenamePhase (IdentUsage _ binder) _) if pred binder ->
x {
  color: darkolivegreen;
}
|]
  where pred bind =
          selectP (\(TraceId id) -> id `mod` nColors == 2) bind

bindUse4 = [S.rule|
x@(Ident @RenamePhase (IdentBinding uniq) _) if pred uniq,
x@(Ident @RenamePhase (IdentUsage _ binder) _) if pred binder ->
x {
  color: purple;
}
|]
  where pred bind =
          selectP (\(TraceId id) -> id `mod` nColors == 3) bind

bindUse5 = [S.rule|
x@(Ident @RenamePhase (IdentBinding uniq) _) if pred uniq,
x@(Ident @RenamePhase (IdentUsage _ binder) _) if pred binder ->
x {
  color: orange;
}
|]
  where pred bind =
          selectP (\(TraceId id) -> id `mod` nColors == 4) bind

bindUse6 = [S.rule|
x@(Ident @RenamePhase (IdentBinding uniq) _) if pred uniq,
x@(Ident @RenamePhase (IdentUsage _ binder) _) if pred binder ->
x {
  color: red;
}
|]
  where pred bind =
          selectP (\(TraceId id) -> id `mod` nColors == 5) bind

stylesheet :: [Rule]
stylesheet = bindUse1
             ++ bindUse2
             ++ bindUse3
             ++ bindUse4
             ++ bindUse5
             ++ bindUse6

main :: IO ()
main = stylesheetMain $ defaultStylish @RenamePhase stylesheet

\end{verbatim}}
\clearpage
\subsection{Type Error (Figure 18)}
\label{sec:listing-type-error}
{\scriptsize
\begin{verbatim}
module Main where

import TinyCommon
import Control.Monad
import Control.Monad.IO.Class
import Control.Monad.State
import Data.Map ( Map )
import Server
import qualified Data.Map as Map
import qualified Stylesheet as S (rule)
import Data.Colour.SRGB

type ColorName = String

data ColorRecord
  = ColorRecord
  { outlineRGB :: Color
  , fillRGB :: Color
  , colorDescription :: String
  }

niceColors :: [ColorRecord]
niceColors =
  [ ColorRecord
    { outlineRGB = sRGB24read "#502D16"
    , fillRGB = sRGB24read "#FFE6D5"
    , colorDescription = "Brown"
    }
  , ColorRecord
    { outlineRGB = sRGB24read "#2C5AA0"
    , fillRGB = sRGB24read "#D7E3F4"
    , colorDescription = "Blue"
    }
  , ColorRecord
    { outlineRGB = sRGB24read "#E5FF80"
    , fillRGB = sRGB24read "#F5FFCF"
    , colorDescription = "Green"
    }
  , ColorRecord
    { outlineRGB = sRGB24read "#F05735"
    , fillRGB = sRGB24read "#F797A7"
    , colorDescription = "Red"
    }
  ]

sExpById :: TraceId -> ColorRecord -> Rule
sExpById target color =
  Rule
  [ PrimitiveSelectorRule
    { targetedStyles =
        [([], [ ("border-color", toHex (outlineRGB color), 1)
              , ("background-color", toHex (fillRGB color), 1)
              , ("border-width", "2", 1)
              , ("padding", "2", 0)
              , ("margin", "2", 0)
              , ("border-radius", "3", 0)
              ])]
    , keepOutPaths = []
    , selector = PatternSelector (-1)
      (select $ \(e::Exp_ TypecheckPhase) -> Just [Binding e]) []
    }
  ]
  $ \case [Binding e] -> idOfExp e == target
          _ -> False

sDeclById :: TraceId -> ColorRecord -> Rule
sDeclById target color =
  Rule
  [ PrimitiveSelectorRule
    { targetedStyles =
        [([], [ ("border-color", toHex (outlineRGB color), 1)
              , ("background-color", toHex (fillRGB color), 1)
              , ("border-width", "2", 1)
              , ("padding", "2", 0)
              , ("margin", "2", 0)
              , ("border-radius", "3", 0)
              ])]
    , keepOutPaths = []
    , selector = PatternSelector (-1)
      (select $ \(d::Decl_ TypecheckPhase) -> Just [Binding d]) []
    }
  ]
  $ \case [Binding d] -> idOfDecl d == target
          _ -> False

sIf = [S.rule|
EIf @TypecheckPhase _ (Located _ t1, _, Located _ t2, _, Located _ t3, _) ->
t1 t2 t3 {
  font-weight: bold;
}
|]

sExpWithTyp :: TraceId -> String -> ColorRecord -> Rule
sExpWithTyp target typ color =
  Rule
  [ PrimitiveSelectorRule
    { targetedStyles = []
    , keepOutPaths = []
    , selector = PatternSelector (-1) (select $ \(e1::Exp_ TypecheckPhase) -> Just [Binding e1]) []
    }
  , PrimitiveSelectorRule
    { targetedStyles =
        [([], [ ("background-color", toHex (fillRGB color), 0)
              , ("border-color", toHex (outlineRGB color), 0)
              , ("border-width", "2", 0)
              , ("padding", "2", 0)
              , ("margin", "2", 0)
              , ("border-radius", "3", 0) ])]
    , keepOutPaths = []
    , selector = PatternSelector (-1) (select $ \(e2::Exp_ TypecheckPhase) -> Just [Binding e2]) []
    }
  ]
  $ \case [Binding e1, Binding e2] ->
            idOfExp e1 == target && selectP ((==typ) . tc_typ_string . getExpAnno @TypecheckPhase) e2
          _ -> False

sDeclWithTyp :: TraceId -> String -> ColorRecord -> Rule
sDeclWithTyp target typ color =
  Rule
  [ PrimitiveSelectorRule
    { targetedStyles = []
    , keepOutPaths = []
    , selector = PatternSelector (-1) (select $ \(d::Decl_ TypecheckPhase) -> Just [Binding d]) []
    }
  , PrimitiveSelectorRule
    { targetedStyles =
        [([], [ ("background-color", toHex (fillRGB color), 0)
              , ("border-color", toHex (outlineRGB color), 0)
              , ("border-width", "2", 0)
              , ("padding", "2", 0)
              , ("margin", "2", 0)
              , ("border-radius", "3", 0) ])]
    , keepOutPaths = []
    , selector = PatternSelector (-1) (select $ \(e::Exp_ TypecheckPhase) -> Just [Binding e]) []
    }
  ]
  $ \case [Binding d, Binding e] ->
            idOfDecl d == target && selectP ((==typ) . tc_typ_string . getExpAnno @TypecheckPhase) e
          _ -> False

data ColoredTcError
  = ColoredUnificationError TraceId ColorRecord (ColorRecord, TcTyp) (ColorRecord, TcTyp)
  | ColoredUnboundVariableError TraceId ColorRecord String Range

nextColor :: String -> State (Int, Map String ColorRecord) ColorRecord
nextColor typeString = do
  table <- gets snd
  case Map.lookup typeString table of
    Just colorName -> pure colorName
    Nothing -> do
      let
        nKeys = length niceColors

      index <- gets fst
      let color = niceColors !! index
      modify (\(idx, map) ->
                ((idx + 1) `mod` nKeys, Map.insert typeString color map))
      pure color

assignColorsToError
  :: TypecheckError
  -> State (Int, Map String ColorRecord) ColoredTcError
assignColorsToError (UnificationError tid ta tb) = do
  let
    taString = pprintType ta defaultNameMap
    tbString = pprintType tb defaultNameMap
  c  <- nextColor (show tid)
  ca <- nextColor taString
  cb <- nextColor tbString
  pure $ ColoredUnificationError tid c (ca, ta) (cb, tb)
assignColorsToError (UnboundVariableError tid nm range) = do
  c <- nextColor ""
  pure $ ColoredUnboundVariableError tid c nm range

pprintError :: ColoredTcError -> IO ()
pprintError (ColoredUnificationError _ c (ca, ta) (cb, tb)) =
  putStrLn $
  "in "
  ++ withANSIColor cRGB cDescription
  ++ " expression, could not unify "
  ++ withANSIColor caRGB taString
  ++ " and "
  ++ withANSIColor cbRGB tbString
  where
    taString = pprintType ta defaultNameMap
    tbString = pprintType tb defaultNameMap

    cDescription = colorDescription c
    cRGB = outlineRGB c
    caRGB = outlineRGB ca
    cbRGB = outlineRGB cb
pprintError (ColoredUnboundVariableError _ c nm range) =
  putStrLn $
  withANSIColor cRGB nm ++ " is unbound at " ++ show range
  where
    cRGB = outlineRGB c

pyretErrors
  :: SourceFile ParsePhase
  -> MsgHandler (RenderableStylishText MsgHandler)
pyretErrors =
  uptoTypechecking $ \(errors, typechecked) -> do
  let
    coloredErrors =
      evalState (mapM assignColorsToError errors) (0, Map.empty)

    stylesheet :: [Rule]
    stylesheet =
      concatMap
      (\case
          (ColoredUnificationError tid c (ca, ta) (cb, tb)) ->
            let
              taString = pprintType ta defaultNameMap
              tbString = pprintType tb defaultNameMap
            in [ sExpById tid c
               , sDeclById tid c
               , sExpWithTyp tid taString ca
               , sDeclWithTyp tid taString ca
               , sExpWithTyp tid tbString cb
               , sDeclWithTyp tid taString cb ]
          (ColoredUnboundVariableError tid c _ _) ->
            [sExpById tid c]) coloredErrors

  liftIO $ forM_ coloredErrors pprintError

  let stylishText =
        applyStyles typechecked (stylesheet ++ sIf)
        $ showStylish []
        $ DefaultStylish typechecked

  pure $ RenderableStylishText stylishText measureString

main :: IO ()
main = stylesheetMainImpl pyretErrors

\end{verbatim}}
\clearpage
\subsection{Projection Boxes (Figure 2)}
\label{sec:listing-projection-boxes}
{\scriptsize
\begin{verbatim}
module Main where

import TinyCommon

css :: String
css =
  ".outer-box {\
  \  position: relative;\
  \  display: flex;\
  \  color: black;\
  \  flex-direction: row;\
  \}\
  \.inner-box {\
  \  max-height: 4em;\
  \  overflow: auto;\
  \  border: 2px solid gray;\
  \  border-radius: 3px;\
  \}\
  \.tooltip {\
  \  position: absolute;\
  \  border-style: solid;\
  \  border-color: transparent gray transparent transparent;\
  \  left: -8px;\
  \  top: calc(0.5em - 4px);\
  \  border-width: 8px;\
  \}\
  \.tooltip-spacer {\
  \  width: 5px;\
  \}\
  \.table {\
  \  border-collapse: collapse;\
  \  border-spacing: 0;\
  \}\
  \.even-col {\
  \  background-color: #D7E3F4;\
  \}\
  \.even-col-header {\
  \  position: sticky;\
  \  top: 0;\
  \  background-color: #D7E3F4;\
  \  box-shadow: 0 2px 0 gray;\
  \}\
  \.odd-col-header {\
  \  position: sticky;\
  \  top: 0;\
  \  background-color: white;\
  \  box-shadow: 0 2px 0 gray;\
  \}"

main :: IO ()
main = stylesheetMain
  $ (projectionBoxesStylish @EvalPhase [])
  { stylishTransform = addCSS css }

\end{verbatim}}
\clearpage
\subsection{Test Runner (Figure 19)}
\label{sec:listing-test-runner}
{\scriptsize
\begin{verbatim}
module Main where

import TinyCommon
import qualified Stylesheet as S (rule)

sPassingTests = [S.rule|
x@(EApp @EvalPhase (EvalExpInfo _ _ _ Passed) xxx) ->
x {
  background-color: #E5FF80;
  border-color: green;
  border-width: 2;
  padding: 2;
  border-radius: 3;
  margin: 2;
}
|]

sFailingTests = [S.rule|
x@(EApp @EvalPhase (EvalExpInfo _ _ _ Failed) xxx) ->
x {
  border-color: #F05735;
  border-width: 2;
  background-color: #F797A7;
  padding: 2;
  margin: 2;
  border-radius: 3;
}
|]

stylesheet :: [Rule]
stylesheet = sPassingTests ++ sFailingTests

css :: String
css =
  ".outer-box {\
  \  position: relative;\
  \  display: flex;\
  \  color: black;\
  \  flex-direction: row;\
  \}\
  \.inner-box {\
  \  max-height: 4em;\
  \  overflow: auto;\
  \  border: 2px solid gray;\
  \  border-radius: 3px;\
  \}\
  \.tooltip {\
  \  position: absolute;\
  \  border-style: solid;\
  \  border-color: transparent gray transparent transparent;\
  \  left: -8px;\
  \  top: calc(0.5em - 4px);\
  \  border-width: 8px;\
  \}\
  \.tooltip-spacer {\
  \  width: 5px;\
  \}\
  \.table {\
  \  border-collapse: collapse;\
  \  border-spacing: 0;\
  \}\
  \.even-col {\
  \  background-color: #D7E3F4;\
  \}\
  \.even-col-header {\
  \  position: sticky;\
  \  top: 0;\
  \  background-color: #D7E3F4;\
  \  box-shadow: 0 2px 0 gray;\
  \}\
  \.odd-col-header {\
  \  position: sticky;\
  \  top: 0;\
  \  background-color: white;\
  \  box-shadow: 0 2px 0 gray;\
  \}"

main :: IO ()
main = stylesheetMain
  $ (projectionBoxesStylish @EvalPhase stylesheet)
  { stylishTransform = addCSS css }

\end{verbatim}}
\clearpage
\subsection{Heat Map (Figure 19)}
\label{sec:listing-heat-map}
{\scriptsize
\begin{verbatim}
module Main where

import Control.Monad
import Data.Colour
import Text.Blaze.Html.Renderer.String ( renderHtml )
import Text.Blaze.Html5 ( (!) )
import TinyCommon
import qualified Data.Colour.Names as Colors
import qualified Stylesheet as S (rule)
import qualified Text.Blaze.Html5 as H
import qualified Text.Blaze.Html5.Attributes as A

import Debug.Trace

sExpByHeat :: (Double, Double) -> Color -> Rule
sExpByHeat (lower, upper) color =
  Rule
  [ PrimitiveSelectorRule
    { targetedStyles = [([], [ ("color", toHex color, 1) ])]
    , keepOutPaths = []
    , selector = PatternSelector (-1) (select $ \(e::Exp_ EvalPhase) -> Just [Binding e]) []
    }
  ]
  $ \case [Binding e] ->
            selectP ((\x -> x > lower && x <= upper) . pct) e
            where
              pct = ev_exp_sample_percentage . getExpAnno @EvalPhase
          _ -> False

sCase = [S.rule|
ECase @EvalPhase _ (Located _ t1, _, Located _ t2, _) ->
t1 t2 {
  font-weight: bold;
}
|]

coolToWarmGradient :: Int -> [Color]
coolToWarmGradient n = map (\t -> blend t warm cool) (linspace n)
  where
    cool = Colors.royalblue
    warm = Colors.orangered

linspace :: Int -> [Double]
linspace n = map (\i -> fromIntegral i * delta) [0..n-1]
  where
    delta = 1 / fromIntegral (n - 1)

ranges :: Int -> Double -> [(Double, Double)]
ranges n max = map (\i -> let f = fromIntegral i
                          in (f * delta, (f + 1) * delta)) [0..n-1]
  where
    delta = max / fromIntegral n

addColorBar :: [Color] -> StylishText -> StylishText
addColorBar gradient (Node path cls sty children) =
  Node path cls sty (colorBar:children)
  where
    colorBar :: StylishText
    colorBar = Node Nothing [] []
      [ HtmlLeaf $ renderHtml $ H.div $
        forM_ gradient $ \color ->
          let style = H.stringValue $ "background-color: "
                      ++ toHex color
                      ++ "; height: 20px;\
                         \width: 38px;\
                         \display: inline-block;"
          in H.span ! A.style style $ pure ()
      , TextLeaf "\n" ]

nRanges :: Int
nRanges = 16

gradient :: [Color]
gradient = coolToWarmGradient nRanges

stylesheet :: [Rule]
stylesheet = zipWith sExpByHeat (ranges nRanges 1) gradient ++ sCase

main :: IO ()
main = stylesheetMain
  $ (defaultStylish @EvalPhase stylesheet)
  { stylishTransform = addColorBar gradient }

\end{verbatim}}
\clearpage

\clearpage
\thispagestyle{empty}
\mbox{}

\vspace{2.50in}
{\noindent\hfill\Huge\textbf{OUTTAKES}\hfill}

\clearpage
\section{Code Style Sheets : Code :: CSS : HTML} 

\emph{``We'll see if this "string stylesheets" idea holds up; if it does, having a simple ADT/DOM analogy seems nice''} (April 12, 2023).
Three earlier attempts at explaining the analogy:

\begin{figure}[h] \includegraphics[width=4.25in]{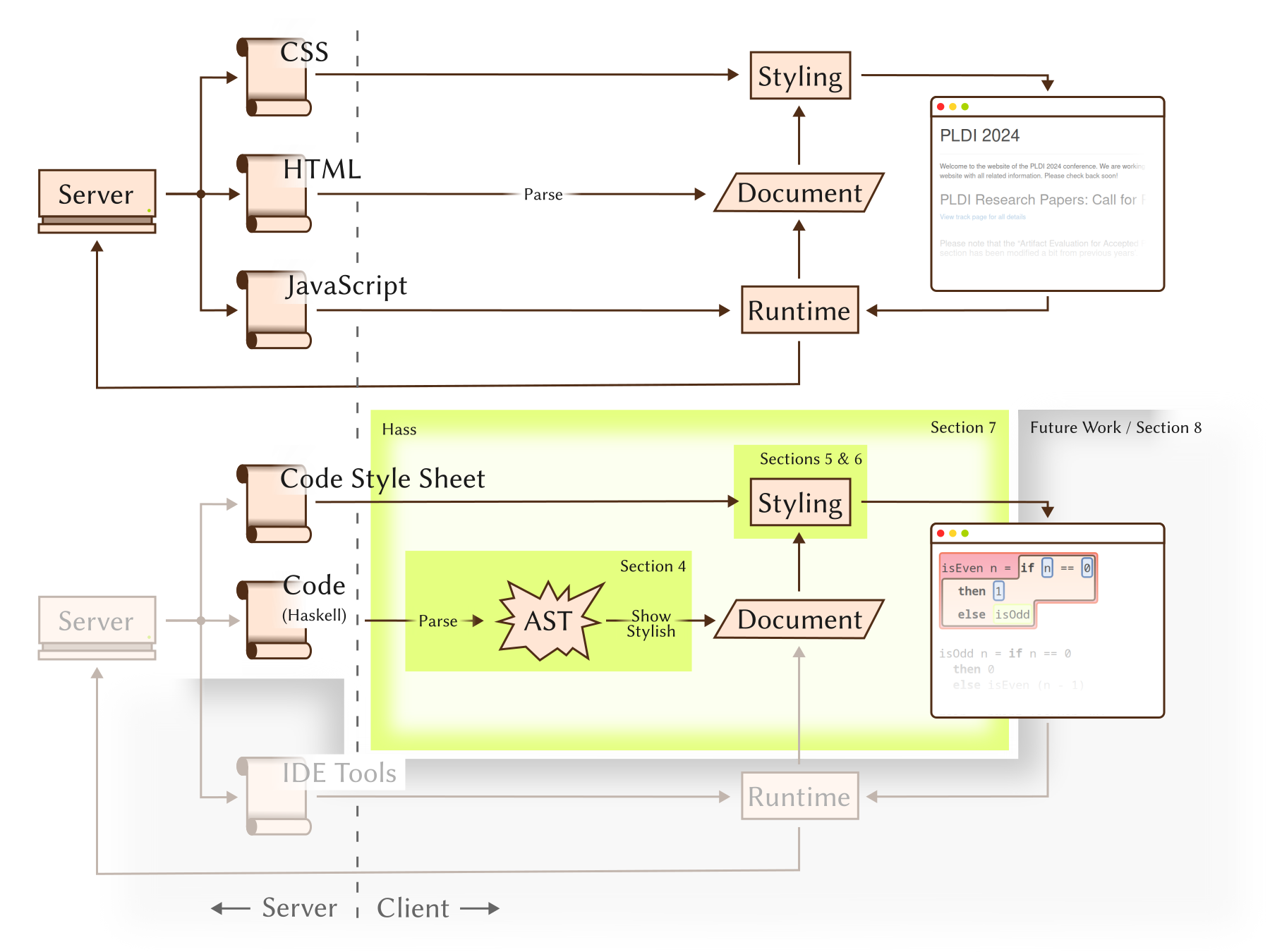}
\caption{Analogy Diagram v1 (November 2023)}
\label{fig:architecture-v1}
\end{figure}
 \begin{figure}[h]

\includegraphics[width=4.25in]{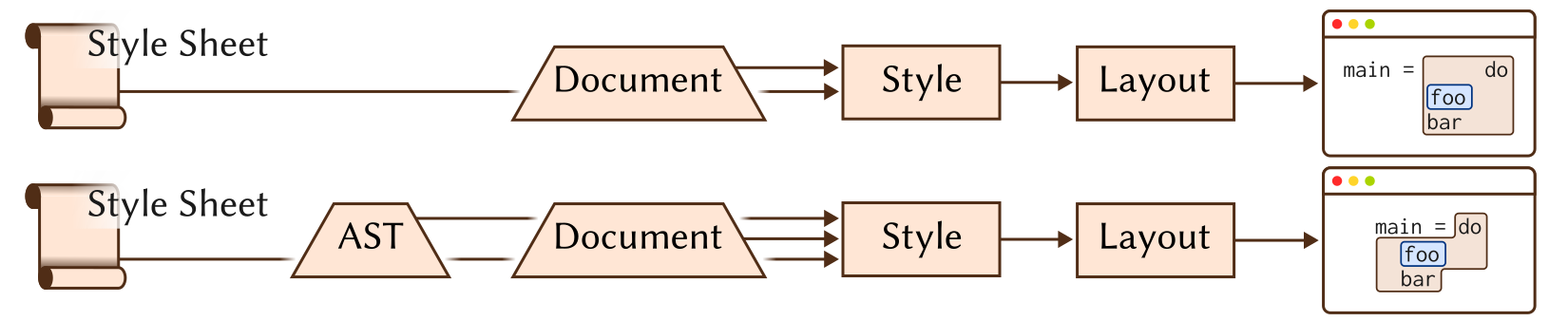}
  \caption{Analogy Diagram v2 (April 2024)}
\label{fig:architecture-v2}
\end{figure}
 \begin{figure}[h]

\includegraphics[width=4.25in]{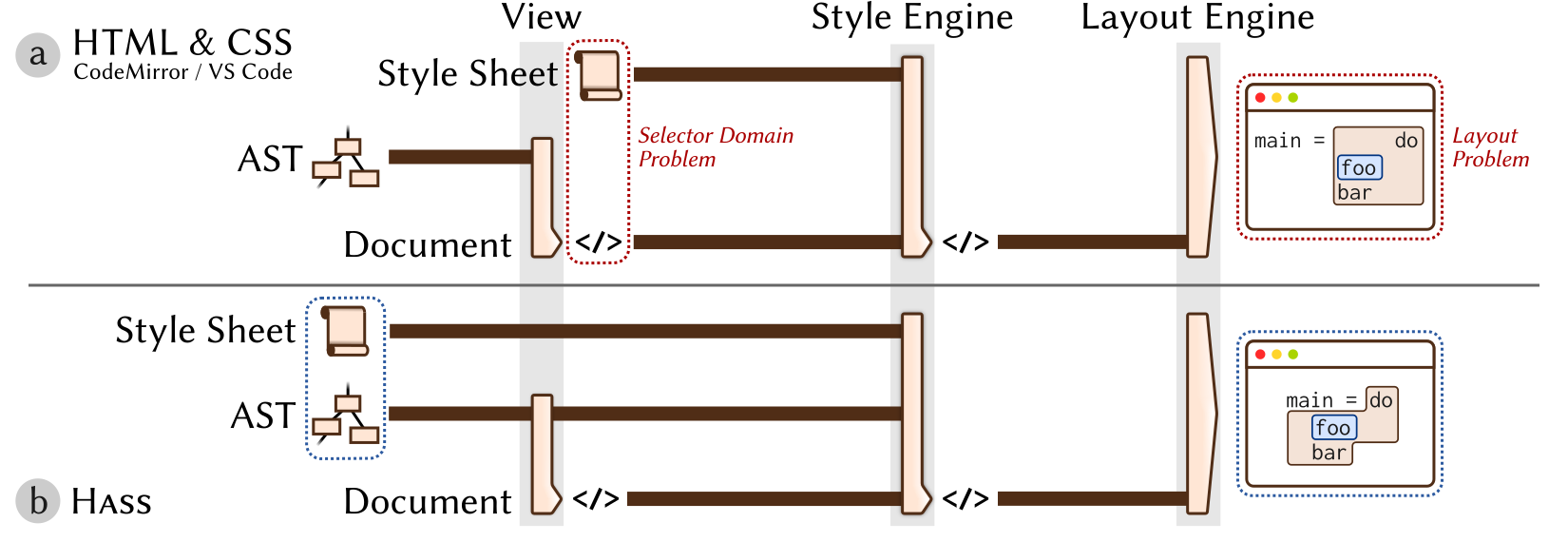}
  \caption{Analogy Diagram v3 (July 2024)}
\label{fig:architecture-v3}
\end{figure}

\clearpage
\parahead{``Code Style'' via Styles}

The term ``code style'' often refers to the structure and formatting of the source code itself.
Lubin and Chugh (2019) sketched a CSS-inspired idea (also called ``code style sheets'') whereby style attribute-value pairs could dictate how the code itself is structured.\footnote{Justin Lubin and Ravi Chugh. 2019.
Type-Directed Program Transformations for the Working Functional Programmer.
\emph{In Workshop on Evaluation and Usability of Programming Languages and Tools (PLATEAU).}
\href{https://doi.org/10.4230/OASIcs.PLATEAU.2019.3}{doi:10.4230/OASIcs.PLATEAU.2019.3}
}
As an example using \hass{} syntax, the rule \verb+tup@(TTuple _) -> tup { max-size: 3; }+ would restrict tuples to have at most three elements---beyond which, the type system might report a warning or the editor might offer a refactoring using records with named fields.
And the rule \verb+t@(TypeAlias _ tup@(TTuple _)) -> tup { newlines: per-component; }+ would force line breaks after every component in a tuple type definition.

In light of our design, such rules might be called ``code-style style sheets.''
Setting a code-style attribute would imply changes to the concrete syntax tree itself (not just the document for display).
Some of these choices might possibly be emulated by pre-populating a display with additional elements to be selectively hidden or displayed based on code-style attributes.
However, more fully and directly implementing code-style attributes a means for performing program transformations is an orthogonal and alluring idea for future work.
 
\clearpage
\section{Pasadena}

\begin{figure}[h]
\includegraphics[height=1.7in]{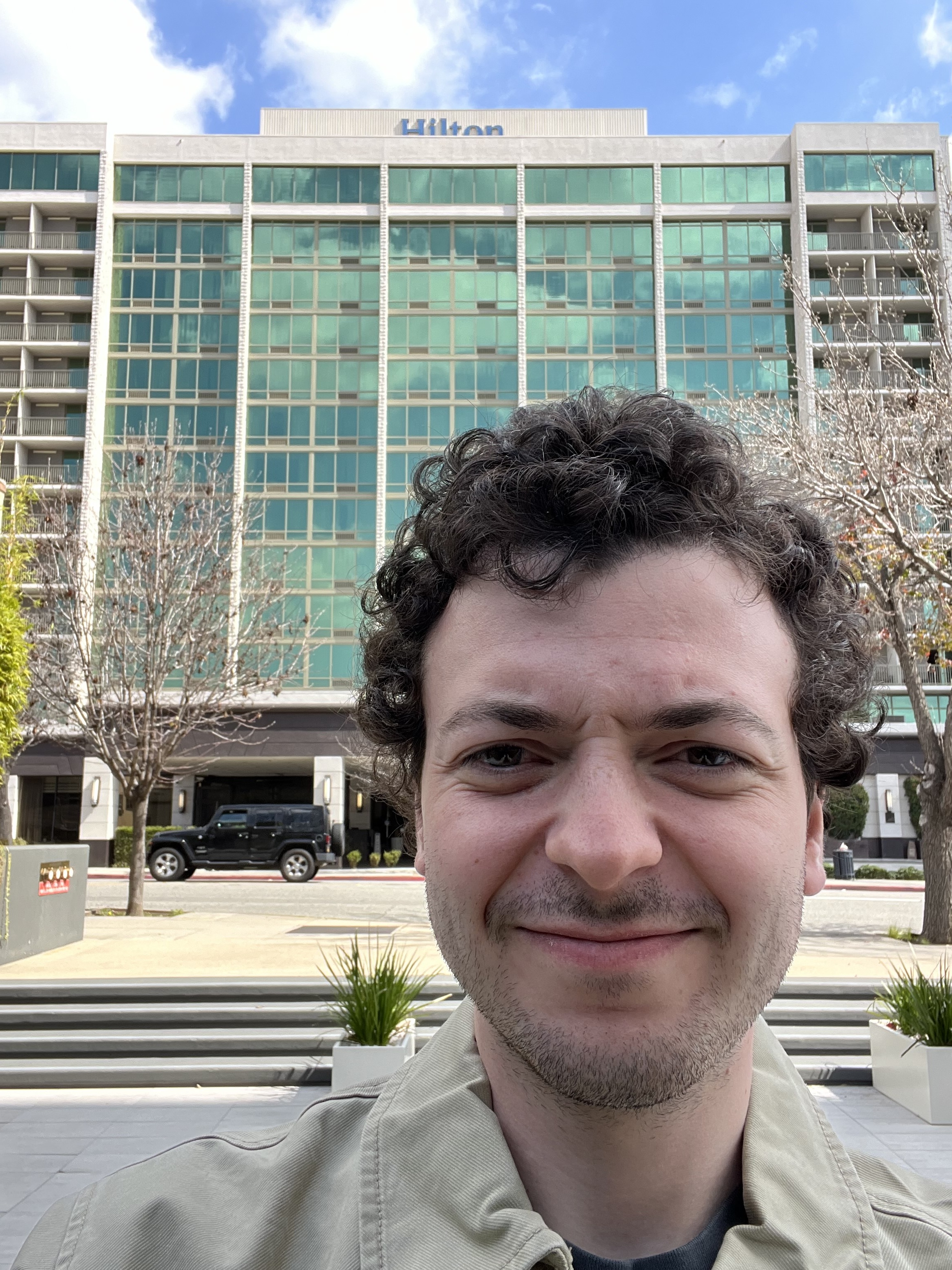}
\includegraphics[height=1.7in]{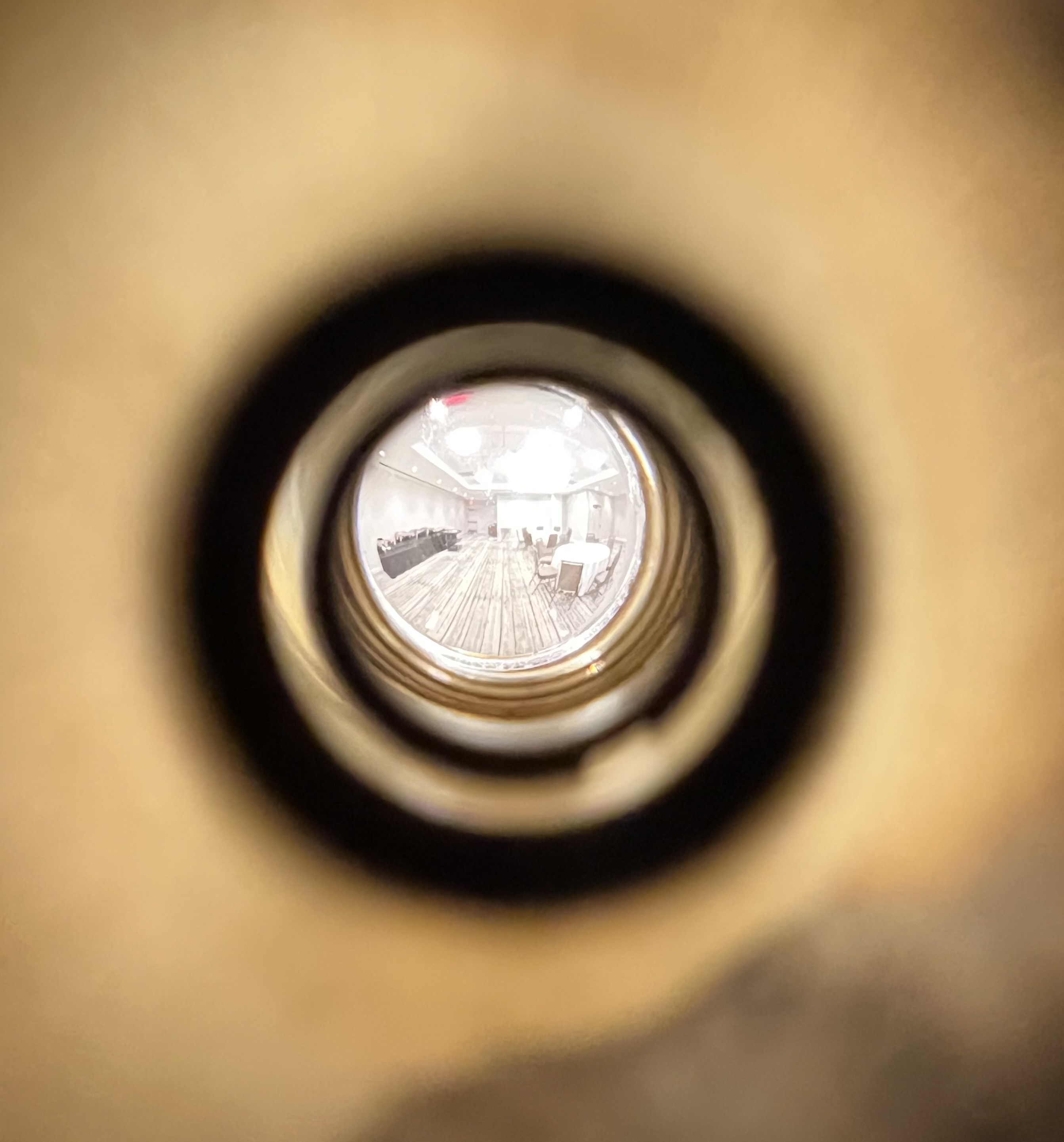}
\includegraphics[height=1.7in]{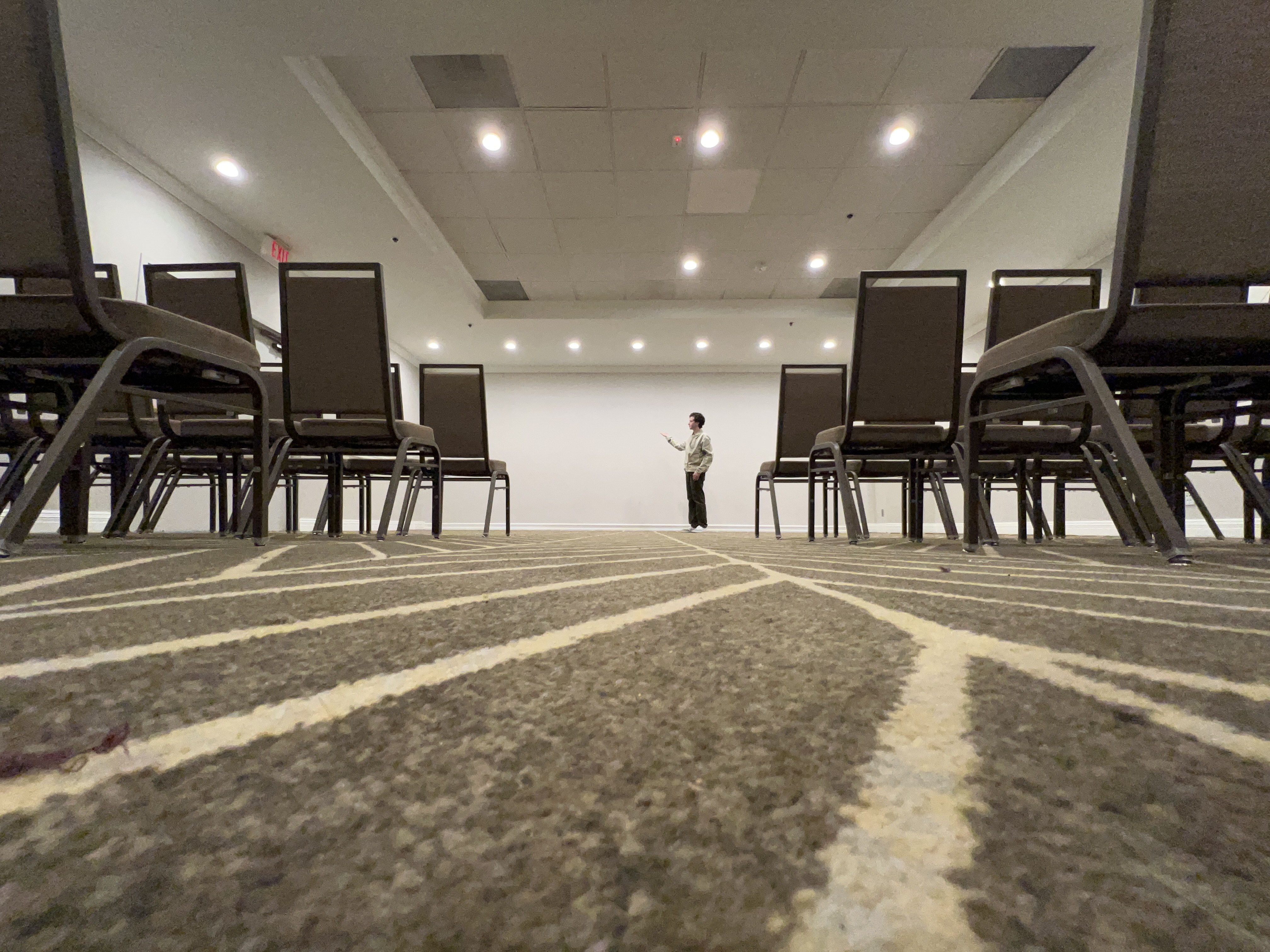}
\caption{Reconnaissance: Hilton Pasadena, March 11, 2024 (Pasadena, CA). Conjuring the whims and vagaries of an OOPSLA 2024 audience. Mission failed.}
\end{figure}
 \section{Gray, Gray, or Gray}

\begin{figure}[h]
\includegraphics[width=\textwidth]{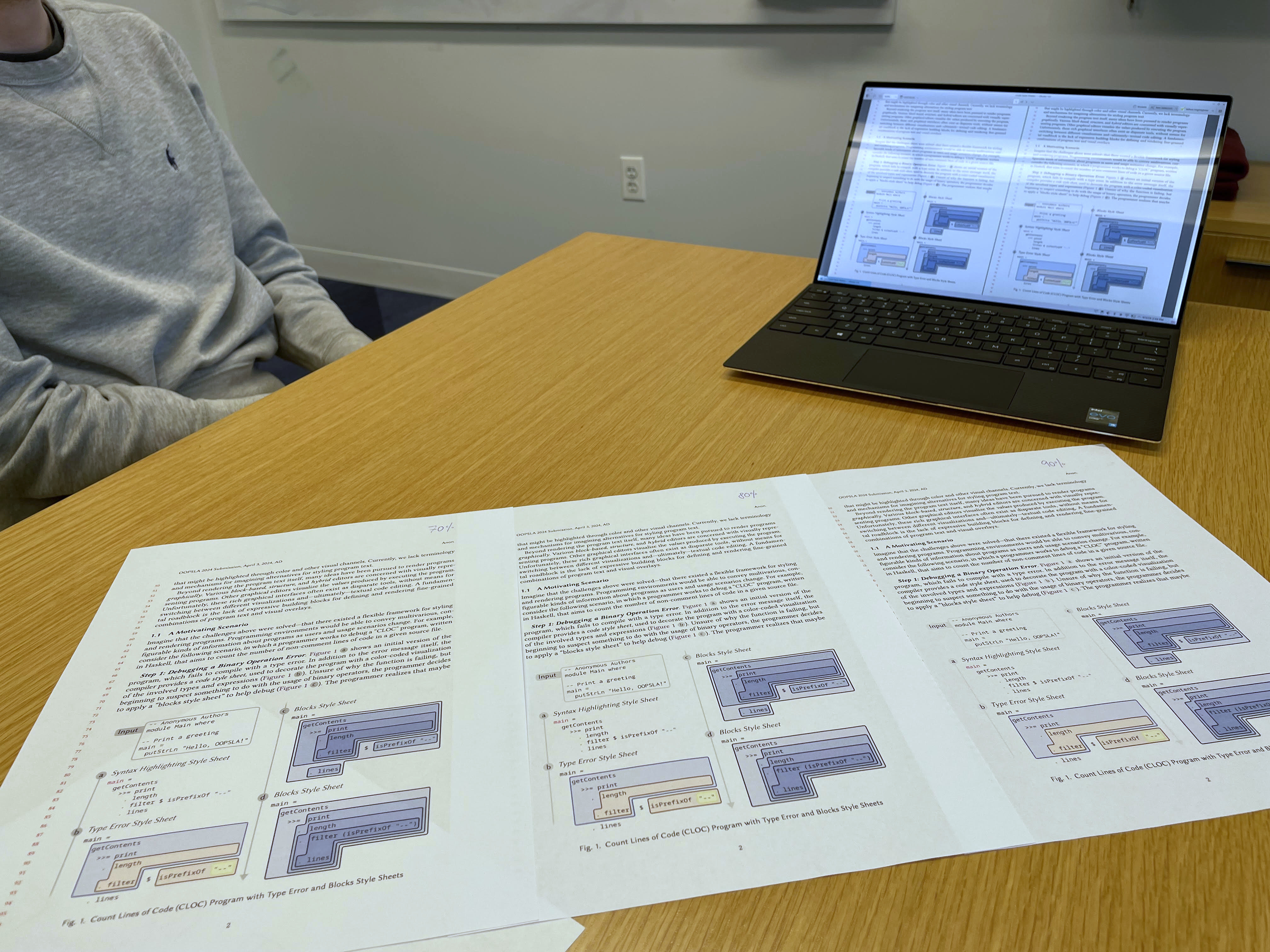}
\caption{Inside the \emph{Major Revisions} studio, comparing shades of gray, April 3, 2024 (Chicago, IL). For all graphic design services, major or minor, contact Sam and Ravi for a quote.}
\end{figure}

\end{document}